\documentclass{aa}

\usepackage{graphicx}
\usepackage{txfonts}
\usepackage{longtable}
\usepackage{siunitx}
\usepackage{float}

\def\swift{{\it Swift~}}

\begin{document} 

\title{A remarkable recurrent nova in M31: Discovery and optical/UV observations of the predicted 2014 eruption}

\author{M.~J. Darnley\inst{1}  \and M. Henze\inst{2} \and I.~A. Steele\inst{1} \and  M.~F. Bode\inst{1} \and V.~A.~R.~M. Ribeiro\inst{3} \and P. Rodr\'\i guez-Gil\inst{4,5} \and A.~W. Shafter\inst{6} \and \\ S.~C. Williams\inst{1} \and D. Baer\inst{6} \and I. Hachisu\inst{7} \and M. Hernanz\inst{8} \and K. Hornoch\inst{9} \and R. Hounsell\inst{10} \and M. Kato\inst{11} \and S. Kiyota\inst{12} \and \\ H. Ku\v{c}\'akov\'a\inst{13} \and H. Maehara\inst{14} \and J.-U. Ness\inst{2} \and A.~S. Piascik\inst{1} \and G. Sala\inst{15,16} \and I. Skillen\inst{17} \and R.~J. Smith\inst{1} \and M. Wolf\inst{13}}

\institute{Astrophysics Research Institute, Liverpool John Moores University, IC2 Liverpool Science Park, Liverpool, L3 5RF, UK \\
\email{M.J.Darnley@ljmu.ac.uk} \and 
European Space Astronomy Centre, P.O.\ Box 78, 28691 Villanueva de la Ca\~{n}ada, Madrid, Spain \and 
Department of Astrophysics/IMAPP, Radboud University, PO Box 9010, NL-6500 GL Nijmegen, The Netherlands \and
Instituto de Astrof\'\i sica de Canarias, V\'\i a L\'actea, s/n, La Laguna, E-38205, Santa Cruz de Tenerife, Spain \and
Departamento de Astrof\'\i sica, Universidad de La Laguna, La Laguna, E-38206, Santa Cruz de Tenerife, Spain \and
Department of Astronomy, San Diego State University, San Diego, CA 92182, USA \and
Department of Earth Science and Astronomy, College of Arts and Sciences, The University of Tokyo, 3-8-1 Komaba, Meguro-ku, Tokyo 153-8902, Japan \and 
Institut de Ci\`{e}ncies de l'Espai (CSIC-IEEC), Campus UAB, Fac. Ci\`{e}ncies, E-08193 Bellaterra, Spain \and 
Astronomical Institute, Academy of Sciences, CZ-251 65 Ond\v{r}ejov, Czech Republic \and 
Space Telescope Science Institute, 3700 San Martin Drive, Baltimore, MD, 21218, USA \and 
Department of Astronomy, Keio University, Hiyoshi, Yokohama 223-8521, Japan \and 
Variable Stars Observers League in Japan (VSOLJ), 7-1 Kitahatsutomi, Kamagaya 273-0126, Japan \and 
Astronomical Institute of the Charles University, Faculty of Mathemathics and Physics, V Hole\v{s}ovi\v{c}k\'ach 2, 180 00 Praha 8, Czech Republic \and
Okayama Astrophysical Observatory, NAOJ, NINS, 3037-5 Honjo, Kamogata, Asakuchi, Okayama 719-0232, Japan \and 
Departament de F\'\i sica i Enginyeria Nuclear, EUETIB, Universitat Polit\`{e}cnica de Catalunya, c/ Compte d'Urgell 187, 08036 Barcelona, Spain \and
Institut d'Estudis Espacials de Catalunya, c/ Gran Capit\`{a} 2-4, Ed. Nexus-201, 08034, Barcelona, Spain \and 
Isaac Newton Group of Telescopes, Apartado de correos 321, Santa Cruz de La Palma, E-38700, Spain
}

\date{Received 5 March 2015 / Accepted 1? June 2015}

\abstract{The Andromeda Galaxy recurrent nova M31N~2008-12a had been caught in eruption eight times.  The inter-eruption period of M31N~2008-12a is $\sim1$ year, making it the most rapidly recurring system known, and a strong single-degenerate Type Ia Supernova progenitor candidate.  Following the 2013 eruption, a campaign was initiated to detect the predicted 2014 eruption and to then perform high cadence optical photometric and spectroscopic monitoring using ground-based telescopes, along with rapid UV and X-ray follow-up with the \swift satellite.  Here we report the results of a high cadence multicolour optical monitoring campaign, the spectroscopic evolution, and the UV photometry.  We also discuss tantalising evidence of a potentially related, vastly-extended, nebulosity.  The 2014 eruption was discovered, before optical maximum, on October 2, 2014.  We find that the optical properties of M31N~2008-12a evolve faster than all Galactic recurrent novae known, and all its eruptions show remarkable similarity both photometrically and spectroscopically.  Optical spectra were obtained as early as 0.26 days post maximum, and again confirm the nova nature of the eruption.  A significant deceleration of the inferred ejecta expansion velocity is observed which may be caused by interaction of the ejecta with surrounding material, possibly a red giant wind.  We find a low ejected mass and low ejection velocity, which are consistent with high mass-accretion rate, high mass white dwarf, and short recurrence time models of novae.  We encourage additional observations, especially around the predicted time of the next eruption, towards the end of 2015.}

\keywords{Galaxies: individual: M31 -- novae, cataclysmic variables -- stars: individual: M31N 2008-12a}

\maketitle

\section{Introduction}

Classical and recurrent novae eruptions number among the most energetic stellar explosions, only Gamma Ray Bursts and Supernovae (SNe) are more luminous.  However, nova eruptions are thousands of times more frequent in a galaxy such as the Milky Way, with Galactic rate estimates of $\sim35$~year$^{-1}$ \citep{1997ApJ...487..226S,2006MNRAS.369..257D}.  Novae are a class of cataclysmic variable \citep{1964ApJ...139..457K} that are characterised by eruptions powered by a thermonuclear runaway on the surface of a white dwarf \citep[WD; the primary;][]{1976IAUS...73..155S}.  In these interacting binary systems the secondary star generally transfers mass to the WD via an accretion disk surrounding the WD.

A classical nova (CN) system typically contains a main sequence secondary \citep[MS-novae;][]{2012ApJ...746...61D}, at least in cases where the progenitor system has been recovered \citep{2012ApJ...746...61D,2014ApJ...788..164P} and theoretical studies predict a recurrence time in the range of a few $\times10^{3}-10^{6}$~years \citep[see][for recent reviews]{2008clno.book.....B,2010AN....331..160B,2014ASPC..490.....W}.  By definition, each CN system has only been observed in eruption once.

Recurrent nova (RN) eruptions share the characteristics of CNe; but have been observed in eruption more than once.  Examination of the Galactic RN systems highlights their differences from CNe.  The secondary star in a RN is typically more evolved, in the sub-giant (SG-nova; also the \object{U Scorpii} group) or red giant \citep[RG-nova; also the \object{RS Ophiuchi} group, see e.g.,][]{2008ASPC..401.....E} stage of evolution; this leads to a high mass accretion rate.  The mass of the WD in RN systems is generally higher, often close to the \citet[$M_{\mathrm{Ch}}$;][]{1931ApJ....74...81C} limit, allowing lower accumulated ignition mass.  The recurrence periods of RNe have been observed to lie between $1-100$ years \citep{2014A&A...563L...9D}, however both ends of this range are likely affected by selection effects.  \citet{2014ApJ...793..136K} predict the true lower recurrence limit to be $\sim2$~months (for a 1.38\,$M_{\sun}$\ WD with a mass accretion rate of $3.6\times10^{-7}$\,$M_{\sun}$\,yr$^{-1}$).

To-date, around 400 novae have been discovered in the Milky Way, yet there are only ten ($\sim2.5$~percent) confirmed Galactic RNe displaying multiple eruptions.  Following careful examination of the eruption properties of Galactic novae, \citet{2014ApJ...788..164P} recently predicted that $9-38$~percent of Galactic novae may have recurrence times $\la100$~years; the majority of which masquerade as CNe until a second eruption is uncovered \citep[as was achieved for the Galactic RN \object{V2487 Ophiuchi;}][]{2009AJ....138.1230P}.  

With almost 1\,000 discovered novae \citep[and on-line database\footnote{\url{http://www.mpe.mpg.de/~m31novae/opt/m31/index.php}}]{2007A&A...465..375P} and a nova eruption rate of $65^{+16}_{-15}$~yr$^{-1}$ \citep{2006MNRAS.369..257D}, the \object{Andromeda Galaxy} (\object{M31}) is the ideal laboratory in which to study a relatively un-biased population of novae.  While the distance to M31 once acted as a barrier, recent campaigns have been able to carry out detailed, multi-wavelength, studies of individual M31 novae \citep[e.g.][]{2009ApJ...690.1148S,2009ApJ...705.1056B,2011A&A...531A..22P}.  A thorough astrometric study of all M31 novae has uncovered 16 likely RNe in M31, and determined that as many as one in three M31 novae may be RNe with recurrence times $\le100$~years \citep{2015Sha}.  Similarly, \citet{2014ApJS..213...10W,2015Wil} found that $\sim40$~percent of M31 novae may contain evolved secondary stars; likely to be red giants (RG-novae).

The remarkable RN \object{M31N 2008-12a} was first discovered undergoing an optical eruption in 2008 Dec \citep{2008Nis}.  Subsequent eruptions were discovered in 2009 Dec \citep{2014ApJ...786...61T}, 2011 Oct \citep{2011Kor}, 2012 Oct \citep{2012Nis}, and 2013 Nov \citep{2013ATel.5607....1T}; strongly indicating a recurrence period of a year (see also Table~\ref{eruption_history}).  For comparison, the shortest recorded inter-eruption period for a Galactic RN was eight years between a pair of eruptions from \object{U Sco} \citep[1979 and 1987;][]{1979IAUC.3341....1B,1987IAUC.4395....1O}.  With the 2013 eruption of \object{M31N 2008-12a} somewhat anticipated, detailed optical and X-ray studies of this eruption were published in \citet[hereafter DWB14]{2014A&A...563L...9D} and \citet[hereafter HND14]{2014A&A...563L...8H}, respectively, with a complementary study published by \citet[hereafter TBW14]{2014ApJ...786...61T}.  Both HND14 and TBW14 connected prior X-ray detections to the nova, indicating additional eruptions in 1992, 1993, and 2001. These had previously been overlooked as no optical counterpart had been discovered at the time.  Archival {\it Hubble Space Telescope} ({\it HST}) observations of the region revealed the presence of the likely progenitor system, with the optical and UV data indicating the presence of a luminous accretion disk \citep[DWB14; TBW14]{2013ATel.5611....1W}.  The existing near-IR (NIR) {\it HST} data were not deep enough to directly confirm the nature of the secondary (either sub-giant or red giant), however, the high quiescent luminosity of the system strongly hints at a RG-nova classification (DWB14).

\begin{table*}
\caption{List of observed eruptions of \object{M31N 2008-12a}.\label{eruption_history}}
\begin{center}
\begin{tabular}{lllll}
\hline
\hline
$t_{\mathrm{max,\,optical}}$\tablefootmark{a} & $t_{\mathrm{max,\,X-ray}}$\tablefootmark{b} & Time since last & Source & References \\
(UT) & (UT) & eruption (days)\tablefootmark{c} \\
\hline
 & 1992 Feb 05 &  & X-ray ({\it ROSAT}) & 1 \\
\hline
 & 1993 Jan 11 & 341 & X-ray ({\it ROSAT}) & 1 \\
\hline
 & 2001 Sep 08 & & X-ray ({\it Chandra}) & 2 \\
\hline
2008 Dec 26 & & & Optical & 3 \\
\hline
2009 Dec 03 & & 342 & Optical (PTF) & 4 \\
\hline
2011 Oct 23.49 & & 689 ($\sim368/337\tablefootmark{\dag})$& Optical & 4, 5, 6, 7 \\
\hline
2012 Oct 19.72 & & 362.2 & Optical & 7, 8, 9 \\
& $<2012$ Nov 06.45 & & X-ray (\swift) & 10 \\
\hline
2013 Nov 28.60 &  & 405.1 & Optical (iPTF) & 4, 7, 11, 12 \\
& 2013 Dec $05.9\pm0.2$ & & X-ray (\swift) & 4, 10 \\
\hline
2014 Oct $03.7\pm0.1$ & & 309.1 & Optical (LT) & 7 \\
& 2014 Oct 13.6  & & X-ray (\swift) & 13 \\
\hline
\end{tabular}
\end{center}
\tablefoot{
An updated version of Table~1 from TBW14.  \tablefoottext{a}{Time of the optical peak.}\tablefoottext{b}{Time of the X-ray peak.}\tablefoottext{c}{Time since last eruption only quoted when consecutive detections are believed to relate to consecutive eruptions.}\tablefoottext{\dag}{Assuming a missed eruption on or around 2010 Oct 20 or 2010 Nov 20 (see \citet{2012ApJ...752..133C} and TBW14 for a detailed discussion about a potential 2010 eruption).}
}
\tablebib{
(1)~\citet{1995ApJ...445L.125W}, (2)~\citet{2004ApJ...609..735W}, (3)~\citet{2008Nis}, (4)~\citet{2014ApJ...786...61T}, (5)~\citet{2011Kor}, (6)~\citet{2011ATel.3725....1B}, (7)~this paper, (8)~\citet{2012Nis}, (9)~\citet{2012ATel.4503....1S}, (10)~\citet{2014A&A...563L...8H}, (11)~\citet{2013ATel.5607....1T}, (12)~\citet{2014A&A...563L...9D}, (13)~\citet{2015Hen}.
}
\end{table*}

\swift X-ray observations of \object{M31N 2008-12a} were first made around 20 days after the 2012 eruption, however no X-ray source was detected.  Following the 2013 eruption, \object{M31N 2008-12a} was clearly detected in the X-ray as a bright supersoft source (SSS) in the first \swift observation, just six days after discovery, indicating that the SSS `turn-on' had been missed.  The SSS-phase lasted for twelve days (turning off at the effective time of the first 2012 observation), with black-body fits to the X-ray spectra indicating a particularly hot source $\sim100$\,eV.  The X-ray emission exhibited significant variation that was anti-correlated with similar fluctuations in the UV (see HND14 and TBW14).  The X-ray properties of the eruption, the rapid turn-on and turn-off times, point towards a high-mass WD and a low ejected mass.  Modelling reported in TBW14 constrained $M_{\mathrm{WD}}>1.3\,M_{\sun}$ and the accretion rate $\dot{M}>1.7\times10^{-7}\,M_{\sun}\,\mathrm{yr}^{-1}$, they also predicted that if \object{M31N 2008-12a} retains 30~percent of the accumulated mass during each eruption then the WD mass would increase towards $M_{\mathrm{Ch}}$ in less than a Myr.  As such, assuming accretion at such a rate is viable for the long-term, the ultimate fate of \object{M31N 2008-12a} is dependant on the composition of the WD.  A O--Ne WD would be expected to undergo electron capture as it surpassed $M_{\mathrm{Ch}}$ \citep[see e.g.,][]{1996ApJ...459..701G}, with a C--O WD experiencing carbon deflagration and exploding as a SN Type Ia \citep[see e.g.,][]{1973ApJ...186.1007W,1999ApJ...519..314H,1999ApJ...522..487H}.

In this Paper we present the fruits of a successful campaign to detect as early as possible the predicted 2014 eruption of \object{M31N 2008-12a}, and the subsequent optical and UV photometric monitoring and optical spectroscopic observations.  In Sect.~\ref{sec:monitor_and_detect} we describe in detail the quiescent monitoring of the system and successful discovery of the 2014 eruption.  In Sect.~\ref{sec:observations} we present our optical photometric and spectroscopic, and UV data sets, and describe their analysis.  In Sect.~\ref{sec:results} we present the results obtained from that analysis, in Sect.~\ref{environment} we present an investigation into the environment of the nova, and we discuss our results in Sect.~\ref{sec:discussion}.  Finally, we present our conclusions, a summary, and a prediction in Sect.~\ref{sec:conclusions}.  The X-ray properties of the 2014 eruption, based on extensive \swift observations, are described in detail in \citep[hereafter HND15]{2015Hen}

\section{Quiescent Monitoring and Detection of the 2014 Eruption}\label{sec:monitor_and_detect}

Following the confirmation of both the recurrent and rapidly evolving nature of \object{M31N 2008-12a} after the 2013 eruption (DWB14, HND14, TBW14), a dedicated monitoring program was put into place to detect subsequent eruptions, including an eruption predicted for the end of 2014 (DWB14, HND14, TBW14).  Starting in late July 2014, an observing campaign began on the 2m, fully-robotic, Liverpool Telescope \citep[LT;][]{2004SPIE.5489..679S} on La Palma, Canary Islands.  The LT obtained nightly (weather permitting) observations centred at the position of \object{M31N 2008-12a} using the IO:O optical CCD camera (a $4\,096\times4\,112$ pixel e2v detector with a $10\arcmin\times10\arcmin$ field of view).  Each observation consisted of a single 60s exposure taken through a Sloan-like $r'$-band filter.  These data were automatically pre-processed by a pipeline running at the LT and were automatically downloaded, typically within minutes of the observation.  An automatic data analysis pipeline \citep[based on a real-time \object{M31} difference image analysis pipeline, see][]{2007ApJ...661L..45D,2010MNRAS.409..247K} then further processed the data and searched for transient objects in real-time.  Any object detected with significance $\ge5\sigma$ within one seeing-disk of the position of \object{M31N 2008-12a} would generate an automatic alert.

On 2014 October 2.904 UT the LT obtained the $51^{\mathrm{st}}$ observation in its nightly cadence monitoring program.  The automated pipeline reported a high significance detection of a new object at $\alpha=0^{\mathrm{h}}45^{\mathrm{m}}28\fs81$ $\delta=41\degr54\arcmin9\farcs9$ (J2000), with a separation of $0\farcs03\pm0\farcs07$ from the position of the 2013 eruption (DWB14).  Preliminary photometry (see Sect.~\ref{lt_photometry}) indicated that this object had a magnitude of $r' = 18.86 \pm 0.02$, roughly consistent with the peak brightness of previous eruptions of \object{M31N 2008-12a} (DWB14, TBW14).  A request for further observations was thus immediately released \citep{2014ATel.6527....1D,2014ATel.6535....1D}.

No object was present at this position in an LT observation obtained the previous day (2014 October 1.909 UT), down to a $5\sigma$ limiting magnitude of $r'>20.4$ \citep{2014ATel.6535....1D}.  An additional $5\sigma$ limiting magnitude of $g>19.5$ was provided by iPTF based on their observations taken on 2014 October 2.468 UT \citep{2014ATel.6532....1C}, $0\fd435$ before the LT detection.

\section{Observations of the 2014 Eruption}\label{sec:observations}

In this Section we describe the observing strategy and data analysis techniques employed for our optical/UV follow-up monitoring campaign for the 2014 eruption of \object{M31N 2008-12a}.

\subsection{Optical Photometry}\label{sec:optical_photometry}

\begin{figure*}
\centering\includegraphics[width=0.94\textwidth]{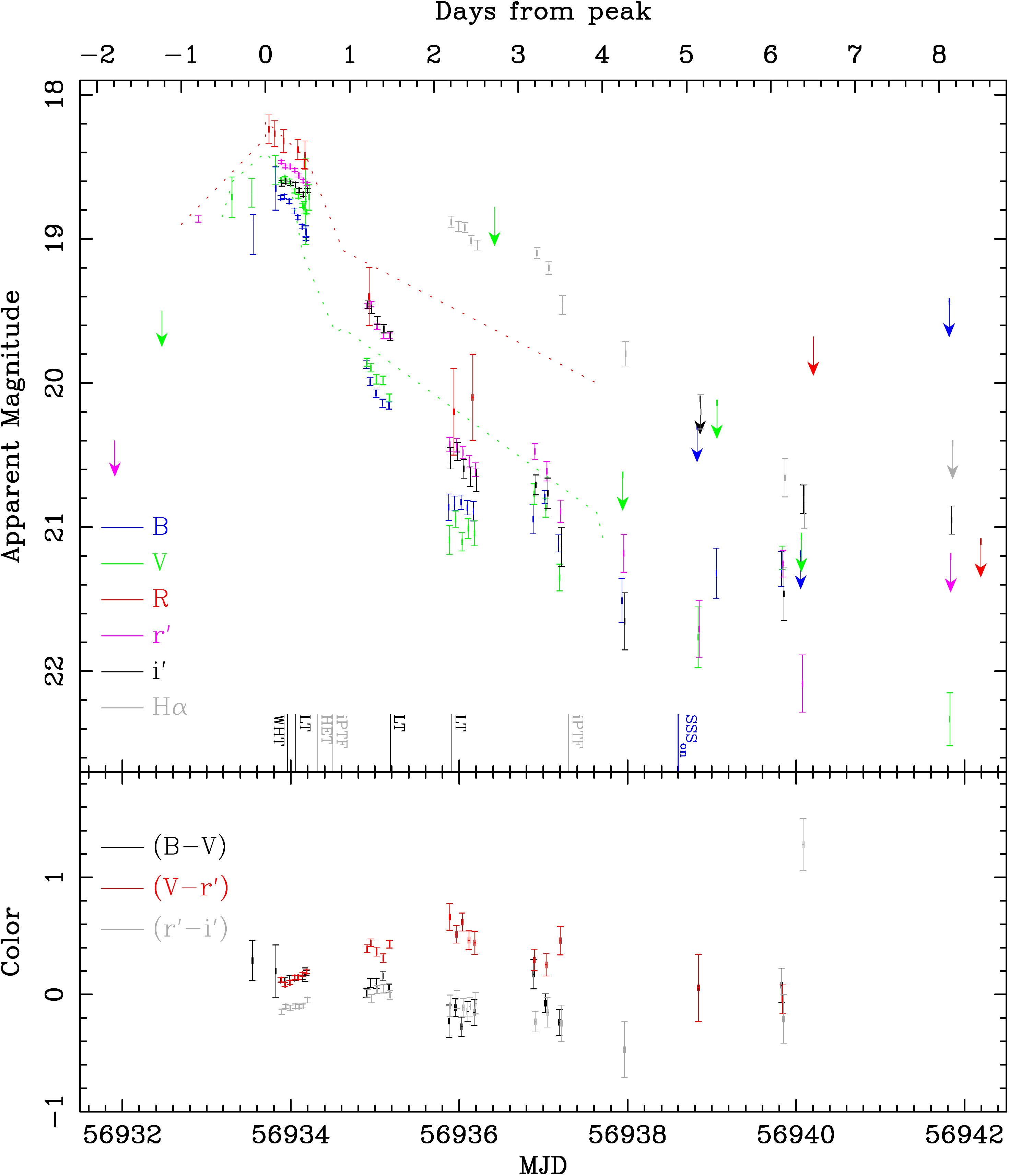}
\caption{Ground-based optical photometry of the 2014 eruption of \object{M31N 2008-12a}, all data are taken from Table~\ref{optical_photometry}.  Epochs of optical spectra from the 2014 eruption (black lines), 2012 eruption (grey line), and the SSS behaviour (blue lines)  are shown for informational purposes. \newline {\bf Top panel:} Optical light curve; the dotted lines indicate a template light curve based on the $V$- (green) and $R/r'$-band (red) observations of the 2008, 2011, 2012, and 2013 eruptions, and we assume $t_{\mathrm{max}}=56\,933.7$ (MJD) or 2014 Oct 3.7 UT (see discussion in Sect.~\ref{maximum_light} and DWB14).  \newline{\bf Bottom panel:} Colour evolution of the 2014 eruption of \object{M31N 2008-12a}.\label{op_lc}.}
\end{figure*}

The 2014 eruption of \object{M31N 2008-12a} was followed photometrically by an array of ground-based optical facilities.  These include the LT, the Mount Laguna Observatory (MLO) 1.0m telescope, the Ond\v{r}ejov Observatory 0.65m telescope, the Danish 1.54m telescope at La Silla, the Kiso Observatory 1.05m, and the iTelescope.net\footnote{\url{http://itelescope.net}} 20-inch telescope (T11) in Mayhill, New Mexico and 24-inch telescope (T24) at the Sierra Remote Observatory, Auberry, California.  The following text describes the data acquisition and reduction for each of these facilities.  The photometric data are presented in Table~\ref{optical_photometry}, and the resulting optical light curve is presented in Fig.~\ref{op_lc} (top panel).  Where near-simultaneous multicolour observations are available from the same facility, the colour data are presented in Table~\ref{colour_tab}, and the resulting colour evolution is shown in the bottom panel of Fig.~\ref{op_lc}.

\subsubsection{Liverpool Telescope Photometry}\label{lt_photometry}

A pre-planned broadband $B$, $V$, $r'$, and $i'$ photometry programme was initiated on the LT immediately following the LT detection of the 2014 eruption of \object{M31N 2008-12a}.  Once \object{M31N 2008-12a} had faded beyond the brightness limit for LT spectroscopic observations, the LT photometric observations were supplemented with narrowband ($\Delta\lambda\sim100$\AA) H$\alpha$ observations.

The LT observing strategy involved taking $3\times120$~s exposures through each of the filters per epoch.  The LT robotic scheduler was initially requested to repeat these observations with a minimum interval (between repeat observations) of 1~hour.  This minimum interval was increased to 1~day from the night beginning 2014 Oct 11 UT.  To counter the signal-to-noise losses as the nova faded, the exposure time was increased to $3\times300$~s from Oct 5.5 UT onwards.  The narrowband H$\alpha$ filter was added to the program from Oct 5.5 UT, and the observations through the broadband filters ceased from the night beginning Oct 11 UT.  LT observations following the 2014 eruption ended at Oct 27.0 UT, and a monitoring campaign for the next eruption immediately began.

The LT data were pre-processed at the telescope and then further processed using standard routines within Starlink \citep{1982QJRAS..23..485D} and IRAF \citep{1993ASPC...52..173T}.  Point-spread function (PSF) fitting was performed using the Starlink {\tt photom} (v1.12-2) package.  Photometric calibration was achieved using 17 stars from \citet{2006AJ....131.2478M} within the IO:O field (see Table~\ref{tab_calib}), transformations from \citet{2005AJ....130..873J} were used to convert these calibration stars from $BVRI$ to $BVr'i'$.  In all cases, uncertainties from the photometric calibration are not the dominant source of error.  The H$\alpha$ data were calibrated by assuming the average H$\alpha$ excess (above their $r'$-band emission) of the 17 calibration stars was $\simeq0$ and by calibrating these stars' H$\alpha$ photometry relative to their $r'$ emission.  Deviations from this assumption were $\le5$~percent.

\begin{table*}
\caption{Photometry calibration stars in the field of \object{M31N 2008-12a} employed with the LT and MLO observations.\label{tab_calib}}
\begin{center}
\begin{tabular}{llllllll|ll}
\hline\hline
\# & LGGS\tablefootmark{\dag} Designation & R.A.\ (J2000) & Dec.\ (J2000) & $B$ & $V$ & $R$ & $I$ & $r'$ & $i'$ \\
\hline
1 & J004511.73+415352.2 & $0^{\mathrm{h}}45^{\mathrm{m}}11\fs73$ & $+41\degr53\arcmin52\farcs2$ & 18.635 & 17.759 & 17.270 & 16.782 & 17.466 & 17.208 \\
2 & J004512.71+415448.5 & $0^{\mathrm{h}}45^{\mathrm{m}}12\fs71$ & $+41\degr54\arcmin48\farcs5$ & 17.711 & 16.873 & 16.423 & 16.010 & 16.598 & 16.413 \\
3 & J004514.35+415505.4 & $0^{\mathrm{h}}45^{\mathrm{m}}14\fs35$ & $+41\degr55\arcmin5\farcs4$ & 19.600 & 18.239 & 17.412 & 16.635 & 17.723 & 17.181 \\
4 & J004515.43+415406.9 & $0^{\mathrm{h}}45^{\mathrm{m}}15\fs43$ & $+41\degr54\arcmin6\farcs9$ & 18.963 & 18.319 & 17.953 & 17.566 & 18.133 & 17.974 \\
5 & J004518.25+415438.3 & $0^{\mathrm{h}}45^{\mathrm{m}}18\fs25$ & $+41\degr54\arcmin38\farcs3$ & 18.933 & 18.200 & 17.778 & 17.353 & 17.973 & 17.776 \\
6 & J004519.69+415605.9 & $0^{\mathrm{h}}45^{\mathrm{m}}19\fs69$ & $+41\degr56\arcmin5\farcs9$ & 17.740 & 17.068 & 16.680 & 16.290 & 16.869 & 16.707 \\
7 & J004522.59+415337.5 & $0^{\mathrm{h}}45^{\mathrm{m}}22\fs59$ & $+41\degr53\arcmin37\farcs5$ & 16.352 & 15.607 & 15.197 & \ldots & 15.374 & \ldots \\
8 & J004522.75+415506.6 & $0^{\mathrm{h}}45^{\mathrm{m}}22\fs75$ & $+41\degr55\arcmin6\farcs6$ & 20.532 & 19.087 & 18.183 & 17.233 & 18.532 & 17.821 \\
9 & J004525.24+415532.6 & $0^{\mathrm{h}}45^{\mathrm{m}}25\fs24$ & $+41\degr55\arcmin32\farcs6$ & 19.121 & 18.233 & 17.742 & 17.278 & 17.935 & 17.700 \\
10 & J004527.48+415530.4 & $0^{\mathrm{h}}45^{\mathrm{m}}27\fs48$ & $+41\degr55\arcmin30\farcs4$ & 17.606 & 16.785 & 16.331 & 15.911 & 16.517 & 16.326 \\
11 & J004528.55+415451.7 & $0^{\mathrm{h}}45^{\mathrm{m}}28\fs55$ & $+41\degr54\arcmin51\farcs7$ & 19.162 & 18.349 & 17.876 & 17.358 & 18.085 & 17.797 \\
12 & J004530.01+415320.9 & $0^{\mathrm{h}}45^{\mathrm{m}}30\fs01$ & $+41\degr53\arcmin20\farcs9$ & 17.991 & 16.945 & 16.318 & \ldots & 16.574 & \ldots \\
13 & J004530.20+415604.8 & $0^{\mathrm{h}}45^{\mathrm{m}}30\fs20$ & $+41\degr56\arcmin4\farcs8$ & 18.362 & 17.640 & 17.230 & 16.833 & 17.418 & 17.249 \\
14 & J004530.50+415511.9 & $0^{\mathrm{h}}45^{\mathrm{m}}30\fs50$ & $+41\degr55\arcmin11\farcs9$ & 15.410 & 14.738 & 14.367 & \ldots & 14.539 & \ldots \\
15 & J004534.14+415504.1 & $0^{\mathrm{h}}45^{\mathrm{m}}34\fs14$ & $+41\degr55\arcmin4\farcs1$ & 18.002 & 17.030 & 16.496 & 15.964 & 16.693 & 16.392 \\
16 & J004539.98+415532.0 & $0^{\mathrm{h}}45^{\mathrm{m}}39\fs98$ & $+41\degr55\arcmin32\farcs0$ & 17.452 & 16.416 & 15.810 & 15.300 & 16.049 & 15.770 \\
17 & J004546.80+415360.0 & $0^{\mathrm{h}}45^{\mathrm{m}}46\fs80$ & $+41\degr54\arcmin0\farcs0$ & 18.074 & 17.375 & 16.949 & 16.546 & 17.163 & 16.989 \\
\hline
 & J004532.50+415443.3 & $0^{\mathrm{h}}45^{\mathrm{m}}32\fs50$ & $+41\degr54\arcmin43\farcs3$ & \multicolumn{6}{c}{PSF Star} \\
\hline
\end{tabular}
\end{center}
\tablefoot{
\tablefoottext{\dag}{Local Group Galaxies Survey.}  Astrometry and $BVRI$ photometry from \citet{2006AJ....131.2478M}, Sloan $r'$ and $i'$ photometry computed via the transformations in \citet[see their Table~1]{2005AJ....130..873J}.
}
\end{table*}

\subsubsection{Mount Laguna Observatory 1.0m Photometry}\label{mlo_photometry}

Photometric observations of M31N 2008-12a were carried out on 2014 October 4 UT (approximately 1.3 days after the discovery of the most recent eruption) using the MLO 1 m reflector. A series of five 120s exposures were taken through each of the Johnson-Cousins $B$, $V$, $R$, $I$ filters \citep[see][]{1990PASP..102.1181B}, and imaged on a Loral $2\,048\times2\,048$ CCD.  The data were initially processed (bias subtracted and flat-fielded) using standard routines in the IRAF software package. The individual images for a given filter were subsequently aligned to a common coordinate system and averaged forming master $B$, $V$, $R$, and $I$-band images. Calibrated $B$, $V$, $R$, and $I$ magnitudes for \object{M31N 2008-12a} were then determined by comparing the instrumental magnitudes for the nova with those of five nearby secondary standard stars ($\#9-\#12$ and \#14; see Table~\ref{tab_calib}) using the IRAF {\tt apphot} package. The resulting magnitudes were reported in \citet{2014ATel.6543....1S} and are given in Table~\ref{optical_photometry}.

\subsubsection{Ond\v{r}ejov Observatory 0.65m and Danish 1.54m Photometry}

Photometric observations at Ond\v{r}ejov started around maximum brightness
of the 2014 eruption of the nova on 2014 October 3.738 UT.
We used the 0.65m telescope at the Ond\v{r}ejov Observatory (operated partly
by Charles University, Prague) equipped with a Moravian Instruments G2-3200 CCD
camera (using a Kodak KAF-3200ME sensor and standard $BVRI$ photometric
filters) mounted at the prime focus. Observations with the Danish 1.54m telescope
at La Silla were obtained remotely from Ond\v{r}ejov during two nights;
when the fading nova was fainter than the $R$-band limit of the DFOSC
instrument. The high airmass of the target observed from La Silla prevented
deeper observations, despite the good observing conditions.
For each epoch, a series of several 90s exposures was taken (see Table~\ref{optical_photometry} for total
exposure times for each epoch).
Standard reduction procedures for raw CCD images were applied (bias and
dark-frame subtraction and flat field correction) using APHOT\footnote{A synthetic
aperture photometry and astrometry software package developed by M.~Velen
and P.~Pravec at the Ond\v{r}ejov Observatory} \citep{1994ExA.....5..375P}.
Reduced images within the same series were co-added to improve the signal-to-noise
ratio and the gradient of the galaxy background was flattened using a spatial
median filter via the SIPS\footnote{\url{http://ccd.mii.cz}} program.
Photometric measurements of the nova were then performed using aperture
photometry in APHOT.  Five nearby secondary standard stars (including \#9 and \#11
listed in Table~\ref{tab_calib}) from \citet{2006AJ....131.2478M} were used to photometrically
calibrate the magnitudes presented in Table~\ref{optical_photometry}.

\subsubsection{iTelescope.net T11 and T24 Photometry}

Photometric observations of M31N 2008-12a were carried out remotely with iTelescope.net. Three 180s exposures were taken through a $V$-band filter using the T24 telescope (Planewave 24-inch CDK Telescope f/6.5 and a FLI PL-9000 CCD camera) at the hosting site in Sierra Remote Observatory (Auberry, CA. USA) on Oct. 03.299 UT. Five 180s exposures were taken through a $V$ filter with same instruments on Oct. 04.214 UT.   The iTelescope.net T11 telescope (Planewave 20-inch CDK telescope and FLI PL-11002M CCD camera) was also used on Oct. 04.777 UT at the New Mexico Skies hosting site near Mayhill, NM, USA.  Exposure times were $5\times180$\,s through a $V$-band filter.  These images were combined and measured with the photometry function of the MIRA pro x64 software\footnote{\url{http://www.mirametrics.com}}. Ten photometric reference stars were chosen from the Fourth U.S.\ Naval Observatory CCD Astrograph Catalog \citep[UCAC4;][]{2013AJ....145...44Z} catalog and used for photometry.

\subsubsection{Kiso Observatory 1.05m Photometry}

Photometry of \object{M31N 2008-12a} was performed on 2014 October 3 and 6 UT using the Kiso 1.05-m Schmidt
telescope, Nagano, Japan, and the KWFC instrument \citep{2012SPIE.8446E..6LS}. We obtained a series of $3\times60$\,s exposures in the $B$- and $V$-bands.
On October 3, we also obtained a single 300\,s exposure and a series of $5\times300$\,s exposures in the $U$-band.
Bias subtraction and flat-fielding were processed using standard routines within IRAF.  Astrometric
calibration and image co-addition were carried out using the SCAMP\footnote{\url{http://www.astromatic.net/software/scamp}} and SWarp\footnote{\url{http://www.astromatic.net/software/swarp}} software, respectively.
Photometry was performed using the aperture photometry routine within SExtractor \citep{1996A&AS..117..393B}.  With photometric calibration
performed by using a number nearby $15-17^{\mathrm{th}}$\,magnitude stars from the SDSS catalog.  We used the transformations 
from \citet{2005AJ....130..873J} to convert the magnitude system of these stars from $u'g'r'$ to $UBV$.

\subsection{Optical Spectroscopy}\label{optical_spectroscopy}

In addition to the photometric monitoring following the detection of the sixth eruption of \object{M31N 2008-12a} within seven years, spectroscopic observations were undertaken by the LT and William Herschel Telescope (WHT), co-sited on La Palma.  A single, early spectrum was obtained by the WHT (hereafter, first epoch).  The LT obtained daily spectra during the first three days post-discovery (second, third, and fourth epochs, respectively).  The following text describes the spectroscopic observations and data reduction, with the spectra themselves presented in Fig.~\ref{spec}.

\subsubsection{William Herschel Telescope ACAM Spectrum}\label{WHT_spectra}

The first epoch spectrum was obtained on 2014 Oct
3.96 UT (exposure mid point; $t-t_{\mathrm{max}}=0.26$~days) at the 4.2m William Herschel Telescope
(WHT) on La Palma.  We used the Auxiliary Port Camera (ACAM) in its
spectroscopy mode with the V400 VPH grism and the $1\farcs5$ slit
oriented at the parallactic angle.  The spectrum was recorded by the
$2\,048 \times 4\,200$ pixel EEV CCD camera under clear conditions with
$0\farcs8$ seeing and an exposure time of 300\,s.  A spectrum of a CuNe arc lamp was taken for
wavelength calibration in the afternoon.  The ACAM spectroscopic setup
provided a spectral resolution of $\sim 20$\AA\ (full-width at
half-maximum, FWHM) in the central part of the
$3\,820-9\,495$\AA\ useful wavelength interval.

We applied standard reduction techniques to the WHT/ACAM spectroscopic
data using IRAF tools.  After bias subtraction and
flat-fielding, we removed the sky background and then extracted the 1D
spectrum with the optimal extraction technique of \citet{1986PASP...98..609H} as
implemented in the {\tt pamela} package. For the wavelength solution, we
fitted a 4$^{\mathrm{th}}$-order polynomial to the pixel-wavelength arc data. The resulting
wavelength scale was corrected for shifts due to instrument flexure
using the \ion{O}{i} (6\,300\AA) night-sky emission line. These
wavelength-related procedures were carried out within the \texttt{molly}
package\footnote{\url{http://deneb.astro.warwick.ac.uk/phsaap/software/
molly/html/INDEX.html}}.  The WHT/ACAM spectrum was later flux calibrated relative to the almost simultaneous LT SPRAT spectrum (second spectral epoch).   Hence, we estimate a total flux uncertainty of $\sim 20$~percent, with an unknown (but likely small) systematic error due to the time difference between these spectra.

\subsubsection{Liverpool Telescope SPRAT Spectra}\label{LT_spectra}

Three epochs of spectroscopy of \object{M31N 2008-12a} were obtained on successive nights from 2014 Oct 3--5 using
the SPRAT spectrograph \citep{2014SPIE.9147E..8HP} in the blue optimised mode on the LT.  A slit width of $1\farcs8$ was used, yielding a spectral resolution of $\sim20${\AA}, and a velocity
resolution of $\sim1\,000$~km~s$^{-1}$ at the central wavelength of 5\,850{\AA}.

These LT spectra represent some of the first fully robotically acquired spectra taken by the newly mounted SPRAT spectrograph during its science commissioning program.

A total of 14 exposures each of duration 600 seconds, were obtained between 2014 Oct 3.95 UT and Oct 4.21
(mean MJD 56\,934.06; $t-t_{\mathrm{max}}=0.32$~days; second epoch).  Four exposures each of duration 1\,200 seconds were obtained
between Oct 5.13 UT and 5.23 (mean MJD 56\,935.18; $t-t_{\mathrm{max}}=1.44$~days; third epoch).  A single 1\,200 second
exposure was obtained at 2014 Oct 5.91 UT ($t-t_{\mathrm{max}}=2.17$~days; fourth epoch).  All of the nights were photometric.

Following bias subtraction, flat fielding, and cosmic ray removal, data reduction was carried out using the Starlink {\tt figaro} \citep[v5.6-6;][]{1988igbo.conf..448C} package.  Sky subtraction was accomplished in the 2D images via a linear fit of the variation of the sky emission in the spatial direction (parallel to the slit).  Following this, a simple extraction of the spectra was carried out.  No trace of residual sky emission could be detected in the extracted spectra. The extracted spectra were then wavelength calibrated using observations of a Xenon arc lamp obtained directly after each exposure, with an rms residual of $\sim1${\AA}.  Following wavelength calibration the spectra were re-binned to a uniform wavelength scale of 6.46{\AA} per pixel between 4\,200 and 7\,500{\AA}.
All of the spectra from a single night were then co-added.

Since no flux
standards were observed on the nights in question, the co-added spectra were
flux calibrated using observations of the spectrophotometric standard \object{BD+28 4211} \citep{1977ApJ...218..767S}
obtained on the photometric night of 2014 Sep 4 (with the same spectrograph configuration and slit width).
The spectra are therefore presented in units of $F_\nu$ (mJy).  Comparison of imaging observations
between the calibration night and the three LT spectra show zero-point differences of $<0.1$ magnitude (i.e.\ $<10$~percent). The greatest uncertainties in the flux calibration will therefore be due to slit losses caused by seeing variations and
misalignment of the object with the slit.  We measure this from our repeated observations of the source
on the same night to be $\sim15$~percent.  We therefore estimate a total flux uncertainty of $\sim 20$~percent.

\subsection{{\em Swift} UV Photometry}\label{uv_photometry}

\begin{figure*}
\centering\includegraphics[width=0.94\textwidth]{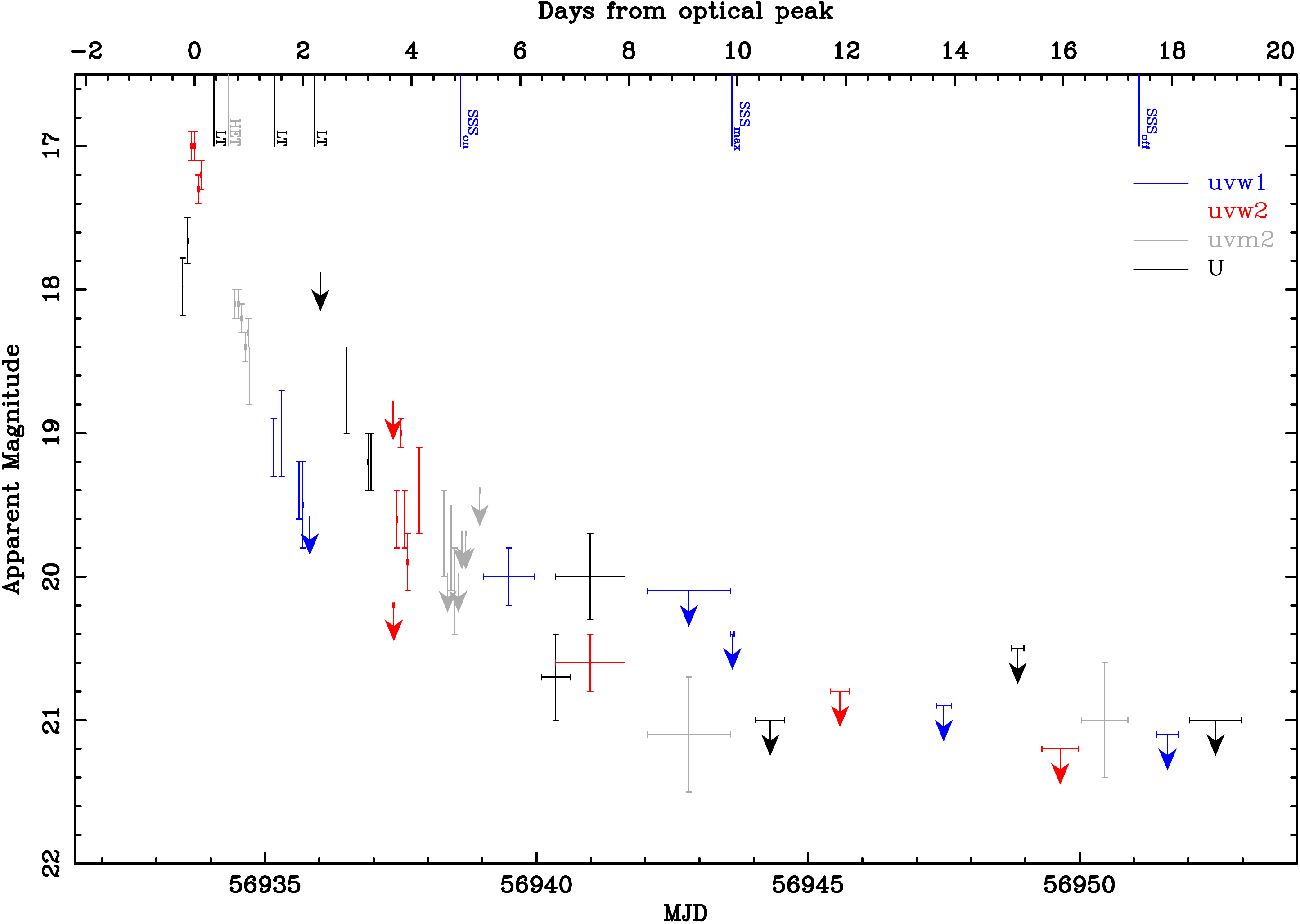}
\caption{\swift UV light curve of the 2014 eruption of \object{M31N 2008-12a}, all data are taken from Table~\ref{uv_tab}.  Here we assume optical maximum light occurred at 2014 Oct 3.7 UT (see discussion in Sect.~\ref{maximum_light}).  Epochs of optical spectra from the 2014 eruption (black lines), 2012 eruption (grey line), and the SSS behaviour (blue lines) are shown for informational purposes.\label{uv_lc}}
\end{figure*}

Immediately after the discovery of the 2014 eruption of M31N 2008-12a we requested a target of opportunity (ToO) program for the daily monitoring of the UV and X-ray emission with the \swift satellite \citep{2004ApJ...611.1005G}. The X-ray data analysis and results are described in detail in HND15. Here, we analysed the \swift UV/optical telescope \citep[UVOT;][]{2005SSRv..120...95R} data and estimated the magnitude of the nova using the \texttt{uvotsource} tool, which performs aperture photometry. We carefully selected source and background regions to minimise any contamination by the irregular background light within the spiral arm. However, we cannot exclude the possibility that unresolved emission in our source region did cause the relatively bright, late upper limits in Fig.~\ref{uv_lc} or the apparent detection in the uvm2 filter around day 17 after maximum. All magnitudes assume the UVOT photometric system \citep{2008MNRAS.383..627P} and have not been corrected for extinction. The UVOT upper limits correspond to 3$\sigma$ confidence.  As \swift is in a low-Earth orbit, a typical observation is a series of `snapshots', while \object{M31N 2008-12a} was bright enough to be detected in single exposures we report the snapshot photometry, otherwise we report the photometry from the combined exposures. The \swift UVOT photometry is presented in Table~\ref{uv_tab} and the UV light curve is shown in Fig.~\ref{uv_lc}.

\section{Results}\label{sec:results}

\subsection{Maximum Light and Template Light Curves}\label{maximum_light}

From inspection of the 2014 optical light curve presented in Fig.~\ref{op_lc}, it appears that the optical maximum of this eruption may have been caught, or at worst marginally missed; the peak magnitudes reached in previous eruptions were $V=18.4$ and $R=18.18$ (see discussion in DWB14), compared to the $V_{\mathrm{max}}=18.5\pm0.1$ and $R_{\mathrm{max}}=18.2\pm0.1$ observed for this eruption \citep[see Table~\ref{optical_photometry} and][]{2014ATel.6546....1H}. As such we assume that maximum light for the 2014 eruption occurred at, or shortly before 2014 Oct $3.7\pm0.1$ UT, and no earlier than Oct 3.3 UT.  For the analysis in this Paper we assume a MJD of maximum light of $t_{\mathrm{max}}=56\,933.7(\pm0.1)$. Note, HND15 define their reference date as the time of eruption, which they assume occurred almost exactly one day before the optical peak (MJD $56\,932.69\pm0.21$\,d). 

In DWB14, a pair of rough `template' $V$- and $R/r'$-band light curves were produced by combining the limited photometric data then available from the 2008, 2011, 2012, and 2013 eruptions.  This approach made use of the supposition by \citet{2010ApJS..187..275S} that all eruptions of a given RN are essentially identical.  Given the large uncertainties in some of the data from these previous eruptions and the significantly improved temporal coverage obtained for the 2014 eruption, these original template light curves do not map well on to the 2014 data.   Hence, we have updated the template light curves initially presented in DWB14 by assuming $t_{\mathrm{max}}$ of 2011 Oct 23.49 UT, 2012 Oct 19.72 UT, and 2013 Oct Nov 28.60 UT for the last three eruptions, these updated templates are shown by the dotted lines in Fig.\ref{op_lc}.  The updated templates (based still only on data from previous eruptions) match the $V$- and $R$-band behaviour seen before and immediately following maximum light in the 2014 eruption very well.  They only begin to deviate after $\sim1$~day post-maximum due to the limited, and uncertain, data available at such times from earlier eruptions.  However, based on data from previous eruptions, we can state that the light curve of the 2014 eruption is indeed remarkably similar to those before it.

\subsection{Light Curve Morphology}
	
The rise from detection ($\la1$~magnitude below peak) to maximum light took just under 1~day; for a nova whose evolution post maximum is so fast, this can be considered a slow rise to peak compared to the majority of Galactic nova light curves \citep[see, for e.g., the Galactic RN light curves presented by][]{2010ApJS..187..275S}.  The decline from maximum then proceeds rapidly, and the light curve decays approximately linearly (that is an exponential decline in the luminosity; as was also noted by TBW14 for previous eruptions) through $BVRr'i'$ filters until around $2-2.5$~days post-maximum.  At this stage the light curve enters a short-lived apparent `plateau' around 2~magnitudes below peak  for approximately 1~day; indeed there may also be a slight re-brightening at this stage, particularly in the $V$-band.  Light curve plateaus have been proposed as an identifying feature of RNe \citep{2008ASPC..401..206H}, \citet{2010ApJS..187..275S} noted that the majority of Galactic RN light curves exhibit plateaus (see also HND15), and \citet{2014ApJ...788..164P} used plateaus as one of a collection of features to identify RNe.  Following this plateau, the light curve again took on an approximately linear decline (at least in the $V$ and $r'$-bands), until varying seeing conditions and the unresolved M31 surface brightness rendered the eruption undetectable (last detected in the $i'$-band 8.148~days post-maximum).  Between days 5 and 7, the $BVr'i'$ emission re-brightened slightly and remained approximately constant for around 1~day, before declining below the detection threshold; the $i'$-band emission remained at this elevated level until at least day~8.  The timing of these `re-brightenings' corresponds to the emergence of the SSS (see HND15) and may be related.  Despite the high cadence nature of these observations, at times significant variation was occurring that may not have been well sampled by our observations.

The narrow-band H$\alpha$ photometric light curve (for which observations commenced around day~2 post-maximum, largely mirrors the $r'$-band light curve, as would be expected if the $r'$-band emission was dominated by the H$\alpha$ line rather than continuum emission, as is indicated in e.g., Fig.~\ref{spec}.  Novae routinely remain bright and detectable in narrow-band H$\alpha$ imaging after they become undetectable through broad-band filters \citep[this has driven a number of extragalactic nova surveys, see e.g.,][]{1987ApJ...318..520C}.  However, in this case the H$\alpha$ emission from \object{M31N 2008-12a} was undetectable $\sim6$~days post maximum.  H$\alpha$ observations continued for just over a month post-eruption but no source was detected at that position (typical $5\sigma$ limiting magnitude of $\sim20$).

Interpretation of the UV light curve is complicated by the large variation in filters (these observations typically used the {\it Swift}/UVOT `filter of the day'); however the general trend closely mirrors that of the optical light curve.  Through a combination of the ground based $U$-band observations and the {\it Swift}/UVOT uvw2 data the time of peak in the UV light curve closely matches that in the optical.  Both the UV light curve and the optical light curve show a decline of $\sim4$ magnitudes within the first eight days post-maximum.  While the sampling of the UV data is too sparse (within a given filter) to see any evidence of a day~$2-2.5$ `plateau', there is tentative evidence (in the $U$-band) for a similar `re-brightening' $\sim6-7$ days post-maximum.  \object{M31N 2008-12a} was undetectable from day~8 post maximum, however there is a {\it Swift}/UVOT uvm2 detection around day~10 (around the time of the maximum SSS emission) and a second at approximately day~16 (just around the SSS turn-off time).  We speculate that this late-time UV emission could be the tail of the SSS emission (see Sect.~4.4 of HND15 for a discussion).

\subsection{Spectroscopic Results}\label{spec_results}

\begin{figure*}
\begin{minipage}{\textwidth}
\includegraphics[width=\textwidth]{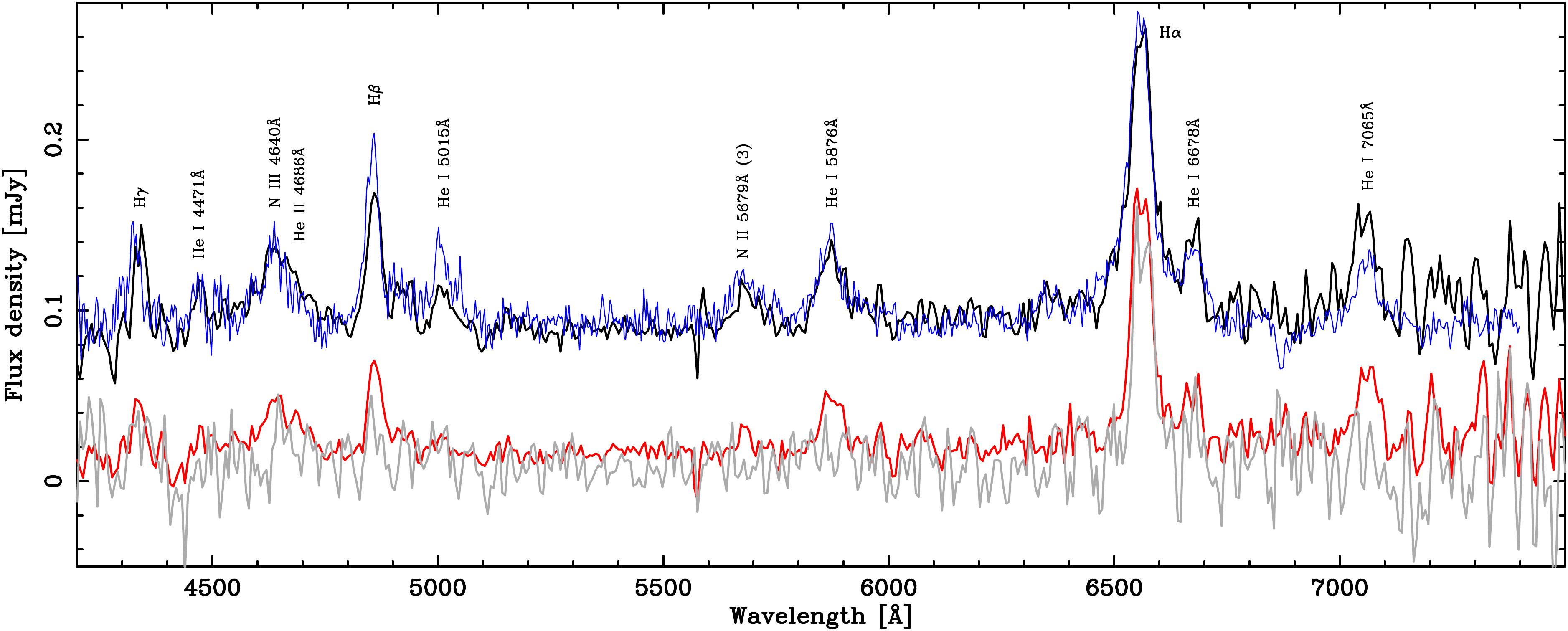}
\caption{Liverpool Telescope SPRAT and William Herschel Telescope ACAM flux calibrated spectra of the 2014 eruption of \object{M31N 2008-12a}.  Blue line: (WHT) taken at 2014 Oct 3.96 UT, $t-t_{\mathrm{max}}=0.26$~days (first epoch).   Black line: (LT) mean time of 2014 Oct 4.06 UT, $t-t_{\mathrm{max}}=0.32\pm0.14$~days (second epoch).  Red line: (LT) mean time of 2014 Oct 5.18 UT, $t-t_{\mathrm{max}}=1.44\pm0.18$~days (third epoch).  Grey line: (LT) taken at 2014 Oct 5.91 UT, $t-t_{\mathrm{max}}=2.17\pm0.01$~days (fourth epoch).\label{spec}}
\end{minipage}
\begin{minipage}[t]{0.49\textwidth}
\includegraphics[width=\textwidth]{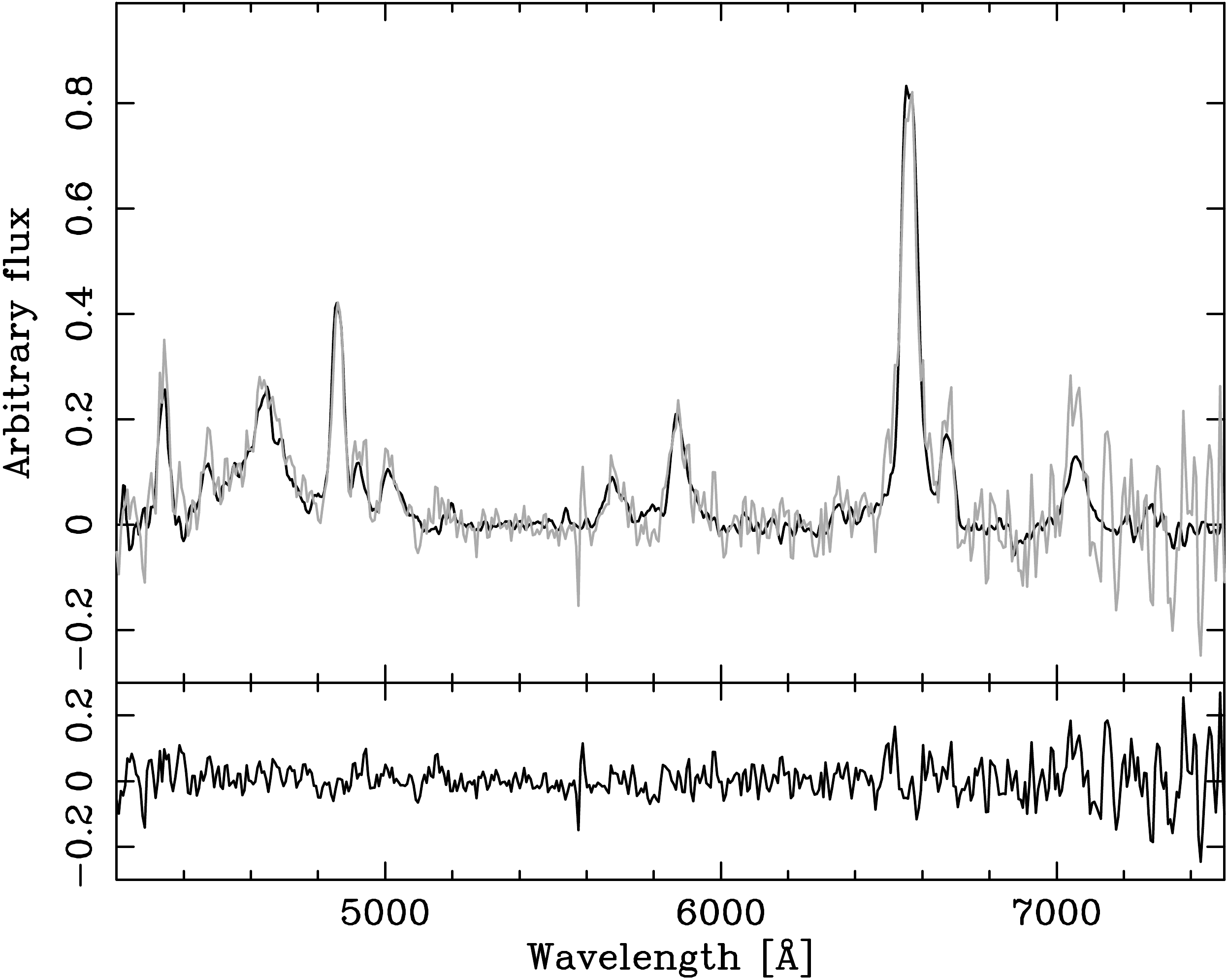}
\caption{Direct comparison between the HET spectrum of the 2012 eruption of \object{M31N 2008-12a} ($t-t_{\mathrm{max}}=0.62\pm0.1$~days; black line; DWB14) and the second epoch spectrum of the 2014 eruption (LT; $t-t_{\mathrm{max}}=0.26$~days; grey line).  The bottom panel shows the residuals following subtraction of the 2012 spectrum from the 2014 spectrum.\label{spec_diff}}
\end{minipage}\hfill
\begin{minipage}[t]{0.49\textwidth}
\includegraphics[width=\textwidth]{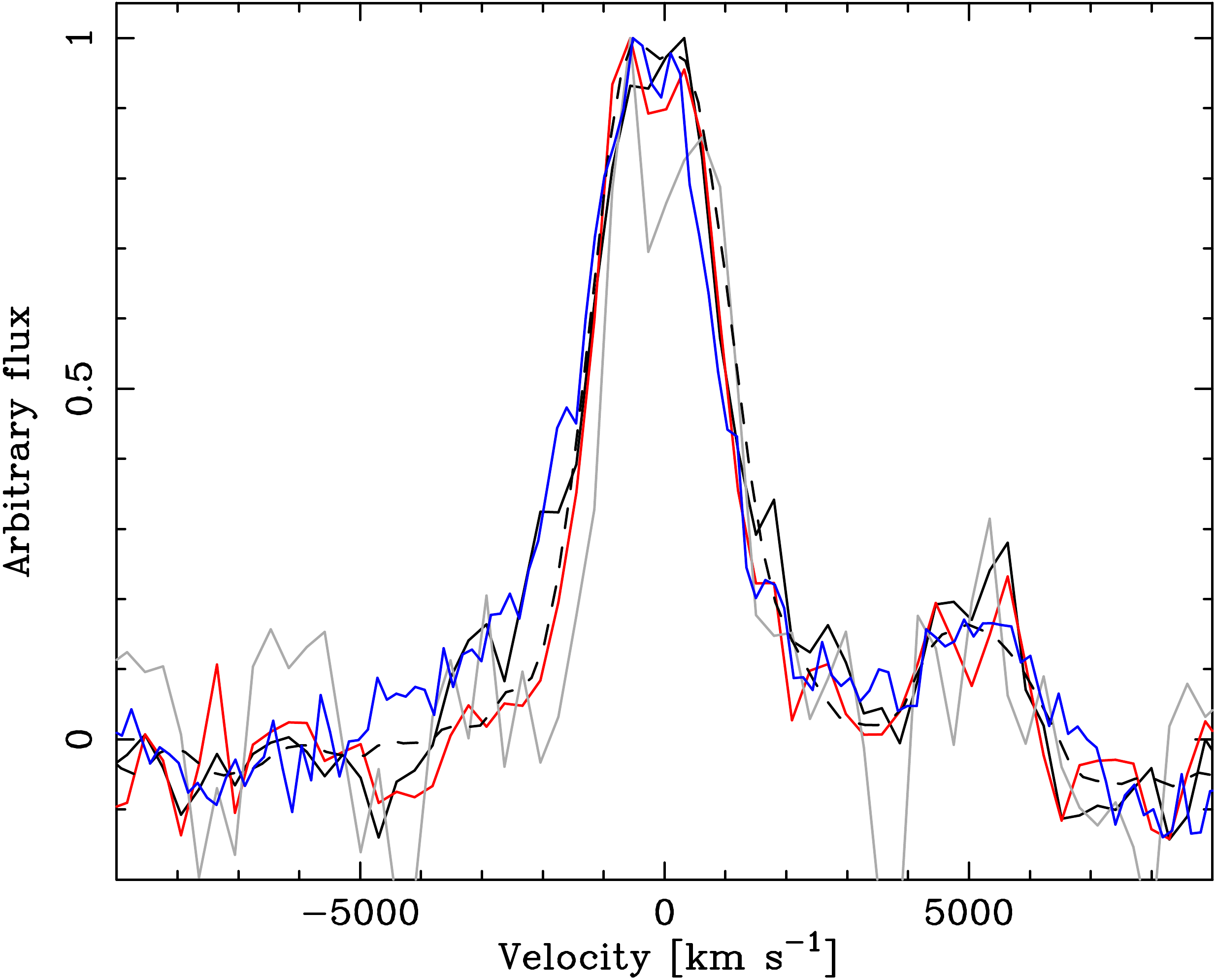}
\caption{Evolution of the H$\alpha$ line profile following the 2014 eruption of \object{M31N 2008-12a}.  Line colors as Fig.~\ref{spec}, dashed-black line indicates the H$\alpha$ profile from the HET spectra following the 2012 eruption ($t-t_{\mathrm{max}}=0.62\pm0.1$~days; see Fig.~\ref{spec_diff}).  The peak redwards of H$\alpha$ is the \ion{He}{i} (6\,678\AA) emission.\label{half}}
\end{minipage}
\end{figure*}

The initial results from the analysis of the second epoch spectrum were reported in \citet{2014ATel.6540....1D}.  The spectra of the 2014 eruption (see Fig.~\ref{spec}) are dominated by the Hydrogen Balmer emission lines (primarily H$\alpha$, but also H$\beta$ and H$\gamma$).  The following emission lines are also clearly present in the first three epochs, \ion{He}{i} (4\,471, 5\,017, 5\,876, 6\,678, and 7\,065\AA), \ion{He}{ii} (4\,686\AA), \ion{N}{ii} (5\,001 and 5\,680\AA), and \ion{N}{iii} (4\,641\AA), due to the significantly decreased exposure time, these lines are not visible in the fourth epoch.  No convincing evidence of \ion{Fe}{ii} emission lines or P~Cygni profiles are seen in these spectra.  The flux of each emission line was calculated by fitting a simple Gaussian profile using the IRAF {\tt fitprofs} command.  The fitted fluxes of the Balmer and redwards \ion{He}{i} lines are shown in Table~\ref{line_list}; here the errors were computed by repeated fitting using different continuum determinations.   The remaining emission lines show complex structures or blends, and typically low signal-to-noise.  As can be seen is Table~\ref{line_list}, there is a significant decrease in the flux from all the emission lines  between the second and third epochs ($t=0.32$\,d and 1.44\,d).  If the line fluxes had remained at the levels seen in these epochs, the \ion{He}{i}, \ion{He}{ii}, \ion{N}{ii}, and \ion{N}{iii} lines could have been detected in the fourth epoch spectrum ($t=2.17$\,d), despite the decreased sensitivity.  The first three spectra are consistent with the eruption of a nova following the initial optically thick fireball phase, and are similar to typical `He/N' spectra previously observed for M31 novae \cite[see the extensive catalogue of][]{2011ApJ...734...12S}.  The morphology of the H$\alpha$ emission line is shown in Fig.~\ref{half}.  There is a well defined central profile but with a clear double peaked structure and some evidence of higher velocity material beyond the central profile.  The morphology is generally consistent with the typically `rectangular' profiles seen in He/N spectra \citep[see e.g.,][]{2012AJ....144...98W}.

\begin{table}
\caption{Selected observed emission lines and fluxes from the three epochs of Liverpool Telescope SPRAT spectra of the 2014 eruption of \object{M31N 2008-12a}.\label{line_list}}
\begin{center}
\begin{tabular}{llll}
\hline\hline
Emission line & \multicolumn{3}{c}{Flux\tablefootmark{a} ($\times10^{-15}$\,W\,m$^{-2}$)} \\
 & $t=0.32$\,d & $t=1.44$\,d & $t=2.17$\,d \\
\hline
H$\alpha$ & $11.4\pm0.7$ & $8.4\pm0.4$ & $7.5\pm0.8$ \\
H$\beta$ & $3.1\pm0.1$ & $2.2\pm0.2$ & $0.6\pm0.3$\\
H$\gamma$ & $2.0\pm0.4$ & $1.5\pm0.2$ & $0.5\pm1.0$ \\
\ion{He}{i} (7\,065\AA) & $3.0\pm0.4$ & $1.9\pm0.3$ & \ldots \\
\ion{He}{i} (6\,678\AA) & $2.0\pm0.4$ & $1.3\pm0.4$ & \ldots \\
\ion{He}{i} (5\,876\AA) & $2.5\pm0.3$ & $1.7\pm0.2$ & \ldots \\

\hline
\end{tabular}
\end{center}
\tablefoot{
\tablefoottext{a}{Line flux is derived from the best-fit Gaussian profile for each emission line and is strongly dependent upon the adopted continuum level.}
}
\end{table}

\begin{figure}
\includegraphics[width=\columnwidth]{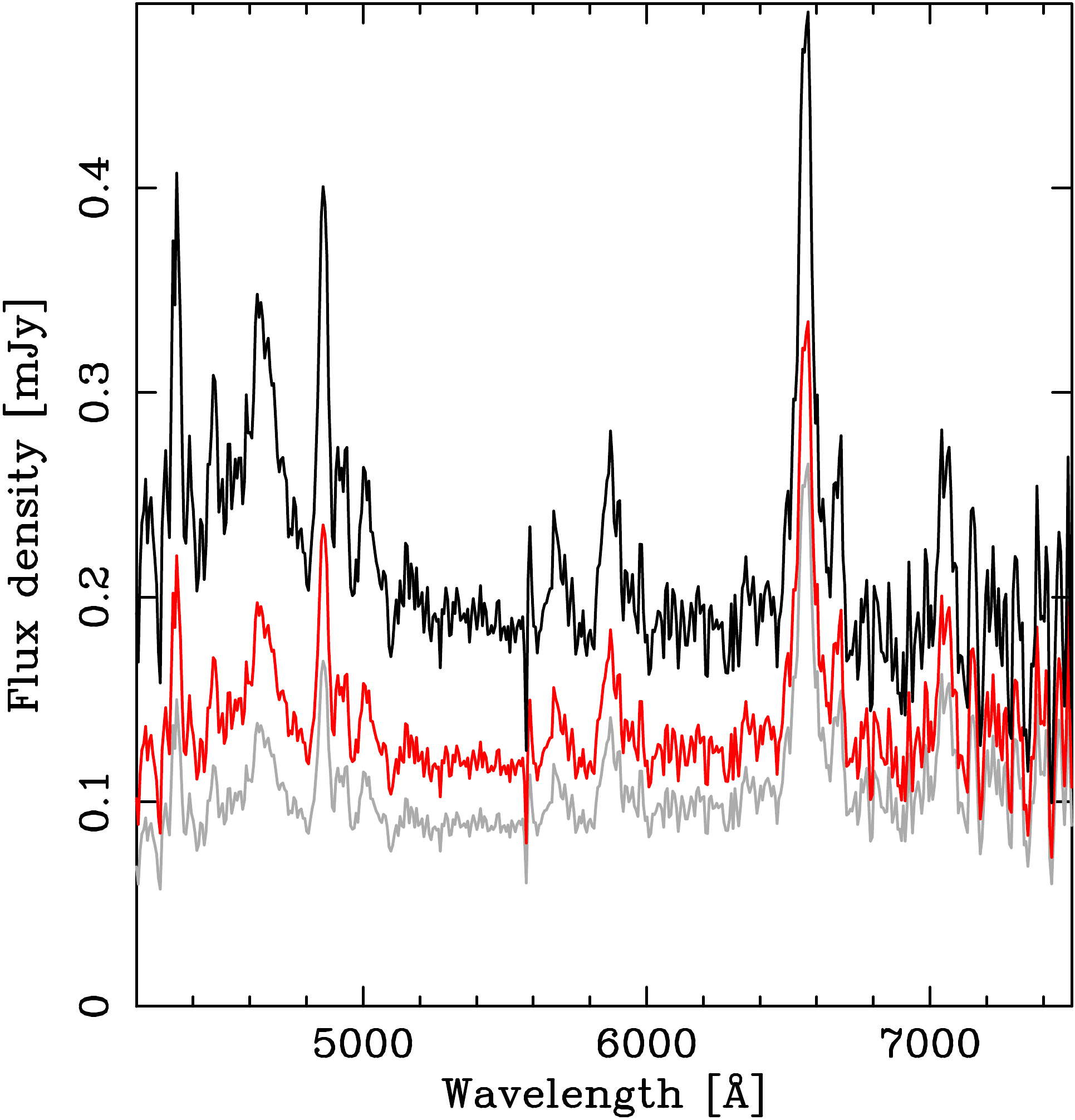}
\caption{Flux calibrated Liverpool Telescope SPRAT spectrum of the 2014 eruption of \object{M31N 2008-12a} taken 0.32~days post-maximum light (second epoch).  Grey: apparent flux, red: deredened spectrum $E_{B-V}=0.1$, black: deredened spectrum $E_{B-V}=0.26$.\label{dered_spec}}
\end{figure}

Continuum emission from the erupting nova is clearly detected in the first three spectra, 0.26\,days, 0.32\,days, and 1.44\,days post-maximum, respectively.  The continuum may also be marginally detected in the fourth epoch spectrum (+2.17\,days).  In these spectra the continuum appears essentially flat, Fig.~\ref{dered_spec} presents the deredened second epoch spectrum assuming $E_{B-V}=0.1$ and $E_{B-V}=0.26$, the lower and upper limits for the extinction towards \object{M31N 2008-12a} (see discussion in Sect.~\ref{MMRD}).  The deredened spectra therefore show a blue continuum, as expected (see discussion in Sect.~\ref{sed_sec}).

The spectra following the 2014 eruption are remarkably similar to those obtained after the 2012 eruption \citep[DWB14]{2012ATel.4503....1S} and after the 2013 eruption (TBW14).  In Fig.~\ref{spec_diff} we present a direct comparison between a spectrum obtained in 2012 from the Hobby Eberly Telescope (HET) and the second epoch spectrum from 2014.  Any continuum emission was subtracted from each spectrum by fitting of a third order polynomial.  These continuum-subtracted spectra were then linearly scaled.  The bottom panel of Fig.~\ref{spec_diff} presents the residuals following a subtraction of these two spectra.  Other than some possible structure around the H$\alpha$ line, the amplitude of the residuals is consistent with the noisier 2014 spectrum.

\subsection{Ejecta Expansion Velocity}

As a proxy for the expansion velocity of the ejecta, we measured the FWHM of the H$\alpha$ emission line by fitting a Gaussian profile to the central portion of the H$\alpha$ line ($\lvert\Delta v\rvert\leq4\,000$~km\,s$^{-1}$).  In all cases a Gaussian profile provided a good fit to the emission line to below the half-maximum flux, although this profile generally did not fit well some of the extended (higher velocity) wings seen in the earlier epochs (see Fig.~\ref{half}).  The computed FWHM velocities for each spectral epoch are shown in Table~\ref{fwhm_tab}, the (weighted) mean indicated FWHM expansion velocity across the four 2014 epochs is $2\,570\pm120$~km\,s$^{-1}$.  Such expansion velocities are consistent with the current sample of M31 He/N spectra \cite[see their Fig.~16]{2011ApJ...734...12S}, but they lie at the lower end of the observed distribution.  The \object{M31N 2008-12a} expansion velocity is unusually low compared to other very-fast and high mass WD novae.  For example, the Galactic RN \object{U Sco} exhibited H$\alpha$ FWHM velocities of $\sim8\,000$~km\,s$^{-1}$ \citep[see e.g.,][]{2013A&A...559A.121A}.  However, such low expansion velocities are in line with predictions from the models of \citet{1995ApJ...445..789P} updated in \citet[see their Table~3]{2005ApJ...623..398Y} who show that all systems with high mass accretion rates have significantly diminished ejection velocities.  Coupling a high mass accretion rate with a high mass white dwarf leads to short inter-eruption timescales.

\begin{table}
\caption{Evolution of the FWHM of the H$\alpha$ profile.\label{fwhm_tab}}
\begin{center}
\begin{tabular}{llll}
\hline\hline
JD & Source & $t-t_{\mathrm{max}}$ & H$\alpha$ FWHM \\
2\,456\,000.5+ && (days) & (km~s$^{-1}$) \\
\hline
933.96 & WHT & 0$.26\pm0.01$ & $2\,740\pm70$ \\
934.06 & LT & $0.32\pm0.14$ & $2\,760\pm110$ \\
220.34\tablefootmark{\dag} & HET & $0.62\pm0.10$ & $2\,470\pm40$ \\
625.4\tablefootmark{\ddag} & Keck & $0.8\pm0.2$ & $2\,600\pm200$ \\
935.18 & LT & $1.44\pm0.18$ & $2\,230\pm90$ \\
935.91 & LT & $2.17\pm0.01$ & $2\,270\pm200$ \\
624.4\tablefootmark{\ddag} & Keck & $3.6\pm0.2$ & $1\,900\pm200$ \\
\hline
\end{tabular}
\end{center}
\tablefoot{
\tablefoottext{\dag}{Spectra of the 2012 eruption of \object{M31N 2008-12a} from DWB14, assuming $t_{\mathrm{max}}=$ 2012 Oct 19.72 UT.}
\tablefoottext{\ddag}{Spectra of the 2013 eruption of \object{M31N 2008-12a} from TBW14, assuming $t_{\mathrm{max}}=$ 2013 Nov 28.60 UT.}
}
\end{table}

In the course of these FWHM determinations we have utilised a number of different approaches to determine the width of the emission lines.  Most of these methods give self-consistent results, we chose the fitting of a Gaussian profile as the simplest method that utilises all the available data.  Typically systematic offsets between methodologies are $\sim100\,\mathrm{km}\,\mathrm{s}^{-1}$, and this value drives our handling of the TBW14 2013 eruption Keck spectra below.

\begin{figure}
\includegraphics[width=\columnwidth]{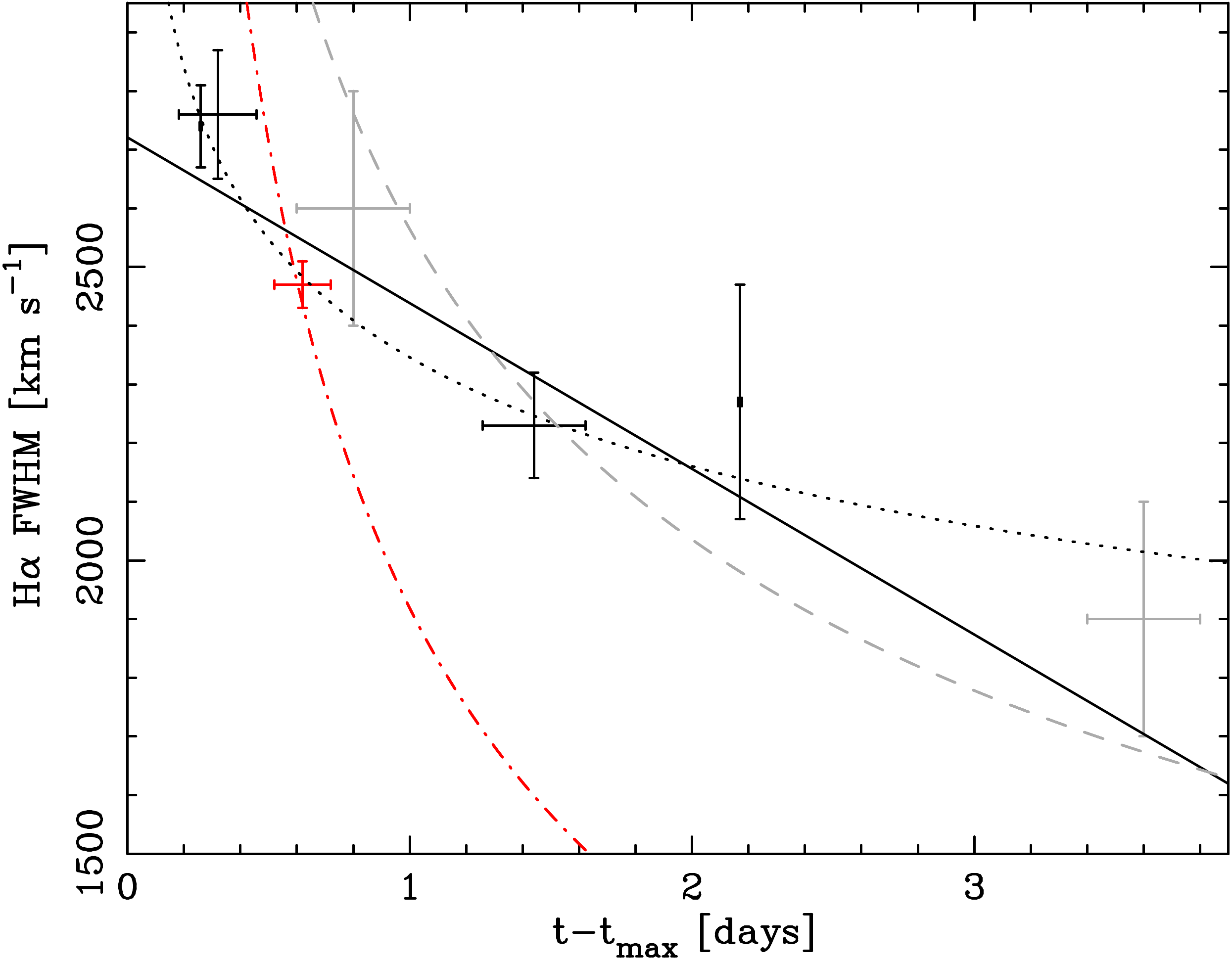}
\caption{The evolution of the FWHM of the H$\alpha$ emission line following the eruption of \object{M31N 2008-12a}.  Black points indicate WHT and LT spectra from the 2014 eruption, grey points are from the 2013 eruption (TBW14), and the red point is from the HET spectra in 2012 (DWB14).  The solid black line shows a simple linear least-squares fit to the 2012, 2013, and 2014 data (gradient $=-280\pm50\,\mathrm{km}\,\mathrm{s}^{-1}\,\mathrm{day}^{-1}$), the grey dashed line a power-law of index $-1/3$, the red dot-dashed line a power-law of index $-1/2$, and the black dotted line the best fit power-law with index $-0.12\pm0.05$, see the text for details.\label{Halpha_plot}}
\end{figure}

A significant `similar implied slowing' by $510\pm110$~km\,s$^{-1}$ is observed between the first and third 2014 epochs, with a `deceleration' by $490\pm140$~km\,s$^{-1}$ observed between the second and third epochs.  A similar implied slowing of the ejecta was noted by TBW14 following the 2013 eruption (see Table~\ref{fwhm_tab}).  For comparison, we have included the H$\alpha$ FWHM velocities following the 2012 and 2013 eruptions in Table~\ref{fwhm_tab}.  Here, we have assumed maximum light times of $t_{\mathrm{max}}=$ 2012 Oct 19.72 UT and $t_{\mathrm{max}}=$ 2013 Nov 26.60 UT for the 2012 and 2013 eruptions, respectively.  To account for any systematic uncertainties in the determination of the maximum light time we have assumed temporal uncertainties of 0.1\,days and 0.2\,days for the 2012 and 2013 spectra, respectively (additional uncertainty for the 2013 spectra is assumed due to lack of knowledge regarding the technicalities of these observations).  Although quoted in DWB14, we have recalculated the 2012 FWHM velocity for consistency with the methodology applied here.  The 2013 FWHM velocities are taken from TBW14 and we have assumed conservative uncertainties of $\pm200$~km\,s$^{-1}$.   The (weighted) mean FWHM velocity across the seven spectral epochs from the 2012, 2013, and 2014 eruptions is $2\,500\pm70$~km\,s$^{-1}$.  In Fig.~\ref{Halpha_plot} we show the evolution with time of the H$\alpha$ profile FWHM velocity using data from the 2012, 2013, and 2014 eruptions.  These data indicate a clear trend of decreasing velocity with time.  These data are consistent with a linear decline of $280\pm50\,\mathrm{km}\,\mathrm{s}^{-1}\,\mathrm{day}^{-1}$ in FWHM velocity with time (the solid black line in Fig.~\ref{Halpha_plot}), but more detailed discussion follows in Sect.~\ref{sec:decel}.  

\section{Environment}\label{environment}

An inspection of H$\alpha$ images taken by the Steward 2.3m (90-inch) Bok Telescope \citep[BT; see][for a full description of the observation strategy]{2008ApJ...686.1261C,2012ApJ...760...13F} revealed extended nebulosity that appeared to surround the position of \object{M31N 2008-12a}.  The Local Group Galaxies Survey \citep[LGGS;][]{2006AJ....131.2478M,2007AJ....134.2474M} narrow-band imaging data confirmed this observation, the nebulosity is visible in their H$\alpha$ and [\ion{S}{ii}] data, but is not seen in [\ion{O}{iii}] imaging.  Extended nebulosity has been seen around Galactic RNe \citep[e.g.\ \object{T Pyxidis;}][]{1997AJ....114..258S,2013ApJ...768...48T} and numerous Galactic CNe \citep[see, e.g.,][]{1995MNRAS.276..353S,1998MNRAS.300..221G,2000AJ....120.2007D}; the remnants of past eruptions.  In addition, nebulosity in the form of planetary nebulae, from previous evolutionary stages of the binaries, has been detected around \object{GK Persei} \citep{1987Natur.329..519B,1989ApJ...344..805S} and \object{V458 Vulpeculae} \citep{2008ApJ...688L..21W}, with extended shell-like \ion{H}{i} emission being detected at even further distances from \object{V458 Vul} possibly due to mass loss during the AGB stage of the primary \citep{2012MNRAS.427L..55R}.  However the apparent size of this feature around \object{M31N 2008-12a}, if directly related to the nova, would be unprecedented for any of the above phenomena.

\subsection{Deep H$\alpha$ imaging}

\begin{figure}
\includegraphics[width=\columnwidth]{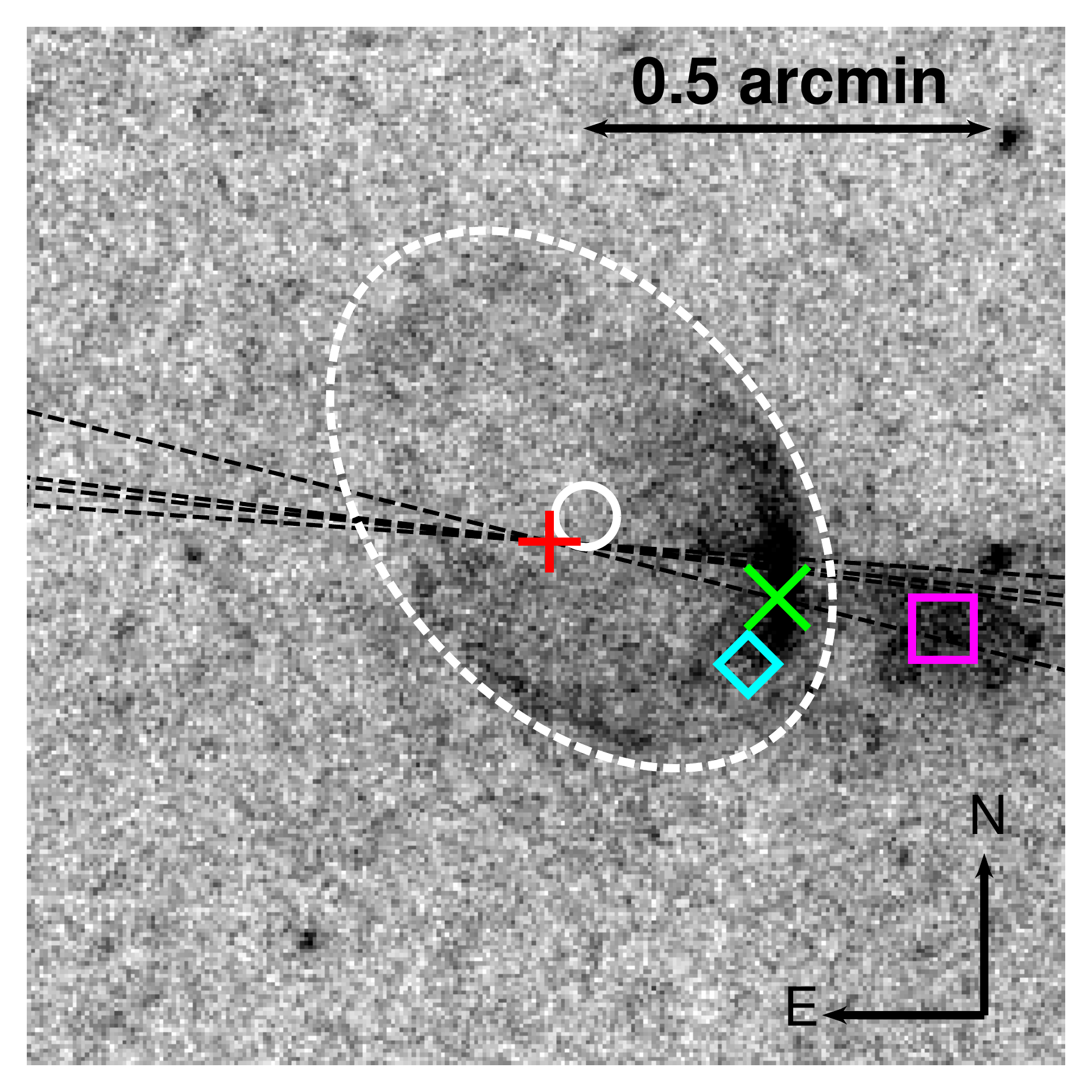}
\caption{LT H$\alpha$ image ($r'$-band continuum subtracted and inverted) showing the $\sim1\farcm3\times1\farcm3$ region around the position of \object{M31N 2008-12a} (indicated by the red cross).  The white circle, green $\times$, and magenta box indicate the catalogue positions of the centre of \ion{H}{ii} regions, and the blue diamond the position of a star cluster (see text in Sect.~\ref{environment} for details).  The white-dashed ellipse (semi-major axis $0\farcm375$) borders the elliptical nebulosity (semi-major axis $0\farcm3$) around \object{M31N 2008-12a} and is to aid the reader.  The position of the South-western knot is also indicated by the green $\times$.  The four black dashed lines indicate the slit position of the four SPRAT spectra that include emission from the SW knot (see Sect.~\ref{sec:knot}).\label{neb_mg}}
\end{figure}

To further investigate this feature, a series of $20\times1\,200$\,s narrow-band H$\alpha$ images of the region around \object{M31N 2008-12a} were taken by the IO:O camera on the LT on the night of 2014 July 30 (UT) in photometric conditions with good seeing.  These images were co-aligned and a median stacked image was produced.  To subtract any contribution from continuum emission from the H$\alpha$ image, the 50 pre-eruption monitoring $r'$-band images (see Sect.~\ref{sec:monitor_and_detect}) were aligned and median stacked, aperture photometry was then performed on 575 point sources common to both images using SExtractor (v2.19.5).  A linear fit was then computed between the $r'$ and H$\alpha$ photometry, allowing the $r'$ image flux to be scaled to that of the H$\alpha$ image, and then subtracted\footnote{We assume that the spectrum of the background is similar to that of the stellar sources; which in the case of the unresolved stellar background of M31 is likely to be reasonable.}.  The $\sim1\farcm3\times1\farcm3$ region around the position of \object{M31N 2008-12a} in the resultant continuum subtracted H$\alpha$ image is shown in Fig.~\ref{neb_mg}.  It is worth noting that all the H$\alpha$ data referred to above would have also included any contribution from [\ion{N}{ii}] (6\,548\AA\ and 6\,584\AA) emission (see also Sect.~\ref{sec:knot}).

The feature appears elliptical in nature with a semi-major axis of $\sim0\farcm3$ and a semi-minor axis of $\sim0\farcm2$ (measured at the approximate midpoint of the shell-like emission; at the distance of M31 these correspond to 67~pc and 45~pc, respectively), the elliptical feature is centred at $\alpha\simeq0^{\mathrm{h}}45^{\mathrm{m}}28\fs6$, $\delta\simeq41\degr54\arcmin13\arcsec$ (J2000) with a position angle on the sky of $\sim40\degr$.  The emission to the south-west of the ellipse is significantly brighter than that in the north-east; this emission peaks in a bright `knot' to the SW (see Sect.~\ref{sec:knot}).  The distance between the geometric centre of the nebulosity and the RN is $\sim4\arcsec$, equivalent to a deprojected separation at the distance of M31 of $\ga13$~pc, this would represent a high but not completely unreasonable range for the proper motion of the RN over a large timescale.

Although larger than the majority of Galactic Supernova Remnants (SNRs), this feature is still somewhat smaller than the largest known Galactic SNR, \object{GSH 138-01-94} that exhibits a radius of $\sim180$~pc \citep{2001ApJ...563..816S} and was discovered by virtue of its 21~cm \ion{H}{i} emission.  Assuming a mean ejecta velocity of 2\,500~km\,s$^{-1}$ (see Sect.~\ref{spec_results}), and zero subsequent deceleration, it would take a single eruption of \object{M31N 2008-12a} between 17\,000 and 27\,000~years to create a remnant of such a size.

If we also assume that the ISM density in the region of M31 around \object{M31N 2008-12a} is similar to that found in the mid-disk of the Milky Way \citep[$\sim0.5$\,M$_{\sun}$\,pc$^{-3}$;][]{1992AJ....103.1552M}, the ISM swept up in this volume by a single ejection event would be $\sim2.8\times10^{5}$\,M$_{\sun}$ \citep[assuming a remnant geometry of a prolate spheroidal shell; see, e.g.][]{1998MNRAS.296..943P}, which is, coincidentally, similar to that swept up by \object{GSH 138-01-94} (although the local density there is much less as the SNR is located out of the Galactic plane).  However, the kinetic energy of a nova ejecta is $\sim10^{8}$ times less than that of a typical SN, and the momentum of the ejecta $\sim10^{8}$ times lower, so it is clear that a single nova eruption could not create such a remnant.  Although perhaps unlikely, the question remains whether multiple nova eruptions over an extended period of time could generate such a phenomenon.

While any possible association between \object{M31N 2008-12a} and the nebulosity is tantalising, it is more likely just a coincidental alignment; indeed the wider region around the RN is littered with filaments and apparent shells of gas (although none so apparently regularly shaped as that in Fig.~\ref{neb_mg}).  There are a number of previously identified \ion{H}{ii} regions apparently associated with this nebulosity.  The sources \object{[AMB2011] HII 3556} \citep[the South-western knot]{2011AJ....142..139A}, \object{[WB92a] 787} \citep{1992A&AS...92..625W}, and \object{[PAV78] 857} \citep{1980A&AS...40...67D} are indicated by the green $\times$, white circle, and magenta box in Fig.~\ref{neb_mg}, respectively.  It seems clear that \object{[WB92a] 787} is an earlier identification of the nebulosity around \object{M31N 2008-12a}, the position, size, and description \citep[``a well defined ring'',][]{1992A&AS...92..625W} are consistent with our observations.  The object \object{[PAV78] 857} is identified as a SNR by \citep{1980A&AS...40...67D}, but as diffuse emission by \citet{1992A&AS...92..625W}.  Finally, \object{[JSD2012] PC 167} (the feature indicated by the blue diamond in Fig.~\ref{neb_mg}) is identified as a cluster of stars \citep{2012ApJ...752...95J}.

\subsection{South-western `Knot' Spectra}\label{sec:knot}

In four of our SPRAT observations of the \object{M31N 2008-12a} (see Sect.~\ref{LT_spectra}), the parallactic alignment of the $\sim80\arcsec\times1\farcs8$ slit happened to lie across the bright knot of SW nebulosity (indicated by the green $\times$ in Fig.~\ref{neb_mg}), and the resulting nebular emission could be clearly seen in the 2d spectral images.  These spectra had sky position angles of $83\degr$, $86\degr$, $256\degr$ and $264\degr$ (see the black dashed lines in Fig.~\ref{neb_mg}).  For each of these spectra where the nebulosity was visible, a 3\arcsec wide region of the 2d image centred on the knot emission was extracted, with sky subtraction accomplished by linear fitting to a 5\arcsec region of sky to either side.  Wavelength calibration was carried out on each individual spectrum using the corresponding arc frame, obtained directly after each exposure, which was solved for the actual rows extracted for that spectrum. The resulting spectra are therefore calibrated in wavelength to the same precision as the nova spectra (1 {\AA} rms).  All eight spectra were then resampled to a uniform pixel scale and co-added, giving a total exposure time of 6\,000 seconds.  By measuring line widths of sky emission lines in the sky extraction regions, the spectral resolution of the data can be estimated as $1\,000\,\mathrm{km}\,\mathrm{s}^{-1}$ (FWHM).

\begin{figure}
\includegraphics[width=\columnwidth]{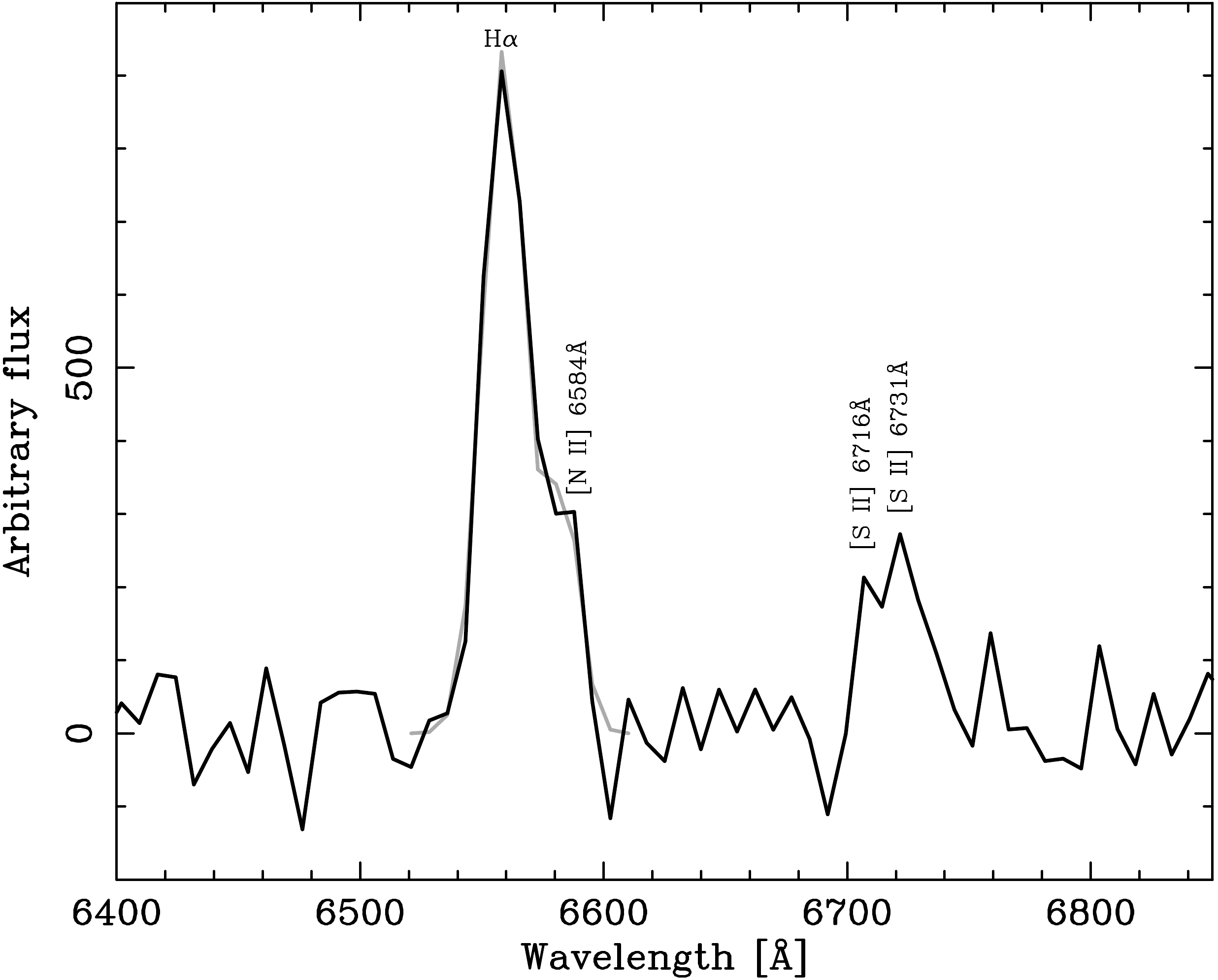}
\caption{Liverpool Telescope SPRAT spectrum of the apparent south-western `knot'  in the extended emission around the position of \object{M31N 2008-12a} (solid black line; see also Fig.~\ref{neb_mg}).  The spectral range has been truncated to show only the region with significant emission; likely H$\alpha$, [\ion{N}{ii}] (6\,584\AA), and [\ion{S}{ii}] (6\,716 and 6\,731\AA) emission is labelled.  The solid grey line indicates a simultaneous fit to the H$\alpha$ and [\ion{N}{ii}] (6\,584\AA) emission lines.\label{nebula_spec}}
\end{figure}

Figure~\ref{nebula_spec} shows the SPRAT spectrum of the SW-knot seen in Fig.~\ref{neb_mg}.  The spectrum is consistent with the LGGS narrow-band imaging in the sense that clear emission from H$\alpha$ and [\ion{S}{ii}] is seen, but [\ion{O}{iii}] is not.  Here the width of the spectral features is dominated by the resolution limit of SPRAT.  Hence, using the morphology of a sky emission line as a template line profile, we were able to separately fit emission from the [\ion{S}{ii}] (6\,716 and 6\,731\AA) doublet, and H$\alpha$ and [\ion{N}{ii}] (6\,584\AA) (see the solid grey line in Fig.~\ref{nebula_spec}) although any emission from [\ion{N}{ii}] (6\,548\AA) was not significantly detected.  The fitted central wavelength of the H$\alpha$ emission from the SW-knot ($6\,559.2\pm0.5$\AA) is consistent with that of the fitted H$\alpha$ emission peak in the spectra of \object{M31N 2008-12a} ($6\,560\pm1$\AA), and these are broadly consistent with expectation, given the blueshift of \object{M31} \citep[see, e.g.,][]{2012AJ....144....4M}.

Emission from [\ion{S}{ii}] is seen in a wide variety of astrophysical phenomena, from Herbig-Haro objects \citep[see e.g.,][]{1975PASP...87..405V} to AGN \citep[see e.g.,][]{1990MNRAS.245..470B}, including, of course, SNRs.  Such emission is often associated with the cooling and recombination zones behind low velocity ($\sim100$\,km\,s$^{-1}$) shocks \cite[see e.g.,][]{1973ApJ...185..441O}.  The [\ion{S}{ii}]/H$\alpha$ ratio is a well established criterion used for the detection of SNRs \citep[see e.g.,][]{1973ApJ...180..725M,1997ApJS..112...49M}, with a ratio [\ion{S}{ii}]/H$\alpha\ga0.45$ often employed for positive identification of a SNR.  We determine this ratio for the SW-knot to be [\ion{S}{ii}]/H$\alpha=0.35\pm0.10$ (here we have corrected for any flux contribution from [\ion{N}{ii}], but we note that this spectrum has not been flux calibrated).  At best, the [\ion{S}{ii}]/H$\alpha$ ratio is inconclusive in this case.  Little, if any, evidence of [\ion{S}{ii}] emission has yet been seen in the spectra of nova remnants \citep[see e.g.,][for the case of \object{T Pyx}]{1982ApJ...261..170W}, although \citet{1997ApJ...475..803C} question whether any sulphur may be depleted by the formation of silicate dust grains.

\subsection{Morpho-kinematical Modelling of the RN Spectra}

Morpho-kinematical modelling of the spectra of erupting novae has been used to investigate the expansion velocity, inclination, and geometry of nova ejecta \citep[see e.g.,][]{2009ApJ...703.1955R,2011MNRAS.412.1701R,2013ApJ...768...49R,2013MNRAS.433.1991R,2013A&A...549A.140S,2013A&A...553A.123S}.   Here we employed the morpho-kinematical code SHAPE \citep[v5.0;][]{2011ITVCG..17..454S} in order to ascertain whether a remnant of similar morphology to that seen in Fig.~\ref{neb_mg} could be produced by, albeit a single eruption of, an erupting nova with a H$\alpha$ profile similar to that of \object{M31N 2008-12a} (as shown in Fig.~\ref{half}).  

To improve the spectra signal-to-noise, we co-added the four 2014 spectra with the 2012 HET spectrum of \object{M31N 2008-12a}.  We introduced an ejecta density profile that varied as $1/r^3$ \citep[as modelled for \object{V959 Monocerotis};][]{2013A&A...553A.123S}, with a Hubble flow velocity distribution; essentially a free expansion.  We constrained the minor to major axis ratio of the ejecta to be $2/3$ (as imposed by the nebulosity; see Fig.~\ref{neb_mg}), and constrained the ejecta model to a thin shell, as we observed a double peak in the eruption spectra (assuming a filling factor of unity).  The inclination angle of the ejecta was also allowed to vary from 0 to 90\,degrees (in steps of 1\,degree; where an inclination 90\,degrees corresponds to the orbital plane being edge-on) and the de-projected velocity varied from 2\,000 to 5\,000\,km\,s$^{-1}$ in steps of 100\,km\,s$^{-1}$.  Based on the central wavelength of the H$\alpha$ emission, the spectra were also corrected for a recessional velocity of $\sim150\,\mathrm{km}\,\mathrm{s}^{-1}$.

\begin{figure}
\includegraphics[width=\columnwidth]{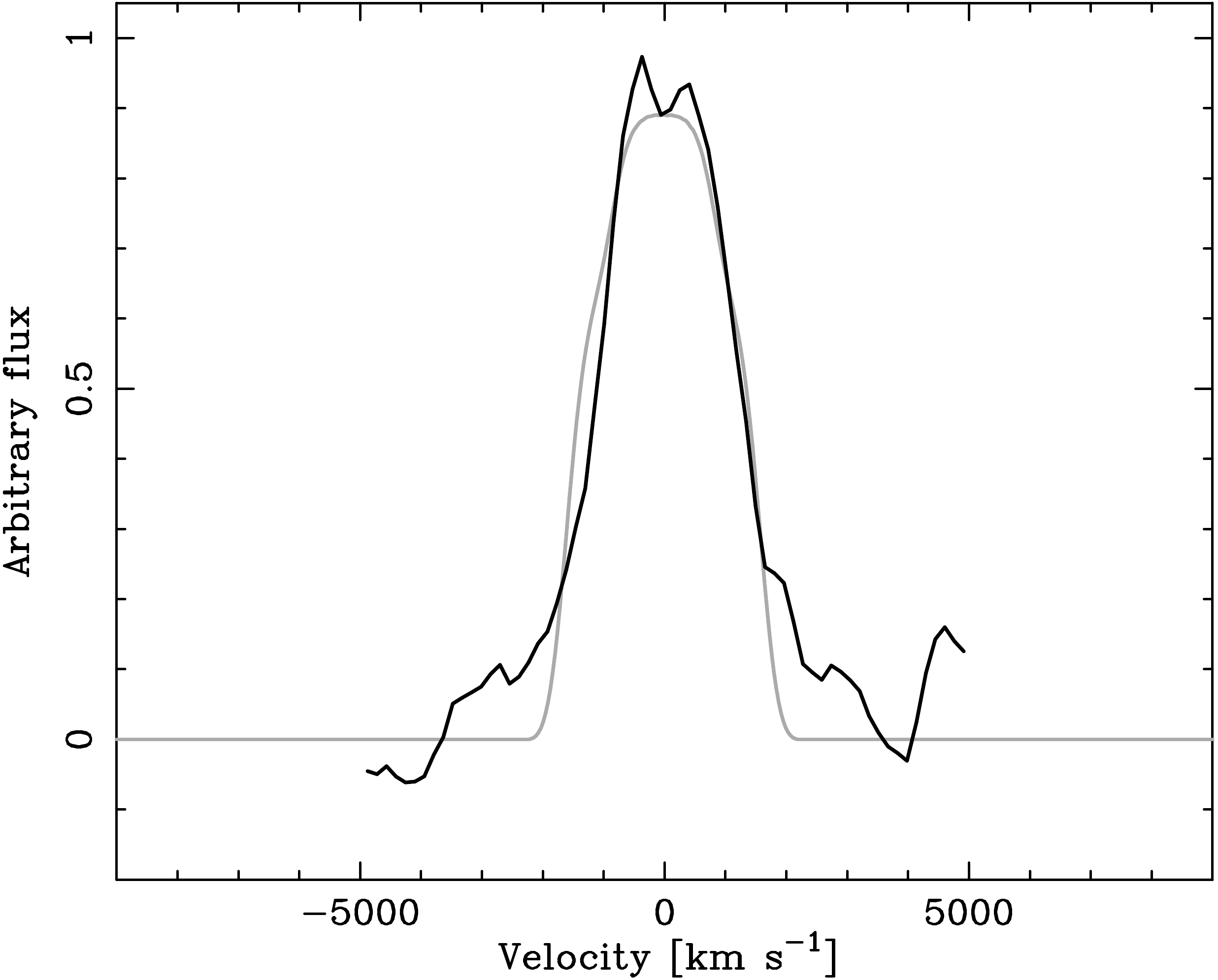}
\caption{The solid black line shows the co-added spectra from the four epochs of 2014 observations and a single epoch of 2012 observations of the erupting \object{M31N 2008-12a}, centred on the H$\alpha$ emission line, (the peak redwards of H$\alpha$ is the \ion{He}{i} (6\,678\AA) emission).  The solid grey line indicates the best-fitting model spectrum expected by a freely expanding ejecta with a bi-polar morphology, see the text for further details and discussion. \label{fig:morph}}
\end{figure}

The results of the morpho-kinematical modelling indeed suggest that a remnant geometry similar to the nebulosity seen around \object{M31N 2008-12a} can be produced by a single nova eruption exhibiting the H$\alpha$ line profile seen in \object{M31N 2008-12a}.  We find that such a remnant would have  a bi-polar geometry, an inclination of $46^{+8}_{-38}$\,degrees, and an expansion velocity of $1\,700\pm200\,\mathrm{km}\,\mathrm{s}^{-1}$.  For the avoidance of doubt, it should be noted here that these are not predictions of the true nature of the \object{M31N 2008-12a} ejecta.  The best-fit model spectra is shown in Fig.~\ref{fig:morph} and compared to the \object{M31N 2008-12a} H$\alpha$ emission.  As is evident in Fig.~\ref{fig:morph}, such a simple model was unable to fit the higher velocity material ($\lvert\Delta v\rvert\geq 2\,000$\,km\,s$^{-1}$).

It is important to note a number of important caveats to this work.  Although we have shown that such a remnant could be produced by a nova with line profiles similar to \object{M31N 2008-12a}, our geometry solution is not likely to be unique.  We imposed some constraints on the possible geometry given our limited knowledge of the system and employed a very simple model of the ejecta.  A more physical model would require consideration of the ejecta interaction with the generally complex circumstellar and circumbinary environments, for example, with any red giant wind (see Section~\ref{sec:decel}) as is required to model the ejecta of \object{RS Oph} \citep[see, for e.g.,][]{2008A&A...484L...9W}.  The development of more complex models based on just these early-time spectra would likely introduce degeneracies in our results.  For example, model fits to \object{V2491 Cyg} at early times replicated reasonably well the observations, however, it was not until the nebular stage that the geometry could be pinned down with some confidence \citep{2011MNRAS.412.1701R}, also \citet{2002A&A...390..155D} discuss the change in line profile due to the termination of a post-eruption wind.  Finally, to produce a remnant of the scale seen in Fig.~\ref{neb_mg} it would require multiple eruptions all exhibiting similar geometries; and we would need to further consider any shaping caused by interaction with the surrounding interstellar medium.

\subsection{Follow Up}

The most plausible explanation for the nebulosity would be a nearby (within M31) SNR, possibly due to an association with the adjacent star cluster, which may also explain the other apparent SNR and the SW-knot; perhaps akin to some of the [\ion{S}{ii}]-bright shell nebulae or `superbubbles' seen in the Large Magellanic Cloud \citep[see e.g.,][]{1977ApJ...212..390L,2014ApJ...792...58Z}.  Of course, such an explanation doesn't rule out \object{M31N 2008-12a} also being associated with this star cluster.  However, none of our investigations have been able to conclusively rule out an origin related to \object{M31N 2008-12a}.  Of course we must also ask ourselves, if there were not a nova near the centre of this nebula, would we still be debating its origin?  However, the question of whether such a rapidly recurring nova can create a remnant as large and visible as would be required here is still an open one, and requires further observations and modelling.

\section{Discussion}\label{sec:discussion}

\subsection{Spectral Energy Distribution}\label{sed_sec}

\begin{figure*}
\includegraphics[width=0.49\textwidth]{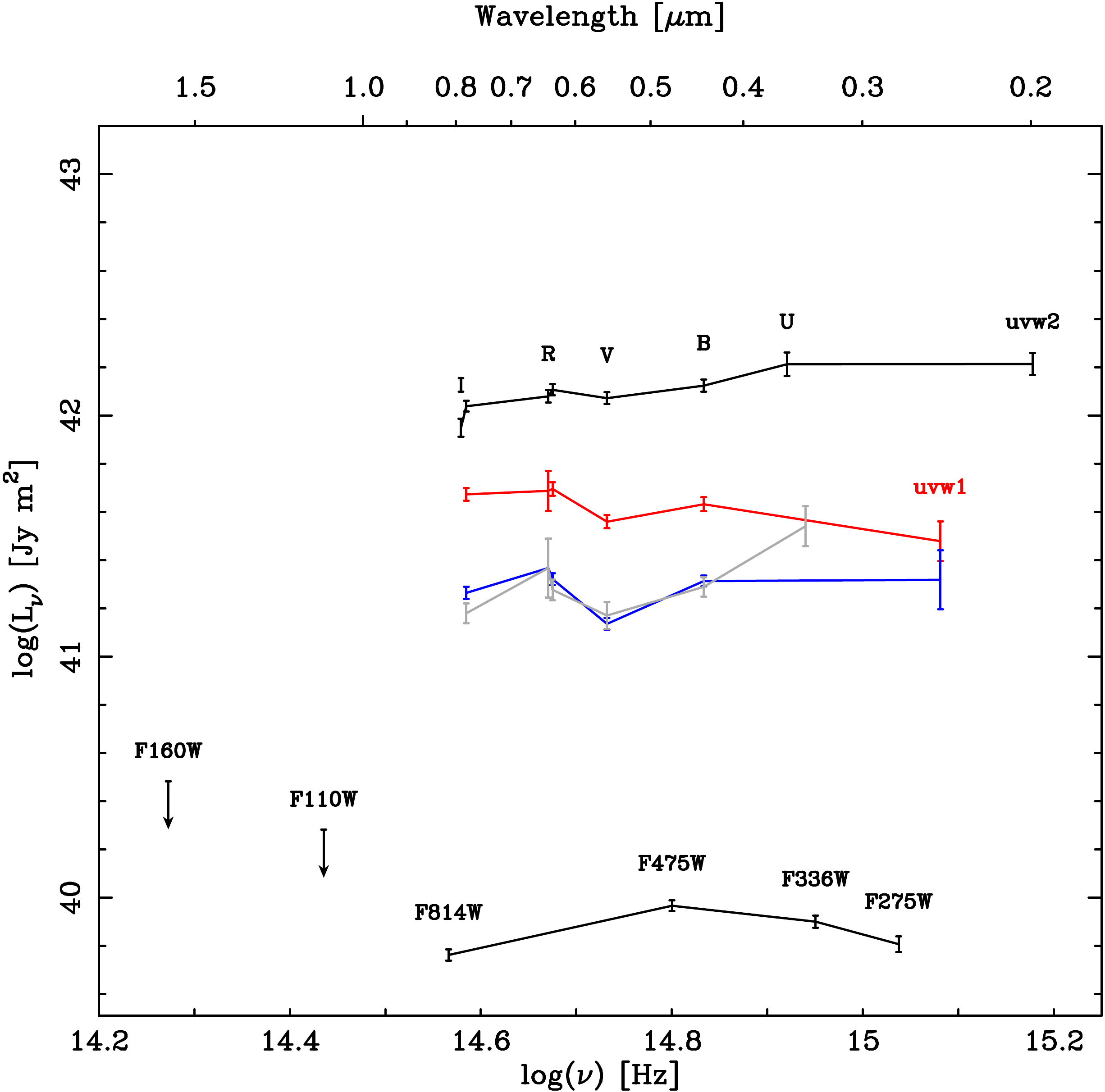}\hfill
\includegraphics[width=0.49\textwidth]{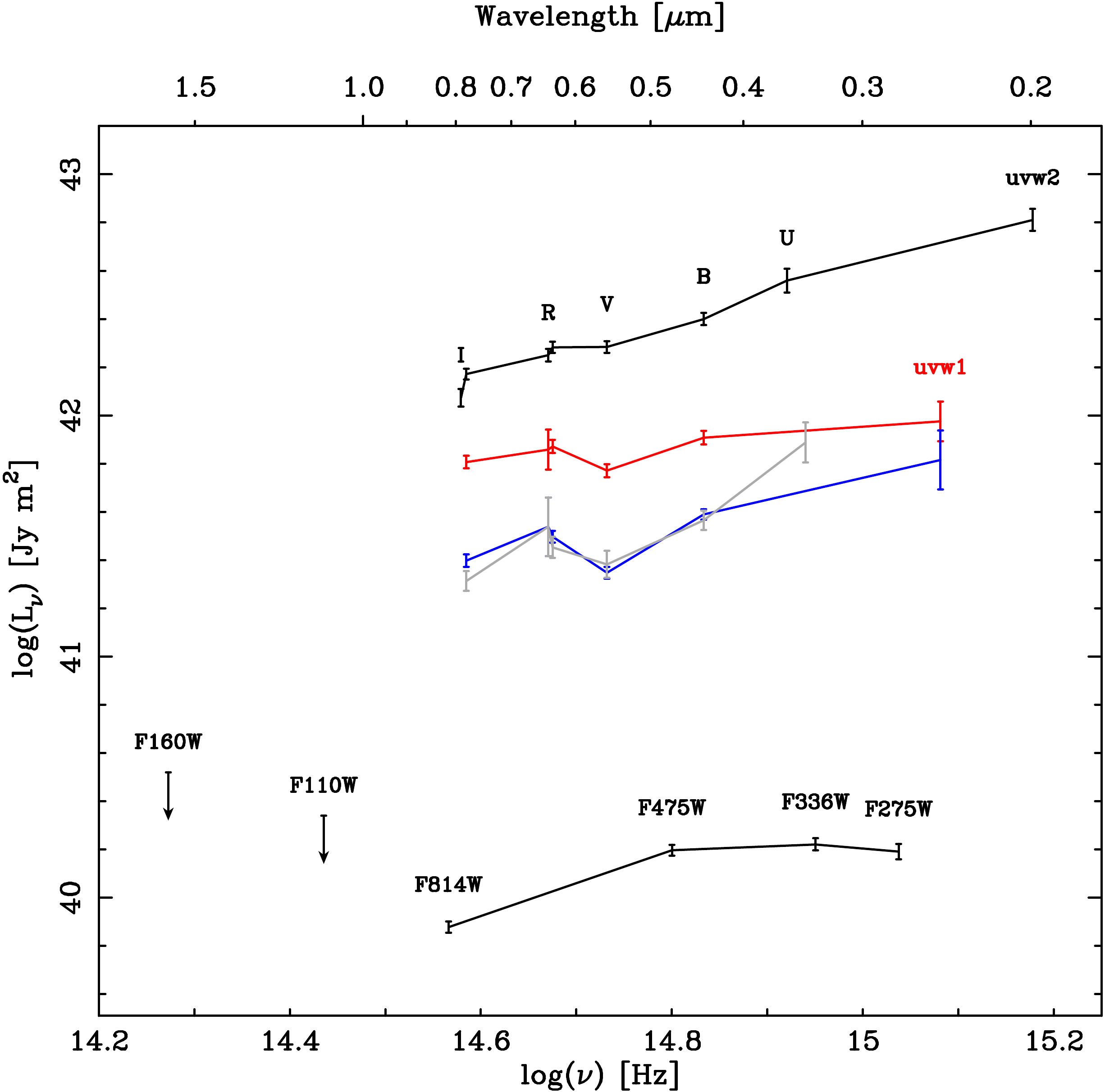}
\caption{Distance and extinction corrected SEDs showing the progenitor system of \object{M31N 2008-12a} (low luminosity black points), compared to the evolving SED of the 2014 eruption (black points, $t\sim0$~days; red points, $t\sim1$~days; blue points, $t\sim2$~days; and grey points $t\sim3$~days).  Units chosen for consistency with similar plots in \citet[see their Fig.~71]{2010ApJS..187..275S} and DWB14 (see their Fig.~4).  The central wavelength locations of the Johnson-Cousins, {\it HST}, and \swift filters are shown to assist the reader.  The left-hand plot is the low extinction scenario, where only the line-of-sight (Galactic) extinction towards M31 is considered \citep[$E_{B-V}^{\mathrm{Galactic}}=0.1$;][]{1992ApJS...79...77S}.  The right-hand plot considers an additional component of extinction internal to M31 \citep[$E_{B-V}^{\mathrm{internal}}=0.16$, $E_{B-V}^{\mathrm{total}}=0.26$;][]{2009A&A...507..283M}.\label{sed}}
\end{figure*}

In Fig.~\ref{sed} we present the evolving spectral evolution during the first three days from the peak of 2014 eruption and at quiescence (DWB14).  The quiescent photometry is taken from the analysis of {\it HST} infrared, optical, and UV imaging data, as presented by DWB14.  For the 2014 data, we have used optical and \swift UV absolute calibrations from \citet{1979PASP...91..589B} and \citet{2010Bre}, respectively.  Here we have assumed a distance to \object{M31} of $770\pm19$~kpc \citep{1990ApJ...365..186F}, a line-of-sight external (Galactic) reddening towards \object{M31} of $E_{B-V}=0.1$ \citep{1992ApJS...79...77S}, and internal (M31) reddening of $E_{B-V}\le0.16$ \citep{2009A&A...507..283M}.  We employ the analytical Galactic mean extinction law of \citet[assuming $R_{V}=3.1$]{1989ApJ...345..245C} to compute the extinction in the regime of the \swift UV filters, with no knowledge of the source spectra we simply evaluate the extinction at the central wavelength of each UV filter.

The optical behaviour of novae around peak has been observed to resemble black body emission \citep{1976ApJ...204L..35G,2008clno.book..167G} and has been modelled in terms of the development of a pseudo-photosphere (PP) in the optically thick ejecta \citep[see][and references therein]{1989clno.conf...61B}. The PP radius $r_{\rm p}$ is greatest, and effective temperature $T_{\rm p}$ at a minimum at optical peak. Thereafter, the mass loss rate from the WD surface declines, and at constant bolometric luminosity, $r_{\rm p}$ shrinks while $T_{\rm p}$ rises, shifting the peak of the emission further into the UV with time past maximum light. In CNe, with relatively high ejected masses, $T_{\rm p} \sim 10\,000$\,K at optical maximum placing the peak emission in the optical part of the spectrum. For a lower mass ejection rate (total mass ejected) for a given ejection velocity, the emission peak may never appear in the optical but only get as far as the UV before moving to ever higher frequencies, ultimately showing up as the SSS when the PP radius has shrunk to scales approaching that of the WD. Thus it may well be the fact that the ejected mass is far lower in an RN such as \object{M31N 2008-12a} that means these objects do not obey the MMRD (see Sect.~\ref{MMRD}).

From Fig.~\ref{sed}, one can see that at the time of optical maximum ($t \sim 0$), the SED does not show an obvious peak in this waveband. For the case where both the foreground Galactic and internal M31 extinctions are included (see the right-hand panel of Fig.~\ref{sed}), there is an indication that the peak emission is at $\nu\ga1.5 \times10^{15}$\,Hz ($\lambda\la2\,000$\,\AA). If emission at this time were that of a black body \citep[as assumed for example in the simplistic models of][although noting the detailed caveats given in HND15]{1989clno.conf...61B}, then from the frequency dependent version of Wien's Law, $T_{\rm p} \ga 2.5 \times 10^4$\,K. Similarly, if we take the dereddened emission at the two highest frequency points for $t \sim 0$ in Table~\ref{uv_tab} (i.e., $U$-band and uvw2 filters) and again assume purely black body emission, then the ratio of monochromatic luminosities ($L_{\nu}$) at the two frequencies gives $T_{\rm p} \ga 3.3 \times 10^4$\,K at this time. 

Taking this effective temperature and $L_{\nu}$ in the uvw2 band\footnote{Note, here we have assumed equivalence between effective temperatures and color temperatures.}, this yields an effective photospheric radius $r_{\rm p} = 1.1 \times 10^{10}$\,m at optical maximum and hence a total bolometric luminosity $L \sim 10^{32}$\,W at this point. For $L_{\rm Edd} = 2 \times 10^{31}$\,W for a WD near $M_{\mathrm{Ch}}$, this implies a highly super-Eddington phase around maximum light. We note that \citet{2008clno.book...77S} find such a short-lived phase for a $1.35\,M_\odot$ WD where $L_{\rm max} \ga 2 \times 10^{32}$\,W. We also note however that to derive a consistent solution for the absolute value of the observed dereddened monochromatic luminosity together with a self consistent bolometric luminosity, assuming purely black body emission, bolometric luminosities even higher than that derived above would be required. Black body fits to the \swift XRT data (HND15) yield $L \sim 3.3 \times 10^{31}$\,W ($\sim 1.8 L_{\rm Edd}$) at later times, suggesting that any highly super-Eddington phase around the optical peak was short lived.  But overall of course, great caution must be taken in over-interpreting results from simple black body assumptions when we know that the spectra at all wavelengths deviate significantly from such simplistic models (again, see discussion in HND15). 

As an alternative explanation, the right panel of Fig.~\ref{sed} shows spectra of optically thin free-free emission ($F_\nu \propto \nu^0$) after the first epoch, with optically thick free-free emission \citep[where e.g., $F_\nu \propto \nu^{2/3}$ would result following][for a constant mass loss rate, constant velocity, fully ionised wind]{1975MNRAS.170...41W} at the peak itself. Classical novae show spectra of optically thin free-free emission in their decay phase \citep[e.g.,][]{1976ApJ...204L..35G}. The brightness of free-free emission is determined mainly by the wind mass-loss rate \citep{1975MNRAS.170...41W}. For a larger envelope mass at ignition, the wind mass-loss rate starts with a large value.  Then the nova becomes fainter as the envelope mass decreases because the wind mass-loss rate decreases with time.  Therefore, for a smaller envelope mass at ignition, the wind mass-loss rate starts with a smaller value and the nova brightness starts with a fainter maximum brightness \citep{2006ApJS..167...59H}.  The recurrence time of 1~year requires a very massive WD ($\gtrsim 1.3 M_\odot$) and very high mass accretion rates, so the hydrogen-rich envelope mass at ignition is small.  As such, this RN would be expected to have a much fainter brightness at maximum compared with CNe.

\subsection{Decline Time and Maximum Magnitude}\label{MMRD}

The speed class of a nova is defined by the time taken for the light curve to decay by two or three magnitudes from peak optical luminosity, $t_{2}$ and $t_{3}$, respectively.  Determining the speed class can be complicated by a complex light curve morphology, or by poor temporal coverage, particularly around peak luminosity.  For example, the slowly evolving RN \object{T Pyx} has a long lived and complex maximum `plateau' \citep[see e.g.,][]{2013ApJ...773...55S,2014AJ....147..107S}, however, the rapidly evolving RN candidate \object{V2491 Cyg} exhibits a secondary maximum in its light curve \citep[see e.g.][and discussions therein]{2011A&A...530A..70D,2011NewA...16..209M,2011MNRAS.412.1701R}.  The faster the decline of a given nova, the more important good determination of the peak time becomes.  DWB14 estimated the decline time of the 2013 eruption to be $t_{2}(B)\simeq5$~days and $t_{2}(V)\simeq4$~days, and \citet{2014ATel.6546....1H} computed $t_{2}(R)=2.7\pm0.5$~days based on preliminary observations of the 2014 eruption.  Using the observations of the 2014 eruption, and the assumed $t_{\mathrm{max}}$ of 2014 Oct $3.7\pm0.1$ UT, we determined the decline times shown in Table~\ref{t2}.  Here, we assumed that the brightest observation available in each filter occurred at, or near, maximum light, and $t_{2}$ and $t_{3}$ times were then determined by simple linear interpolation between observations.  

\begin{table}
\caption{Decline times of the 2014 eruption of \object{M31N 2008-12a}\label{t2}.}
\begin{center}
\begin{tabular}{lll}
\hline\hline
Filter & $t_{2}$ (days) & $t_{3}$ (days) \\
\hline
$B$  & $2.07\pm0.25$ & $4.79\pm0.31$\\
$V$ & $1.77\pm0.13$ & $3.84\pm0.24$ \\
$R$ & $2.40\pm0.51$ & $> 2.46$ \\
$r'$ & $2.30\pm0.11$ & $4.81\pm0.26$ \\
$i'$ & $2.27\pm0.10$ & $4.19\pm0.17$ \\
\hline
\end{tabular}
\end{center}
\tablefoot{Assuming $t_{\mathrm{max}}=56\,933.7\pm0.1$ (MJD) or 2014 Oct $3.7\pm0.1$ UT (see discussion in Sect.~\ref{maximum_light}).}
\end{table}

The decline times for the 2014 eruption are significantly faster than those derived by DWB14 for the 2013 eruption.  This is most likely due to the paucity of light curve data available at the time of that study, rather than a fundamental change in the presentation of the eruption.  The value of $t_{2}(R)=2.4\pm0.5$\,days for the 2014 eruption, which is consistent with that derived using iPTF data following the 2013 eruption ($t_{2}(R)=2.1$\,days; TBW14).  Nova decline times are typically quoted in the $V$-band, partly as the early spectral evolution does not significantly affect the luminosity in this band at this time (for example, the $R/r'$-band $t_{2}$ and $t_{3}$ calculations are significantly affected by the H$\alpha$ emission).  With $t_{2}(V)=1.8\pm0.1$\,days and $t_{3}(V)=3.8\pm0.2$\,days, \object{M31N 2008-12a} declines more rapidly than all known Galactic RNe \citep[surpassing \object{U Sco} with $t_{3}=4.3$~days; see][]{2008ASPC..401...31A}, although the recently confirmed RN \object{M31N 2006-11c} \citep{2015ATel.7116....1H} had a decline time of $t_{3}(R)=3.0\pm0.4$\,days during its 2015 eruption \citep{2015ATel.7142....1H}.

Based on these observations of the 2014 eruption, we can calculate the absolute magnitude at peak luminosity to be in the range $-6.3\ge M_{V}\ge-6.8$, depending upon the amount of extinction internal to \object{M31}.   Here we have again assumed a distance of $770\pm19$~kpc, a Galactic reddening of $E_{B-V}=0.1$, and an internal reddening of $E_{B-V}\le0.16$.  Employing the maximum magnitude -- rate of decline (MMRD) relationships derived for novae in M31 \citep{2011ApJ...734...12S} and Galactically \citep{2000AJ....120.2007D}, we predict the absolute magnitude of a nova as fast as \object{M31N 2008-12a} to be $M_{V}=-9.36\pm0.12$ and $M_{V}=-10.69\pm0.21$, respectively. 

However, the MMRD relationship is known to perform poorly for RNe \citep[see, e.g.][their Fig.~31]{2010ApJS..187..275S,2015ApJ...798...76H}.  Recurrence timescales as short as one year are driven by a combination of a high mass WD and high mass accretion rate, and the accreted, or ignition, mass is small \citep[see, e.g.][]{2005ApJ...623..398Y,2013ApJ...777..136W,2014ApJ...793..136K}.  As such the ejected mass is too small for the PP to expand to the typical red giant size (as in CNe) hence, the effective surface temperature of the PP is always $>10\,000$\,K, the peak of the emission doesn't move into the optical, and the optical luminosity at maximum is therefore significantly diminished (see also discussion in Sect.~\ref{sed_sec}).  \citet{2014A&A...563A...2H} explore a number of different nova eruption parameter correlations for the M31 nova population; their relevance to \object{M31N 2008-12a} is discussed in HND15 (see their Sect.~4.5).

\subsection{Ejecta Deceleration and the Nature of the System}\label{sec:decel}

The marked reduction with time of the width of the H$\alpha$ line profile across data from the 2012, 2013, and 2014 eruptions is presented in Sect.~\ref{spec_results}.  Such a decrease in the inferred velocity of the expanding ejecta (in the case of \object{M31N 2008-12a}, $\sim500$~km\,s$^{-1}$ in a $\sim1.2$~day period) has been observed in RN systems previously.  Narrowing of the emission lines following the eruptions of the Galactic RN \object{RS Oph} has been observed since at least the 1958 eruption \citep{1964AnAp...27..555D}.  \citet[see their Fig.~49]{2009A&A...505..287I} presents a dramatic illustration of the narrowing of the H$\beta$ emission line following the 2006 eruption of \object{RS Oph}.  In \object{RS Oph} the narrowing of the emission lines has been proposed to be caused by the deceleration of the shock driven in to the preexisting red giant wind (RGW) by the high velocity ejecta \citep[first proposed by][]{1967BAN....19..227P}; the associated cooling of the shock was observed in X-rays and modelled by \citet{2006ApJ...652..629B}.  Any shocked X-ray emission from the interaction of the ejecta with surrounding material in \object{M31N 2008-12a} would not be detectable by the \swift XRT at the distance of M31.  Such a shock interaction may also give rise to optical coronal lines (e.g.\ [\ion{Fe}{xiv}] (5\,303\AA) and [\ion{Fe}{x}] (6\,374\AA)), however, in \object{RS Oph} these lines were only detected around thirty days after the eruption \citep{1987rorn.conf....1R}, and perhaps it is not surprisingly there is no evidence of any such emission in the \object{M31N 2008-12a} early-time spectra.

The RGW velocity in \object{RS Oph} is $\sim20$\,km\,s$^{-1}$ (see \citealt{1985MNRAS.217..205B}, and discussion therein; see also \citealt{2009A&A...505..287I}, who determined a velocity of 33\,km\,s$^{-1}$), if we assume a similar red giant secondary in the \object{M31N 2008-12a} system (DWB14) and the RGW is expanding into a region fully cleared by the last eruption (a year previously), then the extent of the RGW would be $\sim4.2$\,AU.  Simplistically, with an initial mean ejecta velocity of $\sim2\,700\,\mathrm{km}\,\mathrm{s}^{-1}$ (see Sect.~\ref{spec_results}), the RGW region would begin to be cleared within $\sim3$~days and we therefore expect a deceleration to at least this time, which covers all of our spectroscopic observations.

The {\it HST} photometry of the progenitor system of \object{M31N 2008-12a} presented by both DWB14 and TBW14 indicated the presence of a luminous accretion disk, on a par with that seen in the \object{RS Oph} system (DWB14, see their Fig.~4).  The quiescent NIR photometry at the location of \object{M31N 2008-12a} was not sufficiently deep  to detect the secondary star and could not rule out, for example, a secondary as luminous as that in \object{RS Oph} (DWB14).  These NIR photometry were only able to exclude a secondary as luminous at that of \object{T Coronae Borealis} \citep[the most NIR luminous quiescent RN system known;][]{2012ApJ...746...61D}.  With the accretion disk in the Galactic recurrent SG-Nova \object{U Sco} being at least an order of magnitude less luminous \citep[see also the discussion regarding the distance to \object{U Sco} in][]{2000ApJ...534L.189H} than that in \object{M31N 2008-12a}, and given the results of \citet{2015Wil}, it seems more likely that \object{M31N 2008-12a} habours a red giant secondary star. 

We can now consider the observed implied deceleration in more detail.  As noted above, Fig.~\ref{Halpha_plot} indicates that the maximum FWHM velocity at $t-t_\mathrm{max} = 0$ was $\sim 2\,700\,\mathrm{km}\,\mathrm{s}^{-1}$ and that the derived velocity of expansion decreased systematically with time to $t-t_{\mathrm{max}}\ga3.6$~days. Also shown in this figure are power-law declines in expansion velocity for strong shocks driven into the pre-existing wind of the evolved (red giant) secondary star of the system by the high velocity ejecta of the eruption. This situation is analogous to that found in \object{RS Oph}, as presented in \citet{1985MNRAS.217..205B} and is described briefly here. 

In the early stages of the interaction, the ejecta are still imparting energy to the shocked wind (Phase I). This is followed by a phase of adiabatic expansion (Phase II) until the shock temperature drops to the point at which the shocked gas becomes well cooled and the momentum-conserving phase of development (Phase III) is established. Plotted on Fig.~\ref{Halpha_plot} are functions corresponding to the expected behaviour of the observed shock velocities when the shock is traversing a wind with a $1/r^2$ density distribution for Phase II ($v_\mathrm{s} \propto t^{-1/3}$; grey dashed line) and Phase III ($v_\mathrm{s} \propto t^{-1/2}$; red dot-dashed line). 

The best fit to the data shown in Fig.\ref{Halpha_plot} (black dotted line) in fact gives $v \propto t^{-0.12 \pm 0.05}$, i.e.\ much shallower than would be expected for either Phase II or Phase III of shocked remnant development. This type of behaviour at early times was however seen in RS Oph. \cite{2006ApJ...652..629B} noted that the change in slope of shock velocities derived from fits to \swift XRT data from the 2006 eruption occurred at $t\sim6$~days, in line with expectations from the \citet{1985MNRAS.217..205B} model for the duration of Phase I of the eruption. Similarly, \citet{2006ApJ...653L.141D} found an abrupt steepening in velocities derived from FWHM and FWZI of infrared emission lines, again from the 2006 eruption, at $t\ga4$~days (see their Fig.~2). We thus conclude that the behaviour of the expansion velocities derived in \object{M31N 2008-12a} up to $t-t_{\mathrm{max}} \sim 3.6$~days implies that the remnant was still in Phase I of development at this time.

From \citet{1985MNRAS.217..205B}, Phase I lasts for a time $t \propto M_{\mathrm{e}}u/\dot{M} v_{\mathrm{e}}$ where $M_{\mathrm{e}}$ and $v_{\mathrm{e}}$ are the ejecta mass and initial velocity respectively, and $\dot{M}$ and $u$ are the mass loss rate and velocity of the RGW. For simplicity, we initially assume that the latter two quantities were the same in \object{M31N 2008-12a} as in \object{RS Oph}. We can also compare the ejection velocities where $v_\mathrm{e} = 5\,100\,\mathrm{km}\,\mathrm{s}^{-1}$ in \object{RS Oph} \citep{2009ApJ...703.1955R} and $v_{\mathrm{e}} = 2\,700\,\mathrm{km}\,\mathrm{s}^{-1}$ in \object{M31N 2008-12a}. Thus with $M_\mathrm{e} = 2 \times 10^{-7}\,M_\odot$ \citep[see e.g.][]{2006Natur.442..279O,2009A&A...493.1049O} and taking the above values for the derived durations of Phase I of $\ga3.6$ and $\sim 6$~days in \object{M31N 2008-12a} and \object{RS Oph} respectively, this implies the total mass ejected in \object{M31N 2008-12a} is $\ga 3\times10^{-8}\,M_\odot$, which is consistent with the models presented by TBW14 and the ejected Hydrogen mass as determined by HND15 (see their Sect.~4.5). There are several caveats here which include the fact that the ejection velocity derived for \object{RS Oph} by \citet{2009ApJ...703.1955R} includes the inclination to the line of sight of the ejection, but above all that the RGW parameters are assumed to be the same in the two objects, which is probably not the case.

The combination of the luminous accretion disk and the ejecta deceleration provide strong evidence in favour of \object{M31N 2008-12a} belonging to the RG-nova sub-class.  However, it should also be noted that the rapidly declining CN \object{CP Puppis} (1942) also exhibited narrowing of its emissions lines \citep{1972SvA....16...32G} and similar effects have been reported in the RNe \object{U Sco} \citep{1999A&A...347L..39M} and \object{CI Aquilae} \citep{2001A&A...366..858K}.  As these are novae containing a main sequence (\object{CP Pup}) or sub-giant (\object{U Sco} and \object{CI Aql}) secondary stars \citep[see][and references therein]{2012ApJ...746...61D} they would not be expected to contain significant circumbinary material \citep[e.g.\ from stellar wind losses; however, see][for a recent discussion about circumbinary material around novae]{2013AJ....146...55W}.

An alternative explanation for the apparent narrowing of nova ejecta emission lines is presented in \citet{1996ApJ...456..717S}.  Given the distribution of velocities within the ejecta during the onset of the eruption \citep[see e.g.,][]{1974ApJS...28..247S}, the ejected material with the highest velocity travels the furthest distance.  As such, the density of this higher velocity material decreases at a faster rate than that of any lower velocity material in the ejecta, and therefore the higher the velocity of ejected material, the faster its emissivity decreases.  As the ejecta expands, the apparent velocity, as derived from the spectral line profile, will decrease.  Therefore, in systems with dense circumbinary material, such as the RGW around RG-novae, one would expect an combination of apparent deceleration and actual shock deceleration of the ejecta.  Differentiating between the two requires more detailed observations.

The early eruption spectra of the Galactic RNe \object{RS Oph} \citep{2009A&A...505..287I} and \object{V745 Scorpii} \citep[also a RG-nova; see e.g.,][]{1990MNRAS.246...78S} both contained \ion{Fe}{ii} emission in addition to the dominant He and N lines.  The \ion{Fe}{ii} emission lines, particularly in \object{RS Oph}, were significantly narrower ($<1\,000$~km\,s$^{-1}$) than the Balmer, He, and N emission, and the line intensities were much lower than the He and N lines \citep[see their Table~2]{2009A&A...505..287I}.  As such, the lower velocity \ion{Fe}{ii} emission seen in \object{RS Oph} and \object{V745 Sco} has been postulated to be due to the large reservoir of circumbinary material that accumulates around those systems as a consequence of their RGWs \cite[see][for a full discussion]{2012AJ....144...98W}.  The detection of lower velocity \ion{Fe}{ii} emission from \object{M31N 2008-12a} could thus provide compelling evidence of a RGW.  However, taking the line intensities from \citet{2009A&A...505..287I}, any such emission would be undetectable in the 2014 eruption spectra.

To unambiguously determine the secondary type in \object{M31N 2008-12a} additional observations are required, be they deeper NIR or IR imaging to detect the secondary directly, the determination of an orbital period, or sufficiently deep and high resolution eruption spectra; but at the distance of M31  the former currently seems the most feasible.

\section{Summary \& Conclusions}\label{sec:conclusions}

In this paper we have described the discovery of the predicted 2014 eruption of \object{M31N 2008-12a}, this system's sixth eruption in seven years.  We have also presented and described the extensive optical and UV follow-up observations.  Here, we summarise our findings.

\begin{enumerate}
\item The predicted 2014 eruption of \object{M31N 2008-12a} was detected, on the rise, at 2014 Oct 2.904 UT by a dedicated automatic pipeline put in place at the LT.

\item The subsequent $BVRr'i'I$ optical photometric monitoring campaign likely caught the point of maximum light at, or shortly before 2014 Oct $3.7\pm0.1$ UT.  The optical light curve is very similar to that of previous eruptions, notably the 2013 eruption.

\item The light curve evolution was monitored with hourly cadence for first three nights post discovery, obtaining detailed multicolor photometry that closely matched updated templates produced exclusively from previous eruptions.  For periods, the photometric cadence matched that obtained for Galactic novae by the SMEI satellite \citep{2010ApJ...724..480H,hou15}, but significant variation may have been occurring on even shorter timescales.

\item The optical decline time was measured as $t_{2}(V)=1.77\pm0.13$\,days and $t_{3}(V)=3.84\pm0.24$\,days, the light curve of \object{M31N 2008-12a} evolves faster than all Galactic RNe yet discovered.  An early optical `plateau', indicative of RNe, and an optical rebrightening, coinciding with the unveiling of the SSS emission, were also observed.

\item Optical spectra were obtained as early as 0.26\,days post-maximum light, and subsequent spectral monitoring showed remarkable similarity to the 2012 and 2013 eruptions.  These spectra confirm the classification of \object{M31N 2008-12a} as an erupting nova within M31.  These observations also included some of the first robotically acquired spectra from the SPRAT instrument on the LT.

\item As in 2012 and 2013, the spectra are dominated by Balmer, He, and N lines, with average expansion velocity of $2\,570\pm120$\,km\,s$^{-1}$.  A significant decrease in width of the H$\alpha$ line is seen over the course of two days ($510\pm110$\,km\,s$^{-1}$).  The spectral evolution is remarkably similar to that seen in previous eruptions.

\item Such a low ejection velocity is consistent with high mass-accretion rate, high mass WD, and short recurrence time models of TNRs.

\item The low peak optical luminosity and the SED at optical maximum are consistent with a low ejected mass, $T_{\mathrm{P}}\ga3.3\times10^{4}\,\mathrm{K}$ at this time, and the luminosity is suggested to be super-Eddington around optical peak.

\item Deep H$\alpha$ imaging of the quiescent system indicates the presence of an elliptical shell roughly centred at the position of the nova.  While larger than even most SNRs, such an apparent alignment gives tantalising hints at an association, and deserves further study.

\item Implied decreasing ejecta velocity may be caused by interaction with existing circumbinary material, most likely a red giant wind -- which would cement the classification of the system as an RS Oph-like RN (RG-nova).

\item Assuming a RG-nova in Phase I of remnant development, by direct analogy to the \object{RS Oph} system we can infer that the ejected mass $\ga 3\times10^{-8}\,M_\odot$, consistent with the TBW14 models.

\item The monitoring campaign to detect the next eruption, predicted to occur at the end of 2015, began on 2014 Oct 27.
\end{enumerate}

\citet{2001MNRAS.327.1323T} identified the Galactic RN \object{U Sco} as the best SN\,Ia progenitor known, estimating that it might explode within $\sim700\,000$\,years.  With its predicted very high mass WD and short inter-eruption timescale \object{M31N 2008-12a} may just have overtaken \object{U Sco} as the best single-degenerate SN\,Ia progenitor candidate known.  

The RN \object{M31N 2008-12a} is a truly unique and remarkable system, and we strongly encourage further observation of the system and its immediate environment.  We eagerly await the next eruption, which we expect to occur between October and December 2015 and for which we plan even more extensive multi-wavelength observations.

\begin{acknowledgements}
The Liverpool Telescope is operated on the island of La Palma by Liverpool John Moores University in the Spanish Observatorio del Roque de
los Muchachos of the Instituto de Astrof\'{i}sica de Canarias with
financial support from STFC.  The William Herschel Telescope is operated on the island of La Palma by the Isaac Newton Group in the Spanish Observatorio del Roque de los Muchachos of the Instituto de Astrof\'{i}sica de Canarias.  Based (in part) on data collected with the Danish 1.54-m telescope at the ESO
La Silla Observatory.  We would like to thank the LT group at LJMU for their help, and patience, in obtaining these data from the newly commissioned SPRAT spectrograph.  We are grateful to the \swift Team for making the ToO observations possible, in particular N.~Gehrels, the duty scientists, as well as the science planners.  The authors would like to specially thank Adam Muzzin and Mauro Stefanon for graciously donating a portion of their WHT time to this study.  We thank Andrew M.~Shafter (Magnolia Science Academy) for assistance with the MLO observations, Karl Misselt for taking and graciously donating the Bok data, and Phil James for advice on continuum subtraction from H$\alpha$ imaging.  This work used the \texttt{pamela} and \texttt{molly} software packages written by Tom Marsh.
PRG acknowledges support from the MINECO under the Ram\'on y Cajal programme
(RYC--2010--05762).  KH was supported by the project RVO:67985815.  MH acknowledges support from an ESA fellowship.  VARMR acknowledges financial support from the Radboud Excellence Initiative.  AWS acknowledges
support from NSF grant AST1009566.  MW was supported by the Czech Science Foundation, grant P209/10/0715.  This research has been supported by the Spanish Ministry of Economy and Competitiveness (MINECO) under the grants ESP2013-41268-R, AYA2011--23102, and AYA2012--38700.  The comments of the referee were appreciated in identifying areas that benefited from additional discussion and clarification.
\end{acknowledgements}

\clearpage
\onecolumn

\begin{longtab}
\begin{longtable}{lS[table-format=2.3]lllllll}
\caption{Optical photometric observations of the 2014 eruption of \object{M31N 2008-12a}.\label{optical_photometry}}\\
\hline\hline
Date & {$t-t_{\mathrm{max}}$} & \multicolumn{2}{c}{JD 2\,456\,000.5+} & Telescope \& & Exposure & Filter & SNR & Photometry \\
(UT)   & {(days)} & Start & End & Instrument & Time (secs)\\
\hline
\endfirsthead
\caption{Optical photometric observations of the 2014 eruption of \object{M31N 2008-12a}, continued.}\\
\hline\hline
Date & {$t-t_{\mathrm{max}}$} & \multicolumn{2}{c}{JD 2\,456\,000.5+} & Telescope  \& & Exposure & Filter & SNR & Photometry \\
(UT)   & {(days)} & Start & End & Instrument & Time (secs) \\
\hline
\endhead
\hline
\endfoot
2014 Oct 3.549 & -0.151 & \multicolumn{2}{c}{933.549} & Kiso 1.05m & $3\times60$ & $B$ & \ldots & $18.97\pm0.14$ \\
2014 Oct 3.822 & 0.122 & \multicolumn{2}{c}{933.822} &  Ond\v{r}ejov 0.65m  & $7\times90$ &  $B$  & \ldots  &  $18.7\pm0.2$\tablefootmark{c} \\ 
2014 Oct 3.878 & 0.178 & 933.875 & 933.880 &  LT IO:O  &  $3\times120$  &  $B$  &  77.6  &  $18.714\pm0.015$ \\ 
2014 Oct 3.924 & 0.223 & 933.921 & 933.926 &  LT IO:O  &  $3\times120$  &  $B$  &  81.8  &  $18.706\pm0.015$ \\ 
2014 Oct 3.980 & 0.280 & 933.978 & 933.983 &  LT IO:O  &  $3\times120$  &  $B$  &  84.2  &  $18.739\pm0.015$ \\ 
2014 Oct 4.038 & 0.338 & 934.036 & 934.041 &  LT IO:O  &  $3\times120$  &  $B$  &  90.3  &  $18.806\pm0.014$ \\ 
2014 Oct 4.083 & 0.384 & 934.081 & 934.086 &  LT IO:O  &  $3\times120$  &  $B$  &  88.4  &  $18.850\pm0.013$ \\ 
2014 Oct 4.133 & 0.433 & 934.131 & 934.135 &  LT IO:O  &  $3\times120$  &  $B$  &  90.7  &  $18.915\pm0.015$ \\ 
2014 Oct 4.179 & 0.480 & 934.175 & 934.184 &  MLO 1.0m  &  $5\times120$  &  $B$  &  \ldots  &  $18.95\pm0.04$\tablefootmark{d} \\ 
2014 Oct 4.181 & 0.481 & 934.179 & 934.183 &  LT IO:O  &  $3\times120$  &  $B$  &  84.8  &  $18.997\pm0.014$ \\ 
2014 Oct 4.898 & 1.198 & 934.896 & 934.900 &  LT IO:O  &  $3\times120$  &  $B$  &  39.2  &  $19.871\pm0.029$ \\ 
2014 Oct 4.943 & 1.243 & 934.941 & 934.945 &  LT IO:O  &  $3\times120$  &  $B$  &  41.8  &  $19.991\pm0.028$ \\ 
2014 Oct 5.013 & 1.312 & 935.010 & 935.015 &  LT IO:O  &  $3\times120$  &  $B$  &  38.7  &  $20.071\pm0.029$ \\ 
2014 Oct 5.090 & 1.390 & 935.088 & 935.092 &  LT IO:O  &  $3\times120$  &  $B$  &  39.3  &  $20.141\pm0.029$ \\ 
2014 Oct 5.163 & 1.464 & 935.161 & 935.166 &  LT IO:O  &  $3\times120$  &  $B$  &  46.2  &  $20.157\pm0.024$ \\ 
2014 Oct 5.875 & 2.175 & 935.871 & 935.878 &  LT IO:O  &  $3\times180$  &  $B$  &  11.8  &  $20.863\pm0.093$ \\ 
2014 Oct 5.946 & 2.246 & 935.940 & 935.951 &  LT IO:O  &  $3\times300$  &  $B$  &  22.7  &  $20.834\pm0.049$ \\ 
2014 Oct 6.020 & 2.320 & 936.015 & 936.025 &  LT IO:O  &  $3\times300$  &  $B$  &  22.2  &  $20.827\pm0.049$ \\ 
2014 Oct 6.096 & 2.395 & 936.090 & 936.101 &  LT IO:O  &  $3\times300$  &  $B$  &  22.0  &  $20.866\pm0.050$ \\ 
2014 Oct 6.168 & 2.468 & 936.163 & 936.174 &  LT IO:O  &  $3\times300$  &  $B$  &  16.4  &  $20.891\pm0.066$ \\ 
2014 Oct 6.875 & 3.175 & 936.870 & 936.881 &  LT IO:O  &  $3\times300$  &  $B$  &  10.7  &  $20.945\pm0.102$ \\ 
2014 Oct 7.016 & 3.316 & 937.011 & 937.022 &  LT IO:O  &  $3\times300$  &  $B$  &  25.2  &  $20.792\pm0.044$ \\ 
2014 Oct 7.181 & 3.481 & 937.176 & 937.186 &  LT IO:O  &  $3\times300$  &  $B$  &  18.4  &  $21.114\pm0.059$ \\ 
2014 Oct 7.930 & 4.231 & 937.925 & 937.936 &  LT IO:O  &  $3\times300$  &  $B$  &  7.1  &  $21.510\pm0.153$ \\ 
2014 Oct 8.824 & 5.124 & 938.819 & 938.829 &  LT IO:O  &  $3\times300$  &  $B$  &  -  &  $>20.3$ \\ 
2014 Oct 9.049 & 5.349 & 939.044 & 939.054 &  LT IO:O  &  $3\times300$  &  $B$  &  6.3  &  $21.320\pm0.174$ \\ 
2014 Oct 9.823 & 6.123 & 939.818 & 939.828 &  LT IO:O  &  $3\times300$  &  $B$  &  8.9  &  $21.293\pm0.122$ \\ 
2014 Oct 10.053 & 6.353 & 940.047 & 940.058 &  LT IO:O  &  $3\times300$  &  $B$  &  -  &  $>21.2$ \\ 
2014 Oct 11.814 & 8.114 & 941.809 & 941.820 &  LT IO:O  &  $3\times300$  &  $B$  &  -  &  $>19.4$ \\ 

\hline
2014 Oct 3.299 & -0.401      & \multicolumn{2}{c}{933.299} & iTelescope T24 & $3\times180$ & $V$ & \ldots & $18.71\pm0.14$ \\
2014 Oct 3.536 & -0.164      & \multicolumn{2}{c}{933.536} & Kiso 1.05m & $3\times60$ & $V$ & \ldots & $18.68\pm0.10$ \\
2014 Oct 3.815 & 0.115 & \multicolumn{2}{c}{933.815} &  Ond\v{r}ejov 0.65m  & $7\times90$  &  $V$  &  \ldots  &  $18.5\pm0.1$\tablefootmark{c} \\ 
2014 Oct 3.882 & 0.183 & 933.880 & 933.885 &  LT IO:O  &  $3\times120$  &  $V$  &  79.1  &  $18.592\pm0.015$ \\ 
2014 Oct 3.928 & 0.228 & 933.926 & 933.931 &  LT IO:O  &  $3\times120$  &  $V$  &  88.8  &  $18.583\pm0.014$ \\ 
2014 Oct 3.986 & 0.285 & 933.983 & 933.988 &  LT IO:O  &  $3\times120$  &  $V$  &  91.9  &  $18.599\pm0.014$ \\ 
2014 Oct 4.043 & 0.343 & 934.041 & 934.046 &  LT IO:O  &  $3\times120$  &  $V$  &  93.4  &  $18.668\pm0.014$ \\ 
2014 Oct 4.089 & 0.388 & 934.086 & 934.091 &  LT IO:O  &  $3\times120$  &  $V$  &  96.9  &  $18.704\pm0.013$ \\ 
2014 Oct 4.138 & 0.438 & 934.136 & 934.140 &  LT IO:O  &  $3\times120$  &  $V$  &  98.3  &  $18.769\pm0.013$ \\ 
2014 Oct 4.158 & 0.458 & 934.152 & 934.163 &  MLO 1.0m  &  $5\times120$  &  $V$  &  \ldots  &  $18.78\pm0.04$\tablefootmark{d} \\ 
2014 Oct 4.177 & 0.477 & \multicolumn{2}{c}{934.177} & iTelescope T11 & $5\times180$ & $V$ & \ldots & $18.74\pm0.30$ \\
2014 Oct 4.186 & 0.486 & 934.184 & 934.188 &  LT IO:O  &  $3\times120$  &  $V$  &  92.7  &  $18.813\pm0.013$ \\ 
2014 Oct 4.214 & 0.514 & \multicolumn{2}{c}{934.214} & iTelescope T24 & $3\times180$ & $V$ & \ldots & $18.74\pm0.09$ \\
2014 Oct 4.903 & 1.203 & 934.901 & 934.905 &  LT IO:O  &  $3\times120$  &  $V$  &  39.6  &  $19.856\pm0.028$ \\ 
2014 Oct 4.948 & 1.248 & 934.946 & 934.950 &  LT IO:O  &  $3\times120$  &  $V$  &  39.3  &  $19.894\pm0.028$ \\ 
2014 Oct 5.017 & 1.317 & 935.015 & 935.020 &  LT IO:O  &  $3\times120$  &  $V$  &  34.4  &  $19.976\pm0.032$ \\ 
2014 Oct 5.095 & 1.395 & 935.093 & 935.097 &  LT IO:O  &  $3\times120$  &  $V$  &  37.2  &  $19.983\pm0.030$ \\ 
2014 Oct 5.168 & 1.469 & 935.166 & 935.171 &  LT IO:O  &  $3\times120$  &  $V$  &  43.5  &  $20.102\pm0.026$ \\ 
2014 Oct 5.881 & 2.182 & 935.878 & 935.885 &  LT IO:O  &  $3\times180$  &  $V$  &  10.8  &  $21.089\pm0.101$ \\ 
2014 Oct 5.957 & 2.257 & 935.952 & 935.963 &  LT IO:O  &  $3\times300$  &  $V$  &  19.2  &  $20.945\pm0.057$ \\ 
2014 Oct 6.032 & 2.331 & 936.026 & 936.037 &  LT IO:O  &  $3\times300$  &  $V$  &  17.0  &  $21.102\pm0.064$ \\ 
2014 Oct 6.106 & 2.407 & 936.101 & 936.112 &  LT IO:O  &  $3\times300$  &  $V$  &  15.7  &  $21.010\pm0.069$ \\ 
2014 Oct 6.179 & 2.480 & 936.174 & 936.185 &  LT IO:O  &  $3\times300$  &  $V$  &  12.6  &  $21.043\pm0.087$ \\
2014 Oct 6.419 & 2.719 & \multicolumn{2}{c}{936.419} & Kiso 1.05m & $3\times60$ & $V$ & \ldots & $>18.8$ \\ 
2014 Oct 6.887 & 3.187 & 936.881 & 936.892 &  LT IO:O  &  $3\times300$  &  $V$  &  15.0  &  $20.771\pm0.073$ \\ 
2014 Oct 7.028 & 3.328 & 937.022 & 937.033 &  LT IO:O  &  $3\times300$  &  $V$  &  16.4  &  $20.865\pm0.067$ \\ 
2014 Oct 7.192 & 3.492 & 937.187 & 937.198 &  LT IO:O  &  $3\times300$  &  $V$  &  11.7  &  $21.350\pm0.093$ \\ 
2014 Oct 7.942 & 4.242 & 937.936 & 937.947 &  LT IO:O  &  $3\times300$  &  $V$  &  -  &  $>20.6$ \\ 
2014 Oct 8.836 & 5.135 & 938.830 & 938.841 &  LT IO:O  &  $3\times300$  &  $V$  &  5.2  &  $21.764\pm0.210$ \\ 
2014 Oct 9.061 & 5.360 & 939.055 & 939.066 &  LT IO:O  &  $3\times300$  &  $V$  &  -  &  $>20.1$ \\ 
2014 Oct 9.834 & 6.134 & 939.829 & 939.840 &  LT IO:O  &  $3\times300$  &  $V$  &  13.6  &  $21.214\pm0.081$ \\ 
2014 Oct 10.064 & 6.364 & 940.059 & 940.070 &  LT IO:O  &  $3\times300$  &  $V$  &  -  &  $>21.1$ \\ 
2014 Oct 11.826 & 8.126 & 941.820 & 941.831 &  LT IO:O  &  $3\times300$  &  $V$  &  5.9  &  $22.334\pm0.183$ \\ 

\hline
2014 Oct 3.738 & 0.038 & \multicolumn{2}{c}{933.738} &  Ond\v{r}ejov 0.65m  & $10\times90$  &  $R$  &  \ldots  &  $18.2\pm0.1$\tablefootmark{c}\\ 
2014 Oct 3.807 & 0.107 & \multicolumn{2}{c}{933.807} &  Ond\v{r}ejov 0.65m  & $7\times90$  &  $R$  &  \ldots  &  $18.27\pm0.09$\tablefootmark{c} \\ 
2014 Oct 3.915 & 0.215 & \multicolumn{2}{c}{933.915} &  Ond\v{r}ejov 0.65m  & $7\times90$  &  $R$  &  \ldots  &  $18.32\pm0.08$\tablefootmark{c} \\ 
2014 Oct 4.083 & 0.383 & \multicolumn{2}{c}{934.083} &  Ond\v{r}ejov 0.65m  & $12\times90$  &  $R$  &  \ldots  &  $18.38\pm0.07$\tablefootmark{c} \\ 
2014 Oct 4.169 & 0.469 & \multicolumn{2}{c}{934.169} &  Ond\v{r}ejov 0.65m  & $7\times90$  &  $R$  &  \ldots  &  $18.4\pm0.1$\tablefootmark{c} \\ 
2014 Oct 4.169 & 0.469 & 934.164 & 934.174 &  MLO 1.0m  &  $5\times120$  &  $R$  &  \ldots  &  $18.48\pm0.03$\tablefootmark{d} \\ 
2014 Oct 4.929 & 1.229 & \multicolumn{2}{c}{934.929} &  Ond\v{r}ejov 0.65m  & $13\times90$  &  $R$  &  \ldots  &  $19.4\pm0.2$\tablefootmark{c} \\ 
2014 Oct 5.935 & 2.235 & \multicolumn{2}{c}{935.935} &  Ond\v{r}ejov 0.65m  & $15\times90$  &  $R$  &  \ldots  &  $20.2\pm0.3$\tablefootmark{c}\\ 
2014 Oct 6.161 & 2.461 & \multicolumn{2}{c}{936.161} &  Ond\v{r}ejov 0.65m  & $20\times90$  &  $R$  &  \ldots  &  $20.1\pm0.3$\tablefootmark{c} \\ 
2014 Oct 10.206 & 6.506 & \multicolumn{2}{c}{940.206} &  Danish 1.54m  & $3\times90$  &  $R$  &  -  &  $>19.7$ \\	 
2014 Oct 12.193 & 8.493 & \multicolumn{2}{c}{942.193} &  Danish 1.54m  & $8\times90$  &  $R$  &  -  &  $>21.1$ \\	 

\hline
2014 Oct 3.750 & 0.050 & \multicolumn{2}{c}{933.750} &  Ond\v{r}ejov 0.65m  & $10\times90$  &  $I$  & \ldots  &  $18.2\pm0.1$ \tablefootmark{c}\\ 
2014 Oct 4.188 & 0.489 & 934.184 & 934.193 &  MLO 1.0m  &  $5\times120$  &  $I$  &  \ldots  &  $18.51\pm0.04$\tablefootmark{d} \\ 

\hline
2014 Oct 2.468 & -1.232 & \multicolumn{2}{c}{932.468} & iPTF & \ldots & $g$ & \ldots & $>19.5$\tablefootmark{b}\\

\hline
2014 Aug 30.984 & -33.717 & 868.057 & 931.910 &  LT IO:O  &  $50\times60$  &  $r'$  &  -  &  $>21.2$ \\ 
2014 Oct 1.909 & -1.790 & 931.909 & 931.910 &  LT IO:O  &  60  &  $r'$  &  -  &  $>20.4$\tablefootmark{a} \\ 
2014 Oct 2.904 & -0.796 & 932.904 & 932.904 &  LT IO:O  &  60  &  $r'$  &  52.4  &  $18.862\pm0.022$\tablefootmark{a} \\ 
2014 Oct 3.888 & 0.188 & 933.885 & 933.890 &  LT IO:O  &  $3\times120$  &  $r'$  &  104.1  &  $18.466\pm0.012$\tablefootmark{a} \\ 
2014 Oct 3.934 & 0.233 & 933.931 & 933.936 &  LT IO:O  &  $3\times120$  &  $r'$  &  120.5  &  $18.498\pm0.010$ \\ 
2014 Oct 3.991 & 0.290 & 933.988 & 933.993 &  LT IO:O  &  $3\times120$  &  $r'$  &  118.5  &  $18.498\pm0.010$ \\ 
2014 Oct 4.049 & 0.348 & 934.046 & 934.051 &  LT IO:O  &  $3\times120$  &  $r'$  &  112.5  &  $18.523\pm0.011$ \\ 
2014 Oct 4.093 & 0.393 & 934.091 & 934.096 &  LT IO:O  &  $3\times120$  &  $r'$  &  117.9  &  $18.560\pm0.011$ \\ 
2014 Oct 4.143 & 0.443 & 934.141 & 934.145 &  LT IO:O  &  $3\times120$  &  $r'$  &  126.7  &  $18.596\pm0.010$ \\ 
2014 Oct 4.191 & 0.491 & 934.189 & 934.193 &  LT IO:O  &  $3\times120$  &  $r'$  &  116.9  &  $18.619\pm0.011$ \\ 
2014 Oct 4.908 & 1.208 & 934.905 & 934.910 &  LT IO:O  &  $3\times120$  &  $r'$  &  61.2  &  $19.466\pm0.019$ \\ 
2014 Oct 4.953 & 1.253 & 934.951 & 934.955 &  LT IO:O  &  $3\times120$  &  $r'$  &  64.0  &  $19.453\pm0.018$ \\ 
2014 Oct 5.023 & 1.322 & 935.020 & 935.025 &  LT IO:O  &  $3\times120$  &  $r'$  &  57.5  &  $19.609\pm0.019$ \\ 
2014 Oct 5.100 & 1.400 & 935.098 & 935.102 &  LT IO:O  &  $3\times120$  &  $r'$  &  53.7  &  $19.672\pm0.021$ \\ 
2014 Oct 5.174 & 1.474 & 935.171 & 935.176 &  LT IO:O  &  $3\times120$  &  $r'$  &  55.8  &  $19.673\pm0.020$ \\ 
2014 Oct 5.889 & 2.189 & 935.885 & 935.892 &  LT IO:O  &  $3\times180$  &  $r'$  &  21.6  &  $20.427\pm0.051$ \\ 
2014 Oct 5.968 & 2.268 & 935.963 & 935.974 &  LT IO:O  &  $3\times300$  &  $r'$  &  26.7  &  $20.432\pm0.047$ \\ 
2014 Oct 6.043 & 2.342 & 936.037 & 936.048 &  LT IO:O  &  $3\times300$  &  $r'$  &  27.8  &  $20.483\pm0.042$ \\ 
2014 Oct 6.118 & 2.418 & 936.112 & 936.123 &  LT IO:O  &  $3\times300$  &  $r'$  &  27.5  &  $20.547\pm0.042$ \\ 
2014 Oct 6.190 & 2.491 & 936.185 & 936.196 &  LT IO:O  &  $3\times300$  &  $r'$  &  23.4  &  $20.602\pm0.048$ \\ 
2014 Oct 6.898 & 3.198 & 936.892 & 936.903 &  LT IO:O  &  $3\times300$  &  $r'$  &  21.7  &  $20.475\pm0.054$ \\ 
2014 Oct 7.038 & 3.339 & 937.033 & 937.044 &  LT IO:O  &  $3\times300$  &  $r'$  &  17.2  &  $20.613\pm0.067$ \\ 
2014 Oct 7.203 & 3.503 & 937.198 & 937.209 &  LT IO:O  &  $3\times300$  &  $r'$  &  14.6  &  $20.890\pm0.077$ \\ 
2014 Oct 7.953 & 4.253 & 937.948 & 937.959 &  LT IO:O  &  $3\times300$  &  $r'$  &  8.3  &  $21.183\pm0.131$ \\ 
2014 Oct 8.846 & 5.146 & 938.841 & 938.852 &  LT IO:O  &  $3\times300$  &  $r'$  &  5.6  &  $21.708\pm0.196$ \\ 
2014 Oct 9.846 & 6.145 & 939.840 & 939.851 &  LT IO:O  &  $3\times300$  &  $r'$  &  12.6  &  $21.254\pm0.093$ \\ 
2014 Oct 10.076 & 6.375 & 940.070 & 940.081 &  LT IO:O  &  $3\times300$  &  $r'$  &  5.5  &  $22.086\pm0.199$ \\ 
2014 Oct 11.837 & 8.137 & 941.832 & 941.842 &  LT IO:O  &  $3\times300$  &  $r'$  &  -  &  $>21.2$ \\ 
2014 Nov 11.892 & 39.192 & 968.909 & 976.875 &  LT IO:O  &  $9\times60$  &  $r'$  &  -  &  $>21.0$ \\ 

\hline
2014 Oct 3.892 & 0.192 & 933.890 & 933.895 &  LT IO:O  &  $3\times120$  &  $i'$  &  65.8  &  $18.614\pm0.019$ \\ 
2014 Oct 3.938 & 0.238 & 933.936 & 933.941 &  LT IO:O  &  $3\times120$  &  $i'$  &  82.6  &  $18.601\pm0.015$ \\ 
2014 Oct 3.995 & 0.295 & 933.993 & 933.998 &  LT IO:O  &  $3\times120$  &  $i'$  &  79.5  &  $18.614\pm0.016$ \\ 
2014 Oct 4.053 & 0.353 & 934.051 & 934.056 &  LT IO:O  &  $3\times120$  &  $i'$  &  82.4  &  $18.623\pm0.016$ \\ 
2014 Oct 4.099 & 0.398 & 934.096 & 934.101 &  LT IO:O  &  $3\times120$  &  $i'$  &  80.1  &  $18.663\pm0.015$ \\ 
2014 Oct 4.148 & 0.448 & 934.146 & 934.150 &  LT IO:O  &  $3\times120$  &  $i'$  &  84.0  &  $18.693\pm0.014$ \\ 
2014 Oct 4.196 & 0.496 & 934.194 & 934.198 &  LT IO:O  &  $3\times120$  &  $i'$  &  78.8  &  $18.664\pm0.015$ \\ 
2014 Oct 4.912 & 1.213 & 934.910 & 934.915 &  LT IO:O  &  $3\times120$  &  $i'$  &  43.8  &  $19.456\pm0.026$ \\ 
2014 Oct 4.958 & 1.258 & 934.956 & 934.960 &  LT IO:O  &  $3\times120$  &  $i'$  &  37.1  &  $19.489\pm0.030$ \\ 
2014 Oct 5.027 & 1.327 & 935.025 & 935.030 &  LT IO:O  &  $3\times120$  &  $i'$  &  40.5  &  $19.567\pm0.028$ \\ 
2014 Oct 5.105 & 1.405 & 935.103 & 935.107 &  LT IO:O  &  $3\times120$  &  $i'$  &  38.8  &  $19.624\pm0.029$ \\ 
2014 Oct 5.178 & 1.479 & 935.176 & 935.181 &  LT IO:O  &  $3\times120$  &  $i'$  &  35.9  &  $19.675\pm0.031$ \\ 
2014 Oct 5.896 & 2.196 & 935.892 & 935.899 &  LT IO:O  &  $3\times180$  &  $i'$  &  14.4  &  $20.522\pm0.076$ \\ 
2014 Oct 5.980 & 2.280 & 935.974 & 935.985 &  LT IO:O  &  $3\times300$  &  $i'$  &  17.5  &  $20.475\pm0.062$ \\ 
2014 Oct 6.053 & 2.354 & 936.048 & 936.059 &  LT IO:O  &  $3\times300$  &  $i'$  &  16.7  &  $20.596\pm0.065$ \\ 
2014 Oct 6.130 & 2.429 & 936.124 & 936.135 &  LT IO:O  &  $3\times300$  &  $i'$  &  16.0  &  $20.651\pm0.068$ \\ 
2014 Oct 6.202 & 2.502 & 936.197 & 936.207 &  LT IO:O  &  $3\times300$  &  $i'$  &  13.8  &  $20.676\pm0.079$ \\ 
2014 Oct 6.909 & 3.209 & 936.904 & 936.914 &  LT IO:O  &  $3\times300$  &  $i'$  &  15.9  &  $20.708\pm0.069$ \\ 
2014 Oct 7.050 & 3.350 & 937.045 & 937.055 &  LT IO:O  &  $3\times300$  &  $i'$  &  10.3  &  $20.766\pm0.106$ \\ 
2014 Oct 7.215 & 3.515 & 937.209 & 937.220 &  LT IO:O  &  $3\times300$  &  $i'$  &  8.1  &  $21.137\pm0.135$ \\ 
2014 Oct 7.965 & 4.264 & 937.959 & 937.970 &  LT IO:O  &  $3\times300$  &  $i'$  &  5.5  &  $21.655\pm0.198$ \\ 
2014 Oct 8.858 & 5.158 & 938.852 & 938.863 &  LT IO:O  &  $3\times300$  &  $i'$  &  -  &  $>20.1$ \\ 
2014 Oct 9.856 & 6.157 & 939.851 & 939.862 &  LT IO:O  &  $3\times300$  &  $i'$  &  5.9  &  $21.464\pm0.186$ \\ 
2014 Oct 10.087 & 6.387 & 940.081 & 940.092 &  LT IO:O  &  $3\times300$  &  $i'$  &  10.9  &  $20.806\pm0.099$ \\ 
2014 Oct 11.849 & 8.148 & 941.843 & 941.854 &  LT IO:O  &  $3\times300$  &  $i'$  &  11.1  &  $20.952\pm0.098$ \\ 

\hline
2014 Jul 30.139 & -65.561 & 868.116 & 868.162 &  LT IO:O  &  $20\times1\,200$  &  H$\alpha$  &  -  &  $>20.8$ \\
2014 Oct 5.902 & 2.203 & 935.899 & 935.906 &  LT IO:O  &  $3\times180$  &  H$\alpha$  &  30.5  &  $18.882\pm0.041$ \\ 
2014 Oct 5.990 & 2.291 & 935.985 & 935.996 &  LT IO:O  &  $3\times300$  &  H$\alpha$  &  40.7  &  $18.914\pm0.034$ \\ 
2014 Oct 6.065 & 2.365 & 936.060 & 936.070 &  LT IO:O  &  $3\times300$  &  H$\alpha$  &  36.5  &  $18.922\pm0.036$ \\ 
2014 Oct 6.140 & 2.440 & 936.135 & 936.146 &  LT IO:O  &  $3\times300$  &  H$\alpha$  &  36.1  &  $19.012\pm0.036$ \\ 
2014 Oct 6.214 & 2.513 & 936.208 & 936.219 &  LT IO:O  &  $3\times300$  &  H$\alpha$  &  36.3  &  $19.043\pm0.035$ \\ 
2014 Oct 6.920 & 3.220 & 936.915 & 936.926 &  LT IO:O  &  $3\times300$  &  H$\alpha$  &  32.8  &  $19.098\pm0.039$ \\ 
2014 Oct 7.062 & 3.361 & 937.056 & 937.067 &  LT IO:O  &  $3\times300$  &  H$\alpha$  &  27.8  &  $19.202\pm0.044$ \\ 
2014 Oct 7.226 & 3.526 & 937.221 & 937.231 &  LT IO:O  &  $3\times300$  &  H$\alpha$  &  17.5  &  $19.459\pm0.066$ \\ 
2014 Oct 7.976 & 4.276 & 937.970 & 937.981 &  LT IO:O  &  $3\times300$  &  H$\alpha$  &  13.1  &  $19.797\pm0.085$ \\ 
2014 Oct 8.869 & 5.169 & 938.863 & 938.874 &  LT IO:O  &  $3\times300$  &  H$\alpha$  &  9.3  &  $20.200\pm0.119$ \\ 
2014 Oct 9.094 & 5.394 & 939.089 & 939.099 &  LT IO:O  &  $3\times300$  &  H$\alpha$  &  -  &  $>16.9$ \\ 
2014 Oct 9.868 & 6.168 & 939.863 & 939.873 &  LT IO:O  &  $3\times300$  &  H$\alpha$  &  8.3  &  $20.658\pm0.133$ \\ 
2014 Oct 10.097 & 6.398 & 940.092 & 940.103 &  LT IO:O  &  $3\times300$  &  H$\alpha$  &  7.3  &  $20.857\pm0.150$ \\ 
2014 Oct 11.860 & 8.159 & 941.854 & 941.865 &  LT IO:O  &  $3\times300$  &  H$\alpha$  &  -  &  $>20.4$ \\ 
2014 Oct 14.955 & 11.254 & 944.949 & 944.960 &  LT IO:O  &  $3\times300$  &  H$\alpha$  &  -  &  $>20.6$ \\ 
2014 Oct 15.947 & 12.247 & 945.942 & 945.953 &  LT IO:O  &  $3\times300$  &  H$\alpha$  &  -  &  $>16.9$ \\ 
2014 Oct 16.930 & 13.230 & 946.925 & 946.936 &  LT IO:O  &  $3\times300$  &  H$\alpha$  &  -  &  $>20.5$ \\ 
2014 Oct 18.026 & 14.326 & 948.021 & 948.031 &  LT IO:O  &  $3\times300$  &  H$\alpha$  &  -  &  $>20.8$ \\ 
2014 Oct 29.978 & 26.278 & 959.973 & 959.984 &  LT IO:O  &  $3\times300$  &  H$\alpha$  &  -  &  $>19.8$ \\ 
2014 Oct 31.953 & 28.253 & 961.948 & 961.959 &  LT IO:O  &  $3\times300$  &  H$\alpha$  &  - &  $>20.0$ \\ 
2014 Nov 4.106 & 31.405 & 965.100 & 965.111 &  LT IO:O  &  $3\times300$  &  H$\alpha$  &  -  &  $>19.8$ \\ 
2014 Nov 5.990 & 33.290 & 966.985 & 966.996 &  LT IO:O  &  $3\times300$  &  H$\alpha$  &  -  &  $>19.8$ \\ 
2014 Nov 6.933 & 34.234 & 967.928 & 967.939 &  LT IO:O  &  $3\times300$  &  H$\alpha$  &  -  &  $>20.0$ \\ 

\end{longtable}
\tablefoot{
\tablefoottext{a}{\citet{2014ATel.6535....1D}},
\tablefoottext{b}{\citet{2014ATel.6532....1C}},
\tablefoottext{c}{\citet{2014ATel.6546....1H}},
\tablefoottext{d}{\citet{2014ATel.6543....1S}}.
}
\end{longtab}

\begin{longtab}
\begin{longtable}{lS[table-format=2.3]lllll}
\caption{Color evolution of the 2014 eruption of \object{M31N 2008-12a}.\label{colour_tab}}\\
\hline\hline
Date & {$t-t_{\mathrm{max}}$} & \multicolumn{2}{c}{JD 2\,456\,000.5+} & Telescope \& & Filters  & Color \\
(UT)   & {(days)} & Start & End & Instrument & \\
\hline
\endfirsthead
\caption{Color evolution of the 2014 eruption of \object{M31N 2008-12a}, continued.}\\
\hline\hline
Date & {$t-t_{\mathrm{max}}$} & \multicolumn{2}{c}{JD 2\,456\,000.5+} & Telescope \& & Filters  & Color \\
(UT)   & {(days)} & Start & End & Instrument &\\
\hline
\endhead
\hline
\endfoot
2014 Oct 3.543 & -0.158 & 933.536 & 933.549 & Kiso 1.05m & $(B-V)$ & $0.29\pm0.17$\\
2014 Oct 3.819 & 0.119 & 933.815 & 933.822 &  Ond\v{r}ejov 0.65m  & $(B-V)$  &  $0.2\pm0.2$\\
2014 Oct 3.880 & 0.180 & 933.875 & 933.885 & LT IO:O	 & $(B-V)$	 & $0.122\pm	0.022$\\
2014 Oct 3.926 & 0.226 & 933.921 & 933.931 & LT IO:O	 & $(B-V)$	 & $0.123\pm	0.020$\\
2014 Oct 3.983 & 0.283 & 933.978 & 933.988 & LT IO:O	 & $(B-V)$	 & $0.140\pm	0.020$\\
2014 Oct 4.041 & 0.341 & 934.036 & 934.046 & LT IO:O	 & $(B-V)$	 & $0.138\pm	0.020$\\
2014 Oct 4.086 & 0.386 & 934.081 & 934.091 & LT IO:O	 & $(B-V)$	 & $0.145\pm	0.018$\\
2014 Oct 4.135 & 0.436 & 934.131 & 934.140 & LT IO:O	 & $(B-V)$	 & $0.146\pm	0.019$\\
2014 Oct 4.168 & 0.468 & 934.157 & 934.178 &  MLO 1.0m  &  $(B-V)$  &  $0.17\pm0.06$ \\
2014 Oct 4.183 & 0.484 & 934.179 & 934.188 & LT IO:O	 & $(B-V)$	 & $0.185\pm	0.019$\\
2014 Oct 4.900 & 1.200 & 934.896 & 934.905 & LT IO:O	 & $(B-V)$	 & $0.014\pm	0.040$\\
2014 Oct 4.946 & 1.246 & 934.941 & 934.950 & LT IO:O	 & $(B-V)$	 & $0.097\pm	0.040$\\
2014 Oct 5.015 & 1.315 & 935.010 & 935.020 & LT IO:O	 & $(B-V)$	 & $0.095\pm	0.043$\\
2014 Oct 5.092 & 1.393 & 935.088 & 935.097 & LT IO:O	 & $(B-V)$	 & $0.158\pm	0.041$\\
2014 Oct 5.166 & 1.466 & 935.161 & 935.171 & LT IO:O	 & $(B-V)$	 & $0.055\pm	0.035$\\
2014 Oct 5.878 & 2.178 & 935.871 & 935.885 & LT IO:O	 & $(B-V)$	 & $-0.227\pm	0.137$\\
2014 Oct 5.952 & 2.252 & 935.940 & 935.963 & LT IO:O	 & $(B-V)$	 & $-0.111\pm	0.075$\\
2014 Oct 6.026 & 2.326 & 936.015 & 936.037 & LT IO:O	 & $(B-V)$	 & $-0.275\pm	0.081$\\
2014 Oct 6.101 & 2.401 & 936.090 & 936.112 & LT IO:O	 & $(B-V)$	 & $-0.144\pm	0.085$\\
2014 Oct 6.174 & 2.474 & 936.163 & 936.185 & LT IO:O	 & $(B-V)$	 & $-0.153\pm	0.109$\\
2014 Oct 6.881 & 3.181 & 936.870 & 936.892 & LT IO:O	 & $(B-V)$	 & $0.174\pm	0.125$\\
2014 Oct 7.022 & 3.322 & 937.011 & 937.033 & LT IO:O	 & $(B-V)$	 & $-0.074\pm	0.080$\\
2014 Oct 7.187 & 3.487 & 937.176 & 937.198 & LT IO:O	 & $(B-V)$	 & $-0.237\pm	0.110$\\
2014 Oct 9.829 & 6.129 & 939.818 & 939.840 & LT IO:O	 & $(B-V)$	 & $0.079\pm	0.147$\\
\hline
2014 Oct 3.885 & 0.185 & 933.880 & 933.890 & 	LT IO:O	 & 	$(V-r')$	 & 	$	0.126	\pm	0.019	$ \\
2014 Oct 3.931 & 0.231 & 933.926 & 933.936 & 	LT IO:O	 & 	$(V-r')$	 & 	$	0.086	\pm	0.017	$ \\
2014 Oct 3.988 & 0.288 & 933.983 & 933.993 & 	LT IO:O	 & 	$(V-r')$	 & 	$	0.101	\pm	0.017	$ \\
2014 Oct 4.046 & 0.346 & 934.041 & 934.051 & 	LT IO:O	 & 	$(V-r')$	 & 	$	0.145	\pm	0.017	$ \\
2014 Oct 4.091 & 0.391 & 934.086 & 934.096 & 	LT IO:O	 & 	$(V-r')$	 & 	$	0.145	\pm	0.017	$ \\
2014 Oct 4.140 & 0.440 & 934.136 & 934.145 & 	LT IO:O	 & 	$(V-r')$	 & 	$	0.173	\pm	0.016	$ \\
2014 Oct 4.188 & 0.489 & 934.184 & 934.193 & 	LT IO:O	 & 	$(V-r')$	 & 	$	0.194	\pm	0.017	$ \\
2014 Oct 4.905 & 1.205 & 934.901 & 934.910 & 	LT IO:O	 & 	$(V-r')$	 & 	$	0.391	\pm	0.034	$ \\
2014 Oct 4.951 & 1.251 & 934.946 & 934.955 & 	LT IO:O	 & 	$(V-r')$	 & 	$	0.441	\pm	0.033	$ \\
2014 Oct 5.020 & 1.320 & 935.015 & 935.025 & 	LT IO:O	 & 	$(V-r')$	 & 	$	0.366	\pm	0.038	$ \\
2014 Oct 5.097 & 1.398 & 935.093 & 935.102 & 	LT IO:O	 & 	$(V-r')$	 & 	$	0.310	\pm	0.037	$ \\
2014 Oct 5.171 & 1.471 & 935.166 & 935.176 & 	LT IO:O	 & 	$(V-r')$	 & 	$	0.429	\pm	0.032	$ \\
2014 Oct 5.885 & 2.185 & 935.878 & 935.892 & 	LT IO:O	 & 	$(V-r')$	 & 	$	0.662	\pm	0.113	$ \\
2014 Oct 5.963 & 2.263 & 935.952 & 935.974 & 	LT IO:O	 & 	$(V-r')$	 & 	$	0.513	\pm	0.074	$ \\
2014 Oct 6.037 & 2.337 & 936.026 & 936.048 & 	LT IO:O	 & 	$(V-r')$	 & 	$	0.619	\pm	0.077	$ \\
2014 Oct 6.112 & 2.412 & 936.101 & 936.123 & 	LT IO:O	 & 	$(V-r')$	 & 	$	0.463	\pm	0.081	$ \\
2014 Oct 6.185 & 2.485 & 936.174 & 936.196 & 	LT IO:O	 & 	$(V-r')$	 & 	$	0.441	\pm	0.099	$ \\
2014 Oct 6.892 & 3.192 & 936.881 & 936.903 & 	LT IO:O	 & 	$(V-r')$	 & 	$	0.295	\pm	0.091	$ \\
2014 Oct 7.033 & 3.333 & 937.022 & 937.044 & 	LT IO:O	 & 	$(V-r')$	 & 	$	0.253	\pm	0.095	$ \\
2014 Oct 7.198 & 3.498 & 937.187 & 937.209 & 	LT IO:O	 & 	$(V-r')$	 & 	$	0.460	\pm	0.121	$ \\										
2014 Oct 8.841 & 5.141 & 938.830 & 938.852 & 	LT IO:O	 & 	$(V-r')$	 & 	$	0.057	\pm	0.287	$ \\
2014 Oct 9.840 & 6.140 & 939.829 & 939.851 & 	LT IO:O	 & 	$(V-r')$	 & 	$	-0.041	\pm	0.123	$ \\
\hline
2014 Oct 3.890 & 0.190 & 933.885 & 933.895 & 	LT IO:O	 & 	$(r'-i')$	 & 	$	-0.147	\pm	0.023	$ \\
2014 Oct 3.936 & 0.236 & 933.931 & 933.941 & 	LT IO:O	 & 	$(r'-i')$	 & 	$	-0.103	\pm	0.018	$ \\
2014 Oct 3.993 & 0.293 & 933.988 & 933.998 & 	LT IO:O	 & 	$(r'-i')$	 & 	$	-0.116	\pm	0.019	$ \\
2014 Oct 4.051 & 0.351 & 934.046 & 934.056 & 	LT IO:O	 & 	$(r'-i')$	 & 	$	-0.100	\pm	0.019	$ \\
2014 Oct 4.096 & 0.396 & 934.091 & 934.101 & 	LT IO:O	 & 	$(r'-i')$	 & 	$	-0.103	\pm	0.018	$ \\
2014 Oct 4.145 & 0.445 & 934.141 & 934.150 & 	LT IO:O	 & 	$(r'-i')$	 & 	$	-0.097	\pm	0.017	$ \\
2014 Oct 4.193 & 0.494 & 934.189 & 934.198 & 	LT IO:O	 & 	$(r'-i')$	 & 	$	-0.045	\pm	0.019	$ \\
2014 Oct 4.910 & 1.210 & 934.905 & 934.915 & 	LT IO:O	 & 	$(r'-i')$	 & 	$	0.010	\pm	0.032	$ \\
2014 Oct 4.956 & 1.256 & 934.951 & 934.960 & 	LT IO:O	 & 	$(r'-i')$	 & 	$	-0.036	\pm	0.035	$ \\
2014 Oct 5.025 & 1.325 & 935.020 & 935.030 & 	LT IO:O	 & 	$(r'-i')$	 & 	$	0.042	\pm	0.034	$ \\
2014 Oct 5.102 & 1.403 & 935.098 & 935.107 & 	LT IO:O	 & 	$(r'-i')$	 & 	$	0.049	\pm	0.036	$ \\
2014 Oct 5.176 & 1.476 & 935.171 & 935.181 & 	LT IO:O	 & 	$(r'-i')$	 & 	$	-0.002	\pm	0.037	$ \\
2014 Oct 5.892 & 2.192 & 935.885 & 935.899 & 	LT IO:O	 & 	$(r'-i')$	 & 	$	-0.095	\pm	0.091	$ \\
2014 Oct 5.974 & 2.274 & 935.963 & 935.985 & 	LT IO:O	 & 	$(r'-i')$	 & 	$	-0.043	\pm	0.078	$ \\
2014 Oct 6.048 & 2.348 & 936.037 & 936.059 & 	LT IO:O	 & 	$(r'-i')$	 & 	$	-0.113	\pm	0.078	$ \\
2014 Oct 6.123 & 2.424 & 936.112 & 936.135 & 	LT IO:O	 & 	$(r'-i')$	 & 	$	-0.104	\pm	0.080	$ \\
2014 Oct 6.196 & 2.496 & 936.185 & 936.207 & 	LT IO:O	 & 	$(r'-i')$	 & 	$	-0.074	\pm	0.092	$ \\
2014 Oct 6.903 & 3.203 & 936.892 & 936.914 & 	LT IO:O	 & 	$(r'-i')$	 & 	$	-0.232	\pm	0.088	$ \\
2014 Oct 7.044 & 3.344 & 937.033 & 937.055 & 	LT IO:O	 & 	$(r'-i')$	 & 	$	-0.153	\pm	0.126	$ \\
2014 Oct 7.209 & 3.509 & 937.198 & 937.220 & 	LT IO:O	 & 	$(r'-i')$	 & 	$	-0.246	\pm	0.155	$ \\
2014 Oct 7.959 & 4.259 & 937.948 & 937.970 & 	LT IO:O	 & 	$(r'-i')$	 & 	$	-0.471	\pm	0.238	$ \\													
2014 Oct 9.851 & 6.151 & 939.840 & 939.862 & 	LT IO:O	 & 	$(r'-i')$	 & 	$	-0.209	\pm	0.207	$ \\
2014 Oct 10.081 & 6.381 & 940.070 & 940.092 & 	LT IO:O	 & 	$(r'-i')$	 & 	$	1.280	\pm	0.222	$ \\

\end{longtable}
\end{longtab}

\begin{longtab}
\begin{longtable}{llllllll}
\caption{Ultraviolet photometry of the 2014 eruption of \object{M31N 2008-12a}.\label{uv_tab}}\\
\hline\hline
Date & {$t-t_{\mathrm{max}}$} & \multicolumn{2}{c}{JD 2\,456\,000.5+} & Telescope \& & Exposure & Filter & Photometry \\
(UT)   & {(days)} & Start & End & Instrument & Time (ks)\\
\hline
\endfirsthead
\caption{Ultraviolet photometry of the 2014 eruption of \object{M31N 2008-12a}, continued.}\\
\hline\hline
Date & {$t-t_{\mathrm{max}}$} & \multicolumn{2}{c}{JD 2\,456\,000.5+} & Telescope \& & Exposure & Filter & Photometry \\
(UT)   & {(days)} & Start & End & Instrument & Time (ks)\\
\hline
\endhead
\hline
\endfoot
2014 Oct 3.643 & -0.058 & 933.633 & 933.652 & {\it Swift} UVOT & 1.6 & uvw2 & $17.0\pm0.1$ \\
2014 Oct 3.704 & 0.004 & 933.688 & 933.719 & {\it Swift} UVOT & 2.5 & uvw2 & $17.0\pm0.1$ \\
2014 Oct 3.769 & 0.069 & 933.754 & 933.785 & {\it Swift} UVOT & 2.5 & uvw2 & $17.3\pm0.1$ \\
2014 Oct 3.832 & 0.132 & 933.824 & 933.840 & {\it Swift} UVOT & 1.5 & uvw2 & $17.2\pm0.1$ \\

2014 Oct 7.356 & 3.656 & 937.356 & 937.356 & {\it Swift} UVOT & 0.1 & uvw2 & $>18.8$ \\
2014 Oct 7.369 & 3.669 & 937.363 & 937.375 & {\it Swift} UVOT & 0.8 & uvw2 & $>20.2$ \\
2014 Oct 7.432 & 3.731 & 937.422 & 937.441 & {\it Swift} UVOT & 1.7 & uvw2 & $19.6\pm0.2$ \\
2014 Oct 7.498 & 3.798 & 937.488 & 937.508 & {\it Swift} UVOT & 1.7 & uvw2 & $19.0\pm0.1$ \\
2014 Oct 7.572 & 3.872 & 937.566 & 937.578 & {\it Swift} UVOT & 1.0 & uvw2 & $19.6\pm0.2$ \\
2014 Oct 7.635 & 3.935 & 937.625 & 937.645 & {\it Swift} UVOT & 1.6 & uvw2 & $19.9\pm0.2$ \\
2014 Oct 7.842 & 4.142 & 937.840 & 937.844 & {\it Swift} UVOT & 0.4 & uvw2 & $19.4\pm0.3$ \\

2014 Oct 10.993 & 7.293 & 940.352 & 941.633 & {\it Swift} UVOT & 6.5 & uvw2 & $20.6\pm0.2$ \\

2014 Oct 15.588 & 11.888 & 945.418 & 945.758 & {\it Swift} UVOT & 1.6 & uvw2 & $>20.6$ \\

2014 Oct 19.643 & 15.943 & 949.309 & 949.977 & {\it Swift} UVOT & 4.1 & uvw2 & $>21.2$ \\

\hline

2014 Oct 4.445 & 0.745 & 934.441 & 934.449 & {\it Swift} UVOT & 0.7 & uvm2 & $18.1\pm0.1$ \\
2014 Oct 4.506 & 0.806 & 934.496 & 934.516 & {\it Swift} UVOT & 1.8 & uvm2 & $18.1\pm0.1$ \\
2014 Oct 4.572 & 0.872 & 934.563 & 934.582 & {\it Swift} UVOT & 1.7 & uvm2 & $18.2\pm0.1$ \\
2014 Oct 4.635 & 0.934 & 934.621 & 934.648 & {\it Swift} UVOT & 2.5 & uvm2 & $18.4\pm0.1$ \\
2014 Oct 4.693 & 0.993 & 934.688 & 934.699 & {\it Swift} UVOT & 0.9 & uvm2 & $18.3\pm0.1$ \\
2014 Oct 4.713 & 1.013 & 934.711 & 934.715 & {\it Swift} UVOT & 0.3 & uvm2 & $18.6\pm0.2$ \\

2014 Oct 8.293 & 4.593 & 938.289 & 938.297 & {\it Swift} UVOT & 0.9 & uvm2 & $19.7\pm0.3$ \\
2014 Oct 8.361 & 4.661 & 938.356 & 938.367 & {\it Swift} UVOT & 1.0 & uvm2 & $>20.0$ \\
2014 Oct 8.430 & 4.730 & 938.426 & 938.434 & {\it Swift} UVOT & 1.0 & uvm2 & $19.8\pm0.3$ \\
2014 Oct 8.496 & 4.796 & 938.488 & 938.504 & {\it Swift} UVOT & 1.5 & uvm2 & $20.1\pm0.3$ \\
2014 Oct 8.561 & 4.861 & 938.555 & 938.566 & {\it Swift} UVOT & 1.0 & uvm2 & $>20.0$ \\
2014 Oct 8.625 & 4.925 & 938.621 & 938.629 & {\it Swift} UVOT & 0.6 & uvm2 & $>19.7$ \\
2014 Oct 8.690 & 4.990 & 938.688 & 938.693 & {\it Swift} UVOT & 0.6 & uvm2 & $>19.7$ \\
2014 Oct 8.955 & 5.255 & 938.953 & 938.957 & {\it Swift} UVOT & 0.4 & uvm2 & $>19.4$ \\

2014 Oct 12.804 & 9.104 & 942.043 & 943.566 & {\it Swift} UVOT & 7.7 & uvm2 & $21.1\pm0.4$ \\

2014 Oct 20.465 & 16.765 & 950.035 & 950.895 & {\it Swift} UVOT & 4.7 & uvm2 & $21.0\pm0.4$ \\

\hline 

2014 Oct 5.160 & 1.460 & 935.156 & 935.164 & {\it Swift} UVOT & 0.9 & uvw1 & $19.1\pm0.2$ \\
2014 Oct 5.307 & 1.607 & 935.305 & 935.309 & {\it Swift} UVOT & 0.3 & uvw1 & $19.0\pm0.3$ \\
2014 Oct 5.627 & 1.927 & 935.621 & 935.633 & {\it Swift} UVOT & 1.0 & uvw1 & $19.4\pm0.2$ \\
2014 Oct 5.693 & 1.993 & 935.688 & 935.699 & {\it Swift} UVOT & 1.1 & uvw1 & $19.5\pm0.3$ \\
2014 Oct 5.828 & 2.128 & 935.824 & 935.832 & {\it Swift} UVOT & 0.6 & uvw1 & $>19.6$ \\

2014 Oct 9.461 & 5.760 & 939.023 & 939.898 & {\it Swift} UVOT & 4.4 & uvw1 & $20.0\pm0.2$ \\

2014 Oct 12.804 & 9.104 & 942.043 & 943.566 & {\it Swift} UVOT & 7.7 & uvw1 & $>20.1$ \\

2014 Oct 13.601 & 9.901 & 943.566 & 943.637 & {\it Swift} UVOT & 1.8 & uvw1 & $>20.4$ \\

2014 Oct 17.502 & 13.802 & 947.363 & 947.641 & {\it Swift} UVOT & 4.8 & uvw1 & $>20.9$ \\

2014 Oct 21.623 & 17.923 & 951.422 & 951.824 & {\it Swift} UVOT & 5.8 & uvw1 & $>21.1$ \\

\hline

2014 Oct 3.484 & -0.216 & \multicolumn{2}{c}{933.484} & Kiso 1.05m & 0.3 & $U$ & $17.98\pm0.20$ \\
2014 Oct 3.575 & -0.125 & \multicolumn{2}{c}{933.575} & Kiso 1.05m & 1.5 & $U$ & $17.66\pm0.16$ \\

2014 Oct 6.023 & 2.323 & 936.023 & 936.023 & {\it Swift} UVOT & 0.02 & $U$ & $>17.9$ \\
2014 Oct 6.500 & 2.800 & 936.500 & 936.500 & {\it Swift} UVOT & 0.1 & $U$ & $18.7\pm0.3$ \\
2014 Oct 8.402 & 4.702 & 939.895 & 936.910 & {\it Swift} UVOT & 1.2 & $U$ & $19.2\pm0.2$ \\
2014 Oct 6.955 & 3.255 & 936.953 & 936.957 & {\it Swift} UVOT & 0.4 & $U$ & $19.2\pm0.2$ \\

2014 Oct 10.356 & 6.655 & 940.086 & 940.625 & {\it Swift} UVOT & 4.4 & $U$ & $20.7\pm0.3$ \\

2014 Oct 10.993 & 7.293 & 940.352 & 941.633 & {\it Swift} UVOT & 6.5 & $U$ & $20.0\pm0.3$ \\

2014 Oct 14.304 & 10.604 & 944.039 & 944.570 & {\it Swift} UVOT & 5.3 & $U$ & $>21.0$ \\

2014 Oct 18.865 & 15.165 & 948.754 & 948.977 & {\it Swift} UVOT & 2.2 & $U$ & $>20.5$ \\

2014 Oct 22.504 & 18.804 & 952.031 & 952.977 & {\it Swift} UVOT & 6.1 & $U$ & $>21.0$ \\

\end{longtable}
\end{longtab}


\begin{thebibliography}{}


\bibitem[Anupama(2008)]{2008ASPC..401...31A} Anupama, G.~C.\ 2008 in ASP Conf. Ser. 401, RS Ophiuchi (2006) and the Recurrent Nova Phenomenon, ed. A. Evans, M. F. Bode, T. J. O'Brien, \& M. J. Darnley (San Francisco, CA: ASP), 31 

\bibitem[Anupama et 
al.(2013)]{2013A&A...559A.121A} Anupama, G.~C., Kamath, U.~S., Ramaprakash, A.~N., et al.\ 2013, \aap, 559, A121 

\bibitem[Azimlu et al.(2011)]{2011AJ....142..139A} Azimlu, M., Marciniak, 
R., \& Barmby, P.\ 2011, \aj, 142, 139 

\bibitem[Barsukova et al.(2011)]{2011ATel.3725....1B} Barsukova, E., 
Fabrika, S., Hornoch, K., et al.\ 2011, ATel, 3725 

\bibitem[Bateson 
\& Hull(1979)]{1979IAUC.3341....1B} Bateson, F.~M., \& Hull, O.\ 1979, \iaucirc, 3341, 1 

\bibitem[Bath 
\& Harkness(1989)]{1989clno.conf...61B} Bath, G.~T., \& Harkness, R.~P.\ 1989, in Classical Novae,~Edited by M.F.~Bode and
A.~Evans,~(Chichester, UK: Wiley), 61 

\bibitem[Bertin 
\& Arnouts(1996)]{1996A&AS..117..393B} Bertin, E., \& Arnouts, S.\ 1996, \aaps, 117, 393 

\bibitem[Bessell(1979)]{1979PASP...91..589B} Bessell, M.~S.\ 1979, \pasp, 
91, 589 

\bibitem[Bessell(1990)]{1990PASP..102.1181B} Bessell, M.~S.\ 1990, \pasp, 
102, 1181 

\bibitem[Bode(2010)]{2010AN....331..160B} Bode, M.~F.\ 2010, AN, 331, 160 

\bibitem[Bode 
\& Kahn(1985)]{1985MNRAS.217..205B} Bode, M.~F., \& Kahn, F.~D.\ 1985, \mnras, 217, 205 

\bibitem[Bode et al.(1987)]{1987Natur.329..519B} Bode, M.~F., Roberts, 
J.~A., Whittet, D.~C.~B., Seaquist, E.~R., 
\& Frail, D.~A.\ 1987, \nat, 329, 519 

\bibitem[Bode 
\& Evans(2008)]{2008clno.book.....B} Bode, M.~F., \& Evans, A.\ 2008,
Classical Novae, 2nd Edition.~Edited by M.F.~Bode and
A.~Evans.~Cambridge Astrophysics Series, No.~43, Cambridge
University Press  

\bibitem[Bode et al.(2006)]{2006ApJ...652..629B} Bode, M.~F., O'Brien, 
T.~J., Osborne, J.~P., et al.\ 2006, \apj, 652, 629 

\bibitem[Bode et al.(2009)]{2009ApJ...705.1056B} Bode, M.~F., Darnley, 
M.~J., Shafter, A.~W., et al.\ 2009, \apj, 705, 1056 

\bibitem[Breeveld(2010)]{2010Bre} Breeveld, A.\ 2010, SWIFT-UVOT-CALDB-16-R01 

\bibitem[Busko 
\& Steiner(1990)]{1990MNRAS.245..470B} Busko, I.~C., \& Steiner, J.~E.\ 1990, \mnras, 245, 470 

\bibitem[Cao et al.(2012)]{2012ApJ...752..133C} Cao, Y., Kasliwal, M.~M., 
Neill, J.~D., et al.\ 2012, \apj, 752, 133  

\bibitem[Cao et al.(2014)]{2014ATel.6532....1C} Cao, Y., Prince, T., 
Kulkarni, S.~R., et al.\ 2014, ATel, 6532 

\bibitem[Cardelli et al.(1989)]{1989ApJ...345..245C} Cardelli, J.~A., 
Clayton, G.~C., \& Mathis, J.~S.\ 1989, \apj, 345, 245  

\bibitem[Campins et al.(1985)]{1985AJ.....90..896C} Campins, H., Rieke, 
G.~H., \& Lebofsky, M.~J.\ 1985, \aj, 90, 896  

\bibitem[Chandrasekhar(1931)]{1931ApJ....74...81C} Chandrasekhar, S.\ 1931, 
\apj, 74, 81 

\bibitem[Ciardullo et al.(1987)]{1987ApJ...318..520C} Ciardullo, R., Ford, 
H.~C., Neill, J.~D., Jacoby, G.~H., \& Shafter, A.~W.\ 1987, \apj, 318, 520 

\bibitem[Coelho et al.(2008)]{2008ApJ...686.1261C} Coelho, E.~A., Shafter, 
A.~W., \& Misselt, K.~A.\ 2008, \apj, 686, 1261 

\bibitem[Cohen(1988)]{1988igbo.conf..448C} Cohen, J.~G.\ 1988, 
Instrumentation for Ground-Based Optical Astronomy, 448 

\bibitem[Contini 
\& Prialnik(1997)]{1997ApJ...475..803C} Contini, M., \& Prialnik, D.\ 1997, \apj, 475, 803 

\bibitem[Darnley et al.(2006)]{2006MNRAS.369..257D} Darnley, M.~J., Bode, 
M.~F., Kerins, E., et al.\ 2006, \mnras, 369, 257 

\bibitem[Darnley et al.(2007)]{2007ApJ...661L..45D} Darnley, M.~J., Kerins, 
E., Newsam, A., et al.\ 2007, \apjl, 661, L45 

\bibitem[Darnley et 
al.(2011)]{2011A&A...530A..70D} Darnley, M.~J., Ribeiro, V.~A.~R.~M., Bode, M.~F., \& Munari, U.\ 2011, \aap, 530, A70 

\bibitem[Darnley et al.(2012)]{2012ApJ...746...61D} Darnley, M.~J., 
Ribeiro, V.~A.~R.~M., Bode, M.~F., Hounsell, R.~A., 
\& Williams, R.~P.\ 2012, \apj, 746, 61 

\bibitem[Darnley et 
al.(2014a)]{2014A&A...563L...9D} Darnley, M.~J., Williams, S.~C., Bode, M.~F., et al.\ 2014a, \aap, 563, L9 (DWB14)  

\bibitem[Darnley et al.(2014b)]{2014ATel.6527....1D} Darnley, M.~J., 
Williams, S.~C., Bode, M.~F., et al.\ 2014b, ATel, 
6527 

\bibitem[Darnley et al.(2014c)]{2014ATel.6535....1D} Darnley, M.~J., 
Williams, S.~C., Bode, M.~F., et al.\ 2014c, ATel, 
6535 

\bibitem[Darnley et al.(2014d)]{2014ATel.6540....1D} Darnley, M.~J., Steele, 
I.~A., Smith, R.~J., et al.\ 2014d, ATel, 6540 

\bibitem[Das et al.(2006)]{2006ApJ...653L.141D} Das, R., Banerjee, 
D.~P.~K., \& Ashok, N.~M.\ 2006, \apjl, 653, L141 

\bibitem[Della Valle et 
al.(2002)]{2002A&A...390..155D} Della Valle, M., Pasquini, L., Daou, D., \& Williams, R.~E.\ 2002, \aap, 390, 155 

\bibitem[Disney 
\& Wallace(1982)]{1982QJRAS..23..485D} Disney, M.~J., \& Wallace, P.~T.\ 1982, \qjras, 23, 485 

\bibitem[Dodorico et 
al.(1980)]{1980A&AS...40...67D} Dodorico, S., Dopita, M.~A., \& Benvenuti, P.\ 1980, \aaps, 40, 67 

\bibitem[Downes 
\& Duerbeck(2000)]{2000AJ....120.2007D} Downes, R.~A., \& Duerbeck, H.~W.\ 2000, \aj, 120, 2007 

\bibitem[Dufay et al.(1964)]{1964AnAp...27..555D} Dufay, J., Bloch, M., 
Bertaud, C., \& Dufay, M.\ 1964, Annales d'Astrophysique, 27, 555 

\bibitem[Evans et al.(2008)]{2008ASPC..401.....E} Evans, A., Bode, M.~F., 
O'Brien, T.~J., 
\& Darnley, M.~J.\ (ed.) 2008, ASP Conf. Ser. 401, RS Ophiuchi (2006) and the Recurrent Nova Phenomenon (San
Francisco, CA: ASP) 

\bibitem[Franck et al.(2012)]{2012ApJ...760...13F} Franck, J.~R., Shafter, 
A.~W., Hornoch, K., \& Misselt, K.~A.\ 2012, \apj, 760, 13 

\bibitem[Freedman 
\& Madore(1990)]{1990ApJ...365..186F} Freedman, W.~L., \& Madore,
B.~F.\ 1990, \apj, 365, 186 

\bibitem[Gallagher 
\& Ney(1976)]{1976ApJ...204L..35G} Gallagher, J.~S., \& Ney, E.~P.\ 1976, \apjl, 204, L35 

\bibitem[Gehrels et al.(2004)]{2004ApJ...611.1005G} Gehrels, N., 
Chincarini, G., Giommi, P., et al.\ 2004, \apj, 611, 1005 

\bibitem[Gehrz(2008)]{2008clno.book..167G} Gehrz, R.~D.\ 2008, in Classical Novae, 2nd Edition.~Edited by M.F.~Bode and A.~Evans.~Cambridge Astrophysics Series, No.~43, Cambridge: Cambridge University Press, 167 

\bibitem[Gill 
\& O'Brien(1998)]{1998MNRAS.300..221G} Gill, C.~D., \& O'Brien, T.~J.\ 1998, \mnras, 300, 221 

\bibitem[Gorbatskii(1972)]{1972SvA....16...32G} Gorbatskii, V.~G.\ 1972, 
\sovast, 16, 32 

\bibitem[Gutierrez et al.(1996)]{1996ApJ...459..701G} Gutierrez, J., 
Garcia-Berro, E., Iben, I., Jr., et al.\ 1996, \apj, 459, 701 

\bibitem[Hachisu et al.(1999a)]{1999ApJ...519..314H} Hachisu, I., Kato, M., 
Nomoto, K., \& Umeda, H.\ 1999a, \apj, 519, 314 

\bibitem[Hachisu et al.(1999b)]{1999ApJ...522..487H} Hachisu, I., Kato, M., 
\& Nomoto, K.\ 1999b, \apj, 522, 487 

\bibitem[Hachisu et al.(2000)]{2000ApJ...534L.189H} Hachisu, I., Kato, M., 
Kato, T., Matsumoto, K., \& Nomoto, K.\ 2000, \apjl, 534, L189 

\bibitem[Hachisu 
\& Kato(2006)]{2006ApJS..167...59H} Hachisu, I., \& Kato, M.\ 2006, \apjs, 167, 59 

\bibitem[Hachisu et al.(2008)]{2008ASPC..401..206H} Hachisu, I., Kato, M., 
Kiyota, S., et al.\ 2008 in ASP Conf. Ser. 401, RS Ophiuchi (2006) and the Recurrent Nova Phenomenon, ed. A. Evans, M. F. Bode, T. J. O'Brien, \& M. J. Darnley (San Francisco, CA: ASP), 206 

\bibitem[Hachisu 
\& Kato(2015)]{2015ApJ...798...76H} Hachisu, I., \& Kato, M.\ 2015, \apj, 798, 76 

\bibitem[Henze et 
al.(2014)]{2014A&A...563A...2H} Henze, M., Pietsch, W., Haberl, F., et al.\ 2014, \aap, 563, A2 

\bibitem[Henze et 
al.(2014)]{2014A&A...563L...8H} Henze, M., Ness, J.-U., Darnley, M.~J., et al.\ 2014, \aap, 563, L8 (HND14)  

\bibitem[Henze et 
al.(2015)]{2015Hen} Henze, M., Ness, J.-U., Darnley, M.~J., et al.\ 2015, \aap, in press, arXiv:1504.06237 (HND15)  

\bibitem[Horne(1986)]{1986PASP...98..609H} Horne, K.\ 1986, \pasp, 98, 609 

\bibitem[Hornoch et al.(2014)]{2014ATel.6546....1H} Hornoch, K., Kucakova, 
H., \& Wolf, M.\ 2014, ATel, 6546 

\bibitem[Hornoch 
\& Shafter(2015)]{2015ATel.7116....1H} Hornoch, K., \& Shafter, A.~W.\ 2015, ATel, 7116 

\bibitem[Hornoch et al.(2015)]{2015ATel.7142....1H} Hornoch, K., Kucakova, 
H., Vrastil, J., et al.\ 2015, ATel, 7142 

\bibitem[Hounsell et al.(2010)]{2010ApJ...724..480H} Hounsell, R., Bode, 
M.~F., Hick, P.~P., et al.\ 2010, \apj, 724, 480 

\bibitem[Hounsell et al.(2015)]{hou15} Hounsell, R., Darnley, M.~J., Bode, 
M.~F., et al.\ 2015, \apj, submitted 

\bibitem[Iijima(2009)]{2009A&A...505..287I} Iijima, T.\ 2009, \aap, 505, 287 

\bibitem[Jester et al.(2005)]{2005AJ....130..873J} Jester, S., Schneider, 
D.~P., Richards, G.~T., et al.\ 2005, \aj, 130, 873 

\bibitem[Johnson et al.(2012)]{2012ApJ...752...95J} Johnson, L.~C., Seth, 
A.~C., Dalcanton, J.~J., et al.\ 2012, \apj, 752, 95 

\bibitem[Jurdana-{\v S}epi{\'c} et 
al.(2012)]{2012A&A...537A..34J} Jurdana-{\v S}epi{\'c}, R., Ribeiro, V.~A.~R.~M., Darnley, M.~J., Munari, U., \& Bode, M.~F.\ 2012, \aap, 537, A34 

\bibitem[Kato et al.(2014)]{2014ApJ...793..136K} Kato, M., Saio, H., 
Hachisu, I., \& Nomoto, K.\ 2014, \apj, 793, 136 

\bibitem[Kerins et al.(2010)]{2010MNRAS.409..247K} Kerins, E., Darnley, 
M.~J., Duke, J.~P., et al.\ 2010, \mnras, 409, 247 

\bibitem[Kiss et 
al.(2001)]{2001A&A...366..858K} Kiss, L.~L., Thomson, J.~R., Ogloza, W., Fur{\'e}sz, G., \& Szil{\'a}di, K.\ 2001, \aap, 366, 858 

\bibitem[Korotkiy \& Elenin(2011)]{2011Kor} Korotkiy, S., \& Elenin, L.\ 2010, CBAT, \url{http://www.cbat.eps.harvard.edu/unconf/followups/J00452885+4154094.html} 

\bibitem[Kraft(1964)]{1964ApJ...139..457K} Kraft, R.~P.\ 1964, \apj, 139, 
457 

\bibitem[Krautter et al.(1996)]{1996ApJ...456..788K} Krautter, J., 
Oegelman, H., Starrfield, S., Wichmann, R., 
\& Pfeffermann, E.\ 1996, \apj, 456, 788 

\bibitem[Lasker(1977)]{1977ApJ...212..390L} Lasker, B.~M.\ 1977, \apj, 212, 
390 

\bibitem[Massey et al.(2006)]{2006AJ....131.2478M} Massey, P., Olsen, 
K.~A.~G., Hodge, P.~W., et al.\ 2006, \aj, 131, 2478 
              
\bibitem[Massey et al.(2007)]{2007AJ....134.2474M} Massey, P., McNeill, 
R.~T., Olsen, K.~A.~G., et al.\ 2007, \aj, 134, 2474 
      
\bibitem[Mathewson 
\& Clarke(1973)]{1973ApJ...180..725M} Mathewson, D.~S., \& Clarke, J.~N.\ 1973, \apj, 180, 725 
      
\bibitem[Matonick 
\& Fesen(1997)]{1997ApJS..112...49M} Matonick, D.~M., \& Fesen, R.~A.\ 1997, \apjs, 112, 49  

\bibitem[McConnachie(2012)]{2012AJ....144....4M} McConnachie, A.~W.\ 2012, 
\aj, 144, 4 
              
\bibitem[Merrifield(1992)]{1992AJ....103.1552M} Merrifield, M.~R.\ 1992, 
\aj, 103, 1552 
                                
\bibitem[Montalto et 
al.(2009)]{2009A&A...507..283M} Montalto, M., Seitz, S., Riffeser, A.,
et al.\ 2009, \aap, 507, 283 

\bibitem[Munari et 
al.(1999)]{1999A&A...347L..39M} Munari, U., Zwitter, T., Tomov, T., et al.\ 1999, \aap, 347, L39 

\bibitem[Munari et al.(2011)]{2011NewA...16..209M} Munari, U., Siviero, A., 
Dallaporta, S., et al.\ 2011, \na, 16, 209 

\bibitem[Nishiyama \& Kabashima(2008)]{2008Nis} Nishiyama, K., \& Kabashima, F.\ 2008, CBAT, \url{http://www.cbat.eps.harvard.edu/iau/CBAT_M31.html#2008-12a} 

\bibitem[Nishiyama \& Kabashima(2012)]{2012Nis} Nishiyama, K., \& Kabashima, F.\ 2012, CBAT, \url{http://www.cbat.eps.harvard.edu/unconf/followups/J00452884+4154095.html} 

\bibitem[O'Brien et al.(2006)]{2006Natur.442..279O} O'Brien, T.~J., Bode, 
M.~F., Porcas, R.~W., et al.\ 2006, \nat, 442, 279 

\bibitem[Orlando et 
al.(2009)]{2009A&A...493.1049O} Orlando, S., Drake, J.~J., \& Laming, J.~M.\ 2009, \aap, 493, 1049 

\bibitem[Osterbrock 
\& Dufour(1973)]{1973ApJ...185..441O} Osterbrock, D.~E., \& Dufour, R.~J.\ 1973, \apj, 185, 441 

\bibitem[Overbeek et al.(1987)]{1987IAUC.4395....1O} Overbeek, D., 
McNaught, R.~H., Whitelock, P., Cragg, T., 
\& Verdenet, M.\ 1987, \iaucirc, 4395, 1 

\bibitem[Osborne et al.(2011)]{2011ApJ...727..124O} Osborne, J.~P., Page, 
K.~L., Beardmore, A.~P., et al.\ 2011, \apj, 727, 124 

\bibitem[Pagnotta et al.(2009)]{2009AJ....138.1230P} Pagnotta, A., 
Schaefer, B.~E., Xiao, L., Collazzi, A.~C., 
\& Kroll, P.\ 2009, \aj, 138, 1230 

\bibitem[Pagnotta 
\& Schaefer(2014)]{2014ApJ...788..164P} Pagnotta, A., \& Schaefer, B.~E.\ 2014, \apj, 788, 164 

\bibitem[Piascik et al.(2014)]{2014SPIE.9147E..8HP} Piascik, A.~S., Steele, 
I.~A., Bates, S.~D., et al.\ 2014, \procspie, 9147, 91478H  

\bibitem[Pietsch et 
al.(2007)]{2007A&A...465..375P} Pietsch, W., Haberl, F., Sala, G., et
al.\ 2007, \aap, 465, 375 

\bibitem[Pietsch et 
al.(2011)]{2011A&A...531A..22P} Pietsch, W., Henze, M., Haberl, F., et al.\ 2011, \aap, 531, A22 

\bibitem[Poole et al.(2008)]{2008MNRAS.383..627P} Poole, T.~S., Breeveld, 
A.~A., Page, M.~J., et al.\ 2008, \mnras, 383, 627 

\bibitem[Porter et al.(1998)]{1998MNRAS.296..943P} Porter, J.~M., O'Brien, 
T.~J., \& Bode, M.~F.\ 1998, \mnras, 296, 943 

\bibitem[Pottasch(1967)]{1967BAN....19..227P} Pottasch, S.~R.\ 1967, \bain, 
19, 227 

\bibitem[Pravec et al.(1994)]{1994ExA.....5..375P} Pravec, P., Hudec, R., 
Sold{\'a}n, J., Sommer, M., 
\& Schenkl, K.~H.\ 1994, Experimental Astronomy, 5, 375 

\bibitem[Prialnik 
\& Kovetz(1995)]{1995ApJ...445..789P} Prialnik, D., \& Kovetz, A.\ 1995, \apj, 445, 789 

\bibitem[Ribeiro et al.(2009)]{2009ApJ...703.1955R} Ribeiro, V.~A.~R.~M., 
Bode, M.~F., Darnley, M.~J., et al.\ 2009, \apj, 703, 1955 

\bibitem[Ribeiro et al.(2011)]{2011MNRAS.412.1701R} Ribeiro, V.~A.~R.~M., 
Darnley, M.~J., Bode, M.~F., et al.\ 2011, \mnras, 412, 1701 

\bibitem[Ribeiro et al.(2013a)]{2013ApJ...768...49R} Ribeiro, V.~A.~R.~M., 
Munari, U., \& Valisa, P.\ 2013a, \apj, 768, 49 

\bibitem[Ribeiro et al.(2013b)]{2013MNRAS.433.1991R} Ribeiro, V.~A.~R.~M., 
Bode, M.~F., Darnley, M.~J., et al.\ 2013b, \mnras, 433, 1991 

\bibitem[Roming et al.(2005)]{2005SSRv..120...95R} Roming, P.~W.~A., 
Kennedy, T.~E., Mason, K.~O., et al.\ 2005, \ssr, 120, 95 

\bibitem[Rosino(1987)]{1987rorn.conf....1R} Rosino, L.\ 1987 in RS Ophiuchi 
(1985) and the Recurrent Nova Phenomenon, ed. M. F. Bode (Utrecht: VNU Science Press), 1 

\bibitem[Roy et al.(2012)]{2012MNRAS.427L..55R} Roy, N., Kantharia, N.~G., 
Eyres, S.~P.~S., et al.\ 2012, \mnras, 427, L55

\bibitem[Sako et al.(2012)]{2012SPIE.8446E..6LS} Sako, S., Aoki, T., Doi, 
M., et al.\ 2012, \procspie, 8446, 84466L 

\bibitem[Schaefer(2010)]{2010ApJS..187..275S} Schaefer, B.~E.\ 2010, \apjs, 
187, 275  

\bibitem[Schaefer et al.(2013)]{2013ApJ...773...55S} Schaefer, B.~E., 
Landolt, A.~U., Linnolt, M., et al.\ 2013, \apj, 773, 55 

\bibitem[Seaquist et al.(1989)]{1989ApJ...344..805S} Seaquist, E.~R., Bode, 
M.~F., Frail, D.~A., et al.\ 1989, \apj, 344, 805

\bibitem[Sekiguchi et al.(1990)]{1990MNRAS.246...78S} Sekiguchi, K., 
Whitelock, P.~A., Feast, M.~W., et al.\ 1990, \mnras, 246, 78 

\bibitem[Shafter(1997)]{1997ApJ...487..226S} Shafter, A.~W.\ 1997, \apj, 
487, 226 

\bibitem[Shafter et al.(2009)]{2009ApJ...690.1148S} Shafter, A.~W., Rau, 
A., Quimby, R.~M., et al.\ 2009, \apj, 690, 1148 

\bibitem[Shafter et al.(2011)]{2011ApJ...734...12S} Shafter, A.~W., 
Darnley, M.~J., Hornoch, K., et al.\ 2011, \apj, 734, 12 

\bibitem[Shafter et al.(2012)]{2012ATel.4503....1S} Shafter, A.~W., 
Hornoch, K., Ciardullo, J.~V.~R., Darnley, M.~J., 
\& Bode, M.~F.\ 2012, ATel, 4503 

\bibitem[Shafter 
\& Darnley(2014)]{2014ATel.6543....1S} Shafter, A.~W., \& Darnley, M.~J.\ 2014, ATel, 6543                    

\bibitem[Shafter et al.(2015)]{2015Sha} Shafter, A.~W., Henze, M., Rector, T.~A., et al.\ 2015, \apjs, 216, 34  

\bibitem[Shara et al.(1997)]{1997AJ....114..258S} Shara, M.~M., Zurek, 
D.~R., Williams, R.~E., et al.\ 1997, \aj, 114, 258 

\bibitem[Shore et al.(1996)]{1996ApJ...456..717S} Shore, S.~N., Kenyon, 
S.~J., Starrfield, S., \& Sonneborn, G.\ 1996, \apj, 456, 717 

\bibitem[Shore et 
al.(2013a)]{2013A&A...549A.140S} Shore, S.~N., Schwarz, G.~J., De Gennaro Aquino, I., et al.\ 2013a, \aap, 549, A140 

\bibitem[Shore et 
al.(2013b)]{2013A&A...553A.123S} Shore, S.~N., De Gennaro Aquino, I., Schwarz, G.~J., et al.\ 2013b, \aap, 553, A123 

\bibitem[Slavin et al.(1995)]{1995MNRAS.276..353S} Slavin, A.~J., O'Brien, 
T.~J., \& Dunlop, J.~S.\ 1995, \mnras, 276, 353 

\bibitem[Stark et al.(1992)]{1992ApJS...79...77S} Stark, A.~A., Gammie, 
C.~F., Wilson, R.~W., et al.\ 1992, \apjs, 79, 77 

\bibitem[Starrfield et al.(1974)]{1974ApJS...28..247S} Starrfield, S., 
Sparks, W.~M., \& Truran, J.~W.\ 1974, \apjs, 28, 247 

\bibitem[Starrfield et al.(1976)]{1976IAUS...73..155S} Starrfield, S., 
Sparks, W.~M., 
\& Truran, J.~W.\ 1976 in IAU Symposium No. 73, Structure and Evolution of Close Binary Systems, ed. P. Eggleton, S. Mitton \& J. Whelan, (Dordrecht: D. Reidel Publishing Co.),155 

\bibitem[Starrfield et al.(2008)]{2008clno.book...77S} Starrfield, S., Iliadis, C., \& Hix, W.~R.\ 2008, in Classical Novae, 2nd Edition.~Edited by M.F.~Bode and A.~Evans.~Cambridge Astrophysics Series, No.~43, Cambridge: Cambridge University Press, 77 

\bibitem[Steele et al.(2004)]{2004SPIE.5489..679S} Steele, I.~A., Smith, 
R.~J., Rees, P.~C., et al.\ 2004, \procspie, 5489, 679 

\bibitem[Steffen et al.(2011)]{2011ITVCG..17..454S} Steffen, W., Koning, 
N., Wenger, S., Morisset, C., 
\& Magnor, M.\ 2011, IEEE Transactions on Visualization and Computer Graphics, Volume 17, Issue 4, 454 

\bibitem[Stil 
\& Irwin(2001)]{2001ApJ...563..816S} Stil, J.~M., \& Irwin, J.~A.\ 2001, \apj, 563, 816 

\bibitem[Stone(1977)]{1977ApJ...218..767S} Stone, R.~P.~S.\ 1977, \apj, 
218, 767 

\bibitem[Surina et al.(2014)]{2014AJ....147..107S} Surina, F., Hounsell, 
R.~A., Bode, M.~F., et al.\ 2014, \aj, 147, 107 

\bibitem[Tang et al.(2013)]{2013ATel.5607....1T} Tang, S., Cao, Y., 
\& Kasliwal, M.~M.\ 2013, ATel, 5607 

\bibitem[Tang et al.(2014)]{2014ApJ...786...61T} Tang, S., Bildsten, L., 
Wolf, W.~M., et al.\ 2014, \apj, 786, 61 (TBW14) 

\bibitem[Thoroughgood et al.(2001)]{2001MNRAS.327.1323T} Thoroughgood, 
T.~D., Dhillon, V.~S., Littlefair, S.~P., Marsh, T.~R., 
\& Smith, D.~A.\ 2001, \mnras, 327, 1323 

\bibitem[Tody(1993)]{1993ASPC...52..173T} Tody, D.\ 1993, Astronomical Data 
Analysis Software and Systems II, 52, 173 

\bibitem[Toraskar et al.(2013)]{2013ApJ...768...48T} Toraskar, J., Mac Low, 
M.-M., Shara, M.~M., \& Zurek, D.~R.\ 2013, \apj, 768, 48 

\bibitem[van den Bergh(1975)]{1975PASP...87..405V} van den Bergh, S.\ 1975, 
\pasp, 87, 405 

\bibitem[Walder et 
al.(2008)]{2008A&A...484L...9W} Walder, R., Folini, D., \& Shore, S.~N.\ 2008, \aap, 484, L9 

\bibitem[Walterbos 
\& Braun(1992)]{1992A&AS...92..625W} Walterbos, R.~A.~M., \& Braun, R.\ 1992, \aaps, 92, 625 

\bibitem[Wesson et al.(2008)]{2008ApJ...688L..21W} Wesson, R., Barlow, 
M.~J., Corradi, R.~L.~M., et al.\ 2008, \apjl, 688, L21 

\bibitem[Williams(1982)]{1982ApJ...261..170W} Williams, R.~E.\ 1982, \apj, 
261, 170 

\bibitem[Williams et al.(2004)]{2004ApJ...609..735W} Williams, B.~F., 
Garcia, M.~R., Kong, A.~K.~H., et al.\ 2004, \apj, 609, 735 

\bibitem[Williams(2012)]{2012AJ....144...98W} Williams, R.\ 2012, \aj, 144, 
98 

\bibitem[Williams(2013)]{2013AJ....146...55W} Williams, R.\ 2013, \aj, 146, 
55 

\bibitem[Williams et al.(2013)]{2013ATel.5611....1W} Williams, S.~C., 
Darnley, M.~J., Bode, M.~F., 
\& Shafter, A.~W.\ 2013, ATel, 5611 

\bibitem[Williams et al.(2014)]{2014ApJS..213...10W} Williams, S.~C., 
Darnley, M.~J., Bode, M.~F., Keen, A., 
\& Shafter, A.~W.\ 2014, \apjs, 213, 10  

\bibitem[Williams et al.(2015)]{2015Wil} Williams, S.~C., 
Darnley, M.~J., Bode, M.~F., \& Shafter, A.~W.\ 2015, \apj, in prep 

\bibitem[Wolf et al.(2013)]{2013ApJ...777..136W} Wolf, W.~M., Bildsten, L., 
Brooks, J., \& Paxton, B.\ 2013, \apj, 777, 136 

\bibitem[Woudt 
\& Ribeiro(2014)]{2014ASPC..490.....W} Woudt, P.~A., \& Ribeiro, V.~A.~R.~M.\ (ed.) 2014, ASP Conf. Ser. 490, Stellar Novae: Past and Future Decades (San
Francisco, CA: ASP) 

\bibitem[Whelan 
\& Iben(1973)]{1973ApJ...186.1007W} Whelan, J., \& Iben, I., Jr.\ 1973, \apj, 186, 1007 

\bibitem[White et al.(1995)]{1995ApJ...445L.125W} White, N.~E., Giommi, P., 
Heise, J., Angelini, L., \& Fantasia, S.\ 1995, \apjl, 445, L125 

\bibitem[Wright 
\& Barlow(1975)]{1975MNRAS.170...41W} Wright, A.~E., \& Barlow, M.~J.\ 1975, \mnras, 170, 41 

\bibitem[Yaron et al.(2005)]{2005ApJ...623..398Y} Yaron, O., Prialnik, D., 
Shara, M.~M., \& Kovetz, A.\ 2005, \apj, 623, 398 

\bibitem[Zacharias et al.(2013)]{2013AJ....145...44Z} Zacharias, N., Finch, 
C.~T., Girard, T.~M., et al.\ 2013, \aj, 145, 44 

\bibitem[Zhang et al.(2014)]{2014ApJ...792...58Z} Zhang, N.-X., Chu, Y.-H., 
Williams, R.~M., et al.\ 2014, \apj, 792, 58 


\end{thebibliography}
\end{document}